\documentclass[useAMS,usenatbib]{mn2e}

\usepackage{graphicx}
\usepackage{subfigure}
\usepackage{paralist}
\usepackage{amssymb}
\usepackage{bm}
\usepackage{bbding}
\usepackage[fleqn]{amsmath}
\usepackage{multirow}
\usepackage[colorlinks,
            linkcolor=blue, 
            anchorcolor=blue, 
            citecolor=blue, 
            ]{hyperref}
\title[Secular resonances]{Quadrupole and octupole order resonances in non-restricted hierarchical planetary systems}
\author[Lei \& Huang]{Hanlun Lei$^{1,2}$\thanks{E-mail: leihl@nju.edu.cn}, Xiumin Huang$^{1}$\\
$^{1}$ School of Astronomy and Space Science, Nanjing University, Nanjing 210023, China\\
$^{2}$ Key Laboratory of Modern Astronomy and Astrophysics in Ministry of Education, Nanjing University, Nanjing 210023, China}

\begin{document}

\date{Accepted. Received; in original form}

\pagerange{\pageref{firstpage}--\pageref{lastpage}} \pubyear{2022}

\maketitle
\label{firstpage}

\begin{abstract}
Nonrestricted hierarchical three-body configurations are common in various scales of astrophysical systems. Dynamical structures of the quadrupole-order resonance (the von Zeipel--Lidov--Kozai resonance) and the octupole-order resonance (the apsidal resonance) under the nonrestricted hierarchical planetary systems are investigated in this work by taking advantage of perturbative treatments. Under the quadrupole-order Hamiltonian model, the distribution of libration and circulation regions as well as the distribution of flipping region are analytically explored in the parameter space spanned by the conserved quantities. The fundamental frequencies of system are produced and then the nominal location of octupole-order resonance is identified. From the viewpoint of perturbative theory, the quadrupole-order Hamiltonian determines the unperturbed dynamical model and the octupole-order Hamiltonian plays an role of perturbation to the quadrupole-order dynamics. The resonant Hamiltonian for octupole-order resonances is formulated by means of averaging theory, giving rise to a new constant of motion. Phase portraits are produced to analyse dynamical structures of octupole-order resonance, including resonant centres, saddle points, dynamical separatrices and islands of libration. By analysing phase portraits, it is found that there are four branches of libration centre and eight libration zones in the considered space. Applications to orbit flips show that there are five flipping regions.
\end{abstract}


\section{Introduction}
\label{Sect1}

Hierarchical three-body systems hold rich dynamical behaviours which are helpful to understand the long-term evolution and stability of astrophysical systems. In the test-particle limit (the inner object is of massless), when the outer perturber is moving around the central body on a circular orbit, the secular dynamics of test particle was studied by \citet{lidov1962evolution} for artificial satellites and \citet{kozai1962secular} for inclined asteroids. Under both the circular and test-particle assumptions, the secular potential is axisymmetric and the particle's vertical angular momentum ($H$) is conserved during the long-term evolution, meaning that orbits cannot flip between prograde and retrograde. \citet{lidov1962evolution} and \citet{kozai1962secular} independently found that there is a resonance between the longitude of pericentre and longitude of ascending node when the mutual inclination is greater than $39.2^{\circ}$ and smaller than $140.8^{\circ}$. The conservation of the vertical angular momentum leads to the coupled oscillations between eccentricity and inclination. Such a dynamical behaviour is coined the standard Lidov--Kozai (or Kozai--Lidov) mechanism. Recently, \citet{ito2019lidov} pointed out that \citet{von1910application} performed a similar analysis for the same problem, thus they suggested to refer to the usual Lidov--Kozai mechanism as ``von Zeiple--Lidov--Kozai" (ZLK) mechanism/effect. \citet{vashkov1999evolution} and \citet{kinoshita2007general} derived explicit expressions for rotating and librating ZLK cycles in terms of elliptic integrals and an infinite Fourier series expansion for the longitude of the ascending node (see also \citeauthor{sidorenko2018eccentric} \citeyear{sidorenko2018eccentric} for an updated version). For a special case where particles are initially at nearly circular orbits, \citet{lubow2021analytic} provided an analytic solution for orbital elements of particles moving along the ZLK cycles at the quadrupole-level approximation.

When the circular assumption is relaxed, the eccentric effect appears when the Hamiltonian is formulated up to the octupole order in semimajor axis ratio. In this context, the standard ZLK effect becomes the eccentric ZLK effect. In the long-term evolution, the ZLK cycles are modulated on longer timescales than the period of ZLK cycle and thus the particle's vertical angular momentum varies, resulting in striking features, including exchanging eccentricity for inclination and the generation of flipping orbit. These striking phenomena are attributed to the eccentric ZLK effect \citep{lithwick2011eccentric}.

To analytically understand the eccentric ZLK effect in the test-particle limit, \citet{katz2011long} averaged the secular equations of motion over the ZLK cycles, giving rise to a new constant of motion during the very long-term evolution. In particular, they provided analytical expressions for the conditions that produce flipping orbits. At the same time, \citet{lithwick2011eccentric} numerically mapped out the conditions where flipping orbits and extreme eccentricities take place in the parameter space with $e_0 < 0.5$ ($e_0$ is the initial eccentricity). Taking advantage of the surfaces of section and the Lyapunov exponent, \citet{li2014} investigated the chaotic and quasi-periodic orbits and they found that the octupole-order resonances can cause orbit flips, extreme eccentricities and even chaotic behaviours. \citet{li2014eccentricity} classified flipping orbits into two types: the low-eccentricity, high-inclination (LeHi) case and the high-eccentricity, low-inclination (HeLi) case. Coplanar flips corresponds to the latter case. The HeLi case is dominated by only the octupole-order resonances, whereas the LeHi case is dominated by the joint effects of quadrupole-order and octupole-order resonances \citep{li2014eccentricity}. \citet{antognini2015timescales} derived analytic expressions for timescales of the ZLK oscillations at both the quadrupole and octupole-order approximations. \citet{will2017orbital} studied the influences of the relativistic effects and third-body effects up to the hexadecapole order upon the orbital flips in hierarchical three-body systems and they found that, for most part, the orbital flips found at the octupole order are robust. From the viewpoint of perturbative treatments, \citet{sidorenko2018eccentric} interpreted the LeHi-type of orbit flips caused by the eccentric ZLK effect as a resonance phenomenon. \citet{Lei_2022} systematically studied the topic of orbit flip from three approaches: Poincar\'e surfaces of section, dynamical system theory (periodic orbits and invariant manifolds) and perturbative treatments. Besides the LeHi case and the HeLi case discussed in \citet{li2014eccentricity}, an addition region of flipping orbit is found in the intermediate-eccentricity space \citep{Lei_2022}. The essence of flipping orbit is achieved: flipping orbits are a type of quasi-periodic (or resonant) trajectories around polar periodic orbits \citep{Lei_2022}. In a recent work \citep{Lei2022dynamical}, it is concluded that the eccentric ZLK effect is dynamically equivalent to the effect of apsidal resonance at the octupole-level approximation, and they pointed out that the behaviour of orbit flip is just one type of dynamical response of the eccentric ZLK effect (or the effect of apsidal resonance).

Regarding non-restricted hierarchical three-body systems, \citet{krymolowski1999studies} and \citet{ford2000secular} presented secular equations of motion (or Hamiltonian) up to the octupole order in semimajor axis ratio by using Hamiltonian perturbation techniques. \citet{lee2003secular} adopted both the octupole-level perturbation theory and direct numerical integrations to investigate the dynamical evolution for coplanar hierarchical planetary systems. In particular, the dynamics of apsidal resonance with critical argument of $\sigma = \varpi_1 - \varpi_2$ ($\varpi_{1,2}$ are the longitudes of pericentre) is studied and applied to some representative exoplanetary systems \citep{lee2003secular}. In a hierarchical planetary system with two comparable-mass planets orbiting a central star, \citet{naoz2011hot} showed that orbits of the inner planet could flip from prograde to retrograde and back again due to the secular planet--planet interaction. Based on this behavior, it becomes possible to form hot Jupiters on retrograde orbits by combining the eccentric ZLK effect and tidal friction \citep{naoz2011hot, naoz2012formation, teyssandier2013extreme, petrovich2015hot, petrovich2016warm, dawson2018origins}. \citet{naoz2013secular} re-derived the secular evolution equations for hierarchical three-body systems at the octupole-level approximation and found that orbital flips of inner planet are possible even at the quadrupole-level approximation. They pointed out that the relation $h_1 - h_2 = \pi$ can be used to simplify the expression of Hamiltonian but the evolutions of $H_1$ and $H_2$ should be derived from the conservation of the total angular momentum rather than from the Hamiltonian canonical relations. \citet{tan2020secular} explored the secular resonances with critical arguments arising in the Hamiltonian under the resonant Hamiltonian model, which is obtained by directly removing those terms involving short-period angles from the octupole-level Hamiltonian (i.e., only the secular and resonant terms are retained). It is of no problem when dealing with the quadrupole-order resonance (the ZLK resonance) because in this case the omitting terms are of octupole order. However, it may be inadequate to formulate the resonant model by directly removing those quadrupole-order periodic terms from the Hamiltonian when studying the octupole-order resonances. \citet{hamers2021properties} performed a semianalytic study about the properties of the ZLK oscillations at the quadrupole-level approximation, including the maximum eccentricities, timescales of eccentricity/inclination oscillation and orbit flips. \citet{naoz2016eccentric} and \citet{shevchenko2016lidov} reviewed various applications of the eccentric ZLK effect to a broad range of astrophysical systems, such as planetary and exoplanetary systems, stellar systems, and galaxies.

However, the eccentric ZLK effect is far from being understood under nonrestricted hierarchial planetary systems. To this end, in this work we analytically explore the dynamical structures of both the quadrupole-order and octupole-order resonances by means of perturbation theory and then we can make clear the dynamical connection between the eccentric ZLK effect and secular resonances at the octupole-level approximation in nonrestricted hierarchical planetary systems.

Under the quadrupole-order approximation, the dynamical model is of one degree of freedom and thus it is integrable. The dynamics of the ZLK resonance are studied in detail. Especially, we obtain analytic expressions of the lower boundary, upper boundary and dynamical separatrix for libration and circulation regions of the ZLK resonance in the parameter space spanned by conserved quantities. It is known that orbit flips are possible at the quadrupole-level approximation, which is different from the restricted case (in the test-particle limit, orbit flip is impossible at the quadrupole-order approximation). Analytic expressions of boundaries for flipping region are derived. The action-angle transformation is introduced to make the quadrupole-order Hamiltonian be independent on the angular coordinates. Under such a canonical transformation, the quadrupole-order Hamiltonian determines the fundamental frequencies (or proper frequencies) of system, which can be used to identify the nominal location of octupole-order resonance. It is found that the secular resonance with critical argument of $\sigma = g_2^* - {\rm sign}(\cos{i_{\rm tot}})g_1^*$ can take place in the considered space (here $g_1$ and $g_2$ are, respectively, the arguments of pericentre for the inner and outer binaries). It is demonstrated that the secular resonance with $\sigma = g_2^* - {\rm sign}(\cos{i_{\rm tot}})g_1^*$ corresponds to the apsidal resonance in the spatial configuration. Thus, octupole-order resonances are dynamically equivalent to apsidal resonances. This is consistent with the relation in the test-particle limit, as shown by \citet{Lei2022dynamical}.

From the viewpoint of perturbative treatments \citep{henrard1986perturbation, henrard1990semi}, the quadrupole-order Hamiltonian determines the unperturbed dynamical model, and the octupole-order Hamiltonian plays an role of perturbation to the quadrupole-order dynamics. Based on this concept, we formulate the resonant Hamiltonian for octupole-order resonances by means of the lowest-order perturbation theory, yielding a new constant of motion. The phase portraits of octupole-order resonance can be used to analyse the dynamical structures, including libration centres, saddle points, dynamical separatrices and island of libration. We find that there are four branches of libration centre and eight libration zones in the considered space for octupole-order resonances. In particular, there are five libration zones that may generate flipping orbits. By analysing phase portraits, flipping regions are produced and different behaviours of flipping orbit are discussed. Analytical results of libration zones causing orbit flips are compared to the numerical distribution of flipping orbits. It is found that the analytical and numerical results are qualitatively consistent.

The remaining part of this work is organized as follows. In Section \ref{Sect2}, the Hamiltonian model is briefly introduced. The quadrupole-order dynamics are analytically discussed in Section \ref{Sect3} and the dynamics of octupole-order resonance are investigated in Section \ref{Sect4} by means of perturbative treatments. Applications to orbit flips are presented in Section \ref{Sect5}. Conclusions are summarised in Section \ref{Sect6}.

\section{Hamiltonian model}
\label{Sect2}

The hierarchical planetary system considered in this work is consisting of a central star with mass $m_0$ and two planetary objects with mass $m_1$ and $m_2$. The central star $m_0$ and the planet $m_1$ constitute the inner tight binary, and the distant planet $m_2$ and the barycentre of the inner binary form the outer binary. For convenience, we adopt the invariant plane-based inertial reference frame, where the origin is at the central star, the $x$-axis is along the nodal line, the $z$-axis is parallel to the total angular momentum vector and the $y$-axis is chosen to complete a right-handed coordinate system. Under the defined reference frame, the orbits of planets are described by using Jacobi coordinates. In particular, $m_0$ is located at the origin, the position vector of $m_1$ relative to $m_0$ is denoted by ${\bm r}_1$, and the position vector of $m_2$ relative to the barycentre of the inner binary is denoted by ${\bm r}_2$. The corresponding orbit elements for describing the orbits of the inner and outer binaries are denoted by the semimajor axis $a_{1,2}$, the eccentricity $e_{1,2}$, the inclination $i_{1,2}$, longitude of ascending node $\Omega_{1,2}$, argument of pericentre $\omega_{1,2}$ and mean anomaly $M_{1,2}$. The angular momentum vector of the inner and outer binaries are denoted by ${\bm G}_{1,2}$, and the total angular momentum vector is denoted by ${\bm G}_{\rm tot}$. The relative angle between ${\bm G}_{1}$ and ${\bm G}_{\rm tot}$ is the inclination of the inner binary $i_1$, the relative angle between ${\bm G}_{2}$ and ${\bm G}_{\rm tot}$ is the inclination of the outer binary $i_2$ and the relative angle between ${\bm G}_{1}$ and ${\bm G}_{2}$ is the mutual inclination between the inner and outer binaries $i_{\rm tot}$ \citep{naoz2013secular}. The orbits of the inner and outer binaries and the invariant plane share the same nodal line, so it is not difficult to get the geometrical relation of inclination: $i_{\rm tot} = i_1 + i_2$. In the whole manuscript, we use subscript ``1'' to stand for the variables of the inner binary and ``2'' for the variables of the outer binary unless otherwise specified.

Due to the hierarchical configuration, the semimajor axis ratio between the inner and outer binaries $\alpha = a_1/a_2$ is a small parameter. Thus, it is possible to expand the Hamiltonian function of system as a power series in $\alpha$ \citep{harrington1968dynamical, harrington1969stellar}. It is called quadrupole-level approximation when the Hamiltonian is truncated at the second order in $\alpha$ and octupole-level approximation when the Hamiltonian is truncated at the third order in $\alpha$. In order to study long-term evolutions, it is usual to perform double averages of the Hamiltonian function over the orbital periods of the inner and outer binaries \citep{harrington1968dynamical, harrington1969stellar, ford2000secular, naoz2013secular, naoz2016eccentric}. Second-order corrections to the standard double-averaged Hamiltonian can be found in different contexts \citep{krymolowski1999studies, luo2016double, lei2018modified, hamers2019analytica, hamers2019analyticb, lei2019semi, will2021higher}.

The conservation of the total angular momentum implies that the relation $\Omega_1 - \Omega_2 = \pi$ (also called `elimination of nodes') always holds \citep{naoz2013secular, naoz2016eccentric}. This relation can be used to simplify the expression of secular Hamiltonian but it cannot be used to derive the equations of motion for evolution of inclination, as pointed out by \citet{naoz2013secular}. Up to the octupole order in semimajor axis ratio, the double-averaged Hamiltonian reads \citep{krymolowski1999studies, ford2000secular, blaes2002kozai, naoz2013secular, naoz2016eccentric}
\begin{equation}\label{Eq1}
{\cal H} =  - {C_0}\left( {{F_2} + \epsilon {F_3}} \right)
\end{equation}
where the parameters $C_0$ and $\epsilon$ are given by
\begin{equation*}
{C_0} = \frac{{{\cal G}{m_0}{m_1}{m_2}}}{{16\left( {{m_0} + {m_1}} \right)}}\frac{{a_1^2}}{{a_2^3}},\quad \epsilon  = \frac{{{m_0} - {m_1}}}{{{m_0} + {m_1}}}\frac{{{a_1}}}{{{a_2}}}
\end{equation*}
with ${\cal G}$ as the universal gravitational constant, the quadrupole-order term is
\begin{equation*}
\begin{aligned}
{F_2} = &\frac{1}{{{{\left( {1 - e_2^2} \right)}^{3/2}}}}\left[ {\left( {2 + 3e_1^2} \right)\left( {3{{\cos }^2}{i_{\rm tot}} - 1} \right)} \right.\\
&\left. { + 15e_1^2 {\left(1-{\cos}^2{i_{\rm tot}}\right)}\cos 2{\omega _1}} \right]
\end{aligned}
\end{equation*}
and the octupole-order term is
\begin{equation*}
\begin{aligned}
{F_3} &= \frac{{15}}{{32}}\frac{{{e_1}{e_2}}}{{{{\left( {1 - e_2^2} \right)}^{5/2}}}}\left\{ {\left( {4 + 3e_1^2} \right)} \right.\\
&\times \left[ {\left( { - 1 + 11\cos {i_{\rm tot}} + 5{{\cos }^2}{i_{\rm tot}} - 15{{\cos }^3}{i_{\rm tot}}} \right)\cos \left( {{\omega _1} + {\omega _2}} \right)} \right.\\
&\left. { + \left( { - 1 - 11\cos {i_{\rm tot}} + 5{{\cos }^2}{i_{\rm tot}} + 15{{\cos }^3}{i_{\rm tot}}} \right)\cos \left( {{\omega _1} - {\omega _2}} \right)} \right]\\
&+ 35e_1^2\left[ {\left( {1 - \cos {i_{\rm tot}} - {{\cos }^2}{i_{\rm tot}} + {{\cos }^3}{i_{\rm tot}}} \right)\cos \left( {3{\omega _1} + {\omega _2}} \right)} \right.\\
&\left. {\left. { + \left( {1 + \cos {i_{\rm tot}} - {{\cos }^2}{i_{\rm tot}} - {{\cos }^3}{i_{\rm tot}}} \right)\cos \left( {3{\omega _1} - {\omega _2}} \right)} \right]} \right\}
\end{aligned}
\end{equation*}
The coefficient $C_0$ is a constant in the long-term evolution (because $a_1$ and $a_2$ are unchanged in the long-term evolution), so that the Hamiltonian can be normalised (scaled by $C_0$) as
\begin{equation}\label{Eq2}
{\cal H} =  - \left( {{F_2} + \epsilon {F_3}} \right).
\end{equation}
To describe the long-term evolutions, let us introduce the following set of normalised Delaunay's variables,
\begin{equation}\label{Eq3}
\begin{aligned}
{G_1} &= \sqrt {1 - e_1^2},\quad {g_1} = {\omega _1}\\
{G_2} &= \beta \sqrt {1 - e_2^2},\quad {g_2} = {\omega _2}\\
{H_1} &= {G_1}\cos {i_1},\quad {h_1} = {\Omega _1}\\
{H_2} &= {G_2}\cos {i_2},\quad {h_2} = {\Omega _2}
\end{aligned}
\end{equation}
where the parameter $\beta$ is given by
\begin{equation*}
\beta  = \frac{{\left( {{m_0} + {m_1}} \right){m_2}}}{{{m_0}{m_1}}}\sqrt {\frac{{\left( {{m_0} + {m_1}} \right)}}{{\left( {{m_0} + {m_1} + {m_2}} \right)}}\frac{{{a_2}}}{{{a_1}}}}.
\end{equation*}

In terms of Delaunay variables, the Hamiltonian represented by equation (\ref{Eq2}) can be denoted by
\begin{equation}\label{Eq4}
\begin{aligned}
{\cal H} &= {\cal H}_2 + {\cal H}_3\\
&= -{F_2}\left( {{G_{\rm tot}};{G_1},{G_2},{g_1}} \right) - \epsilon {F_3}\left( {{G_{\rm tot}};{G_1},{G_2},{g_1},{g_2}} \right)
\end{aligned}
\end{equation}
which shows that the dynamical model is of two degrees of freedom. The total angular momentum vector of system can be expressed as
\begin{equation*}
{\bm G}_{\rm tot} = {\bm G}_{1} + {\bm G}_{2},
\end{equation*}
and its magnitude is given by
\begin{equation*}
G_{\rm tot} = G_1 \cos{i_1} + G_2 \cos{i_2} = H_1 + H_2.
\end{equation*}
The mutual inclination can be expressed by
\begin{equation*}
\cos {i_{\rm tot}} = \frac{{G_{\rm tot}^2 - G_1^2 - G_2^2}}{{2{G_1}{G_2}}}.
\end{equation*}
Additionally, Hamiltonian canonical relations yield the equations of motion as follows \citep{morbidelli2002modern}:
\begin{equation}\label{Eq3_1}
\begin{aligned}
\frac{{{\rm d}{g_1}}}{{{\rm d}t}} &= \frac{{\partial {\cal H}}}{{\partial {G_1}}},\quad \frac{{{\rm d}{G_1}}}{{{\rm d}t}} =  - \frac{{\partial {\cal H}}}{{\partial {g_1}}},\\
\frac{{{\rm d}{g_2}}}{{{\rm d}t}} &= \frac{{\partial {\cal H}}}{{\partial {G_2}}},\quad \frac{{{\rm d}{G_2}}}{{{\rm d}t}} =  - \frac{{\partial {\cal H}}}{{\partial {g_2}}}.
\end{aligned}
\end{equation}
It should be noted that, although $h_1$ and $h_2$ are absent from the Hamiltonian due to the substitution of $h_1 - h_2 = \pi$, it does not mean their conjugate momenta $H_1$ and $H_2$ are constant in the long-term evolution \citep{naoz2013secular}. In practice, the variations of $H_1$ and $H_2$ are produced by the following geometrical relations,
\begin{equation*}
{H_1} = \frac{{G_{\rm tot}^2 + G_1^2 - G_2^2}}{{2{G_{\rm tot}}}},\quad {H_2} = \frac{{G_{\rm tot}^2 + G_2^2 - G_1^2}}{{2{G_{\rm tot}}}}.
\end{equation*}

\begin{figure*}
\centering
\includegraphics[width=0.48\textwidth]{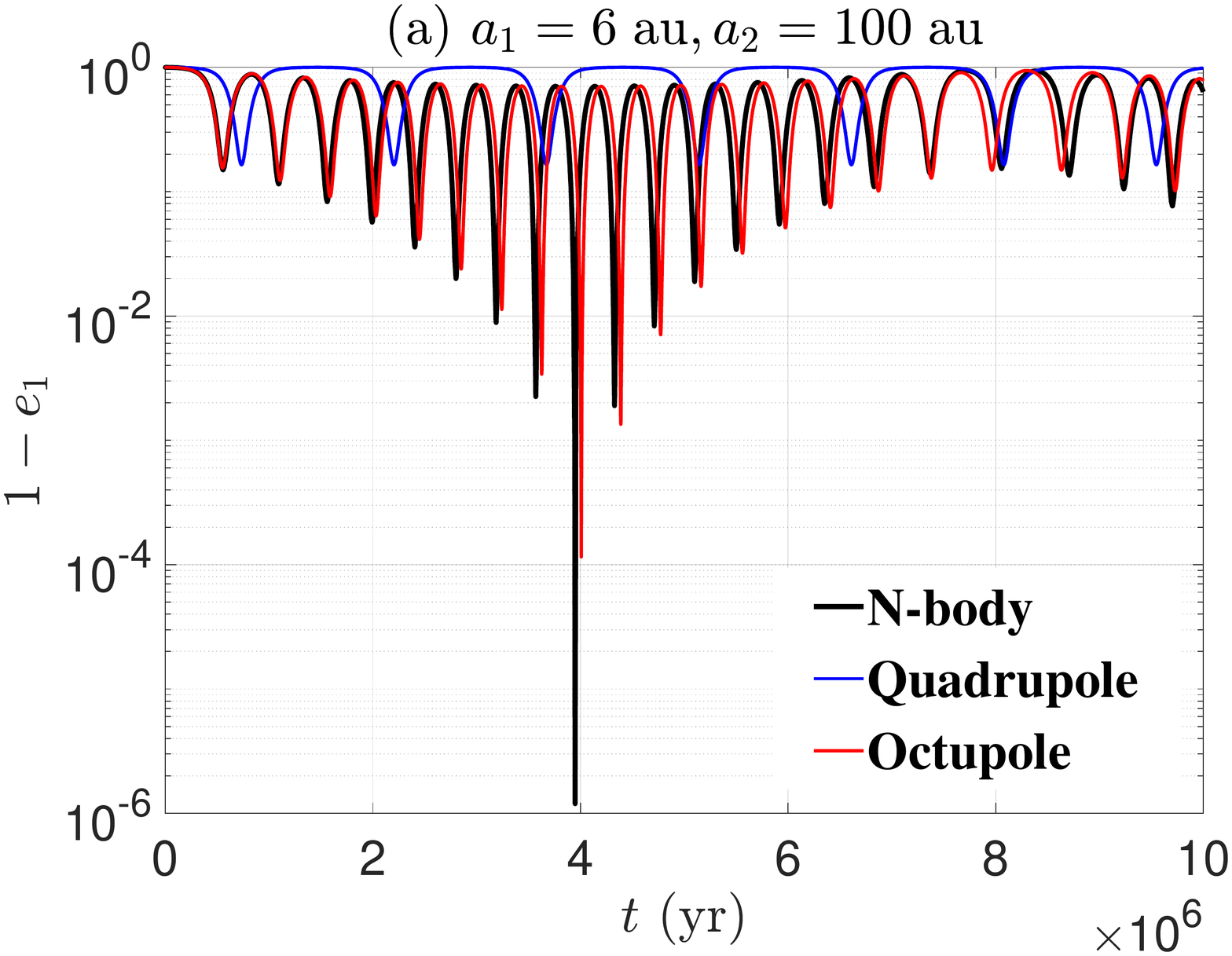}
\includegraphics[width=0.48\textwidth]{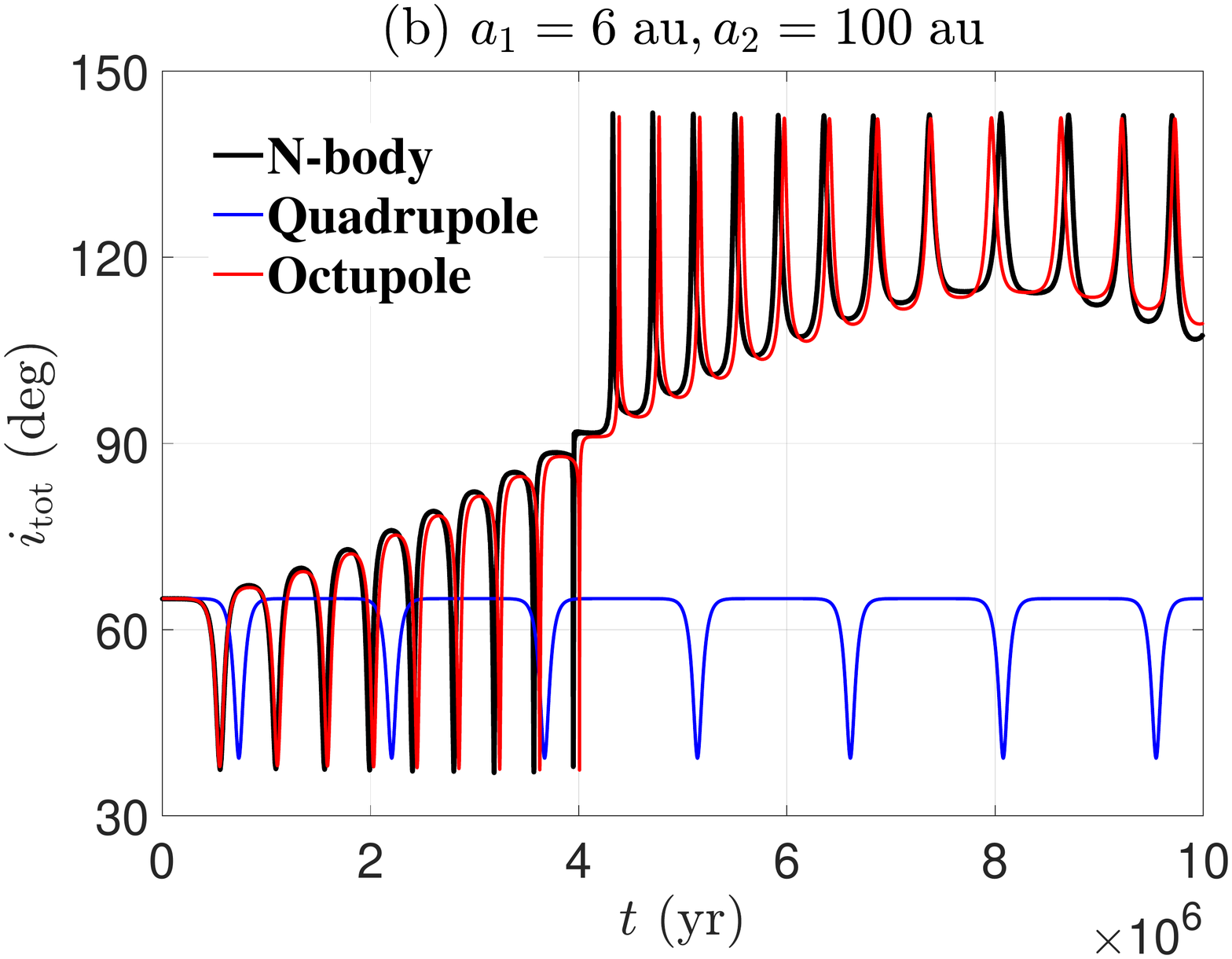}\\
\includegraphics[width=0.48\textwidth]{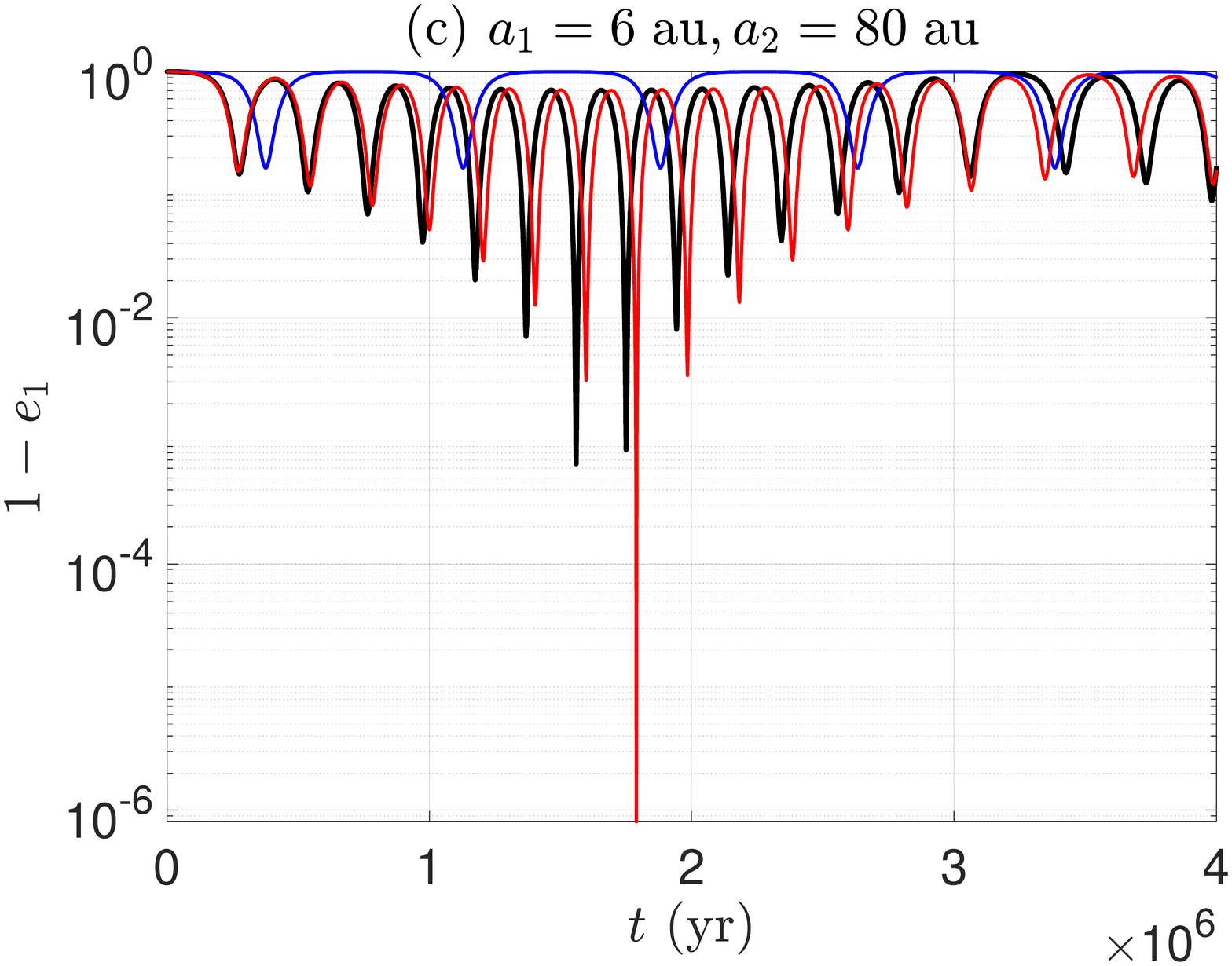}
\includegraphics[width=0.48\textwidth]{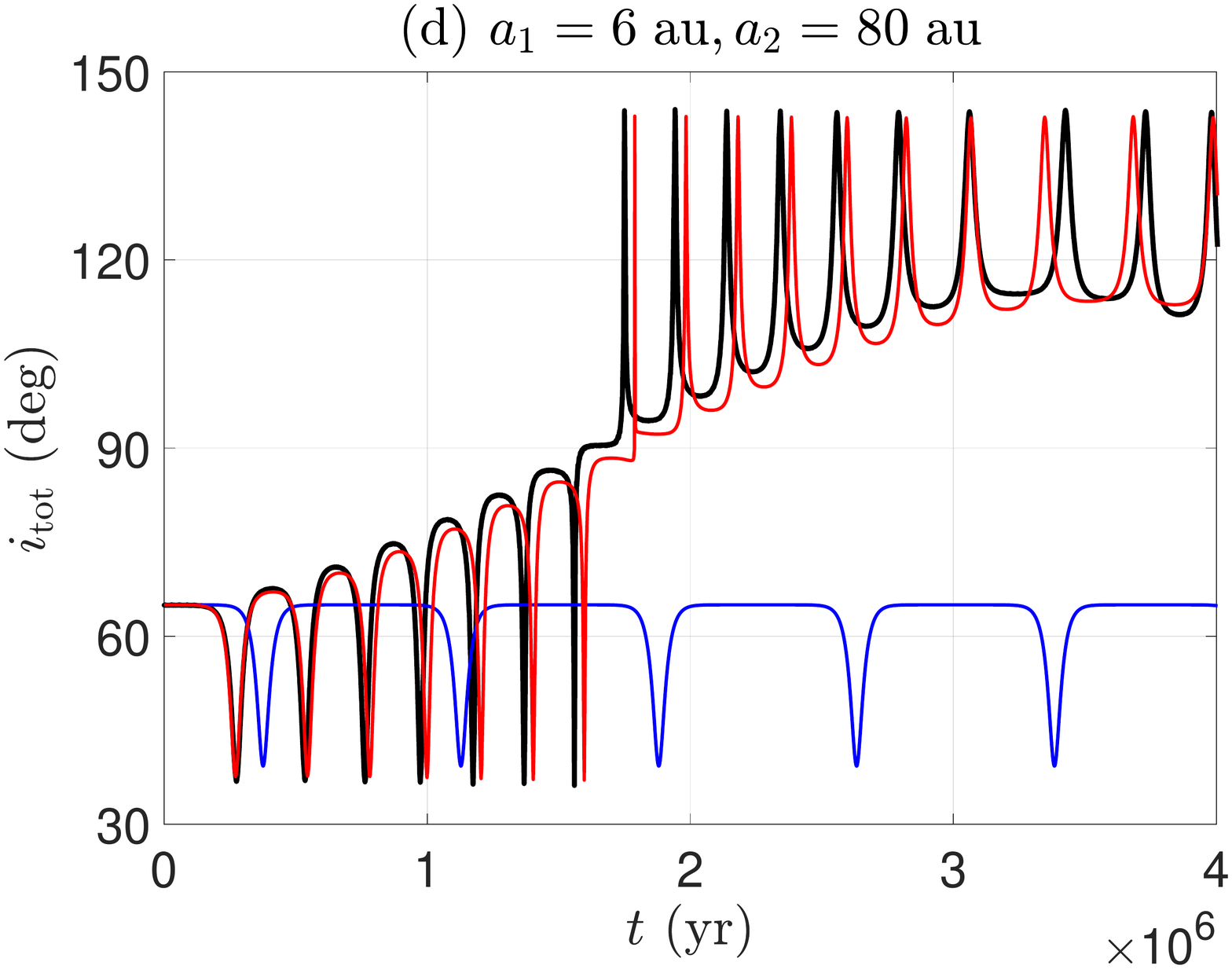}\\
\includegraphics[width=0.48\textwidth]{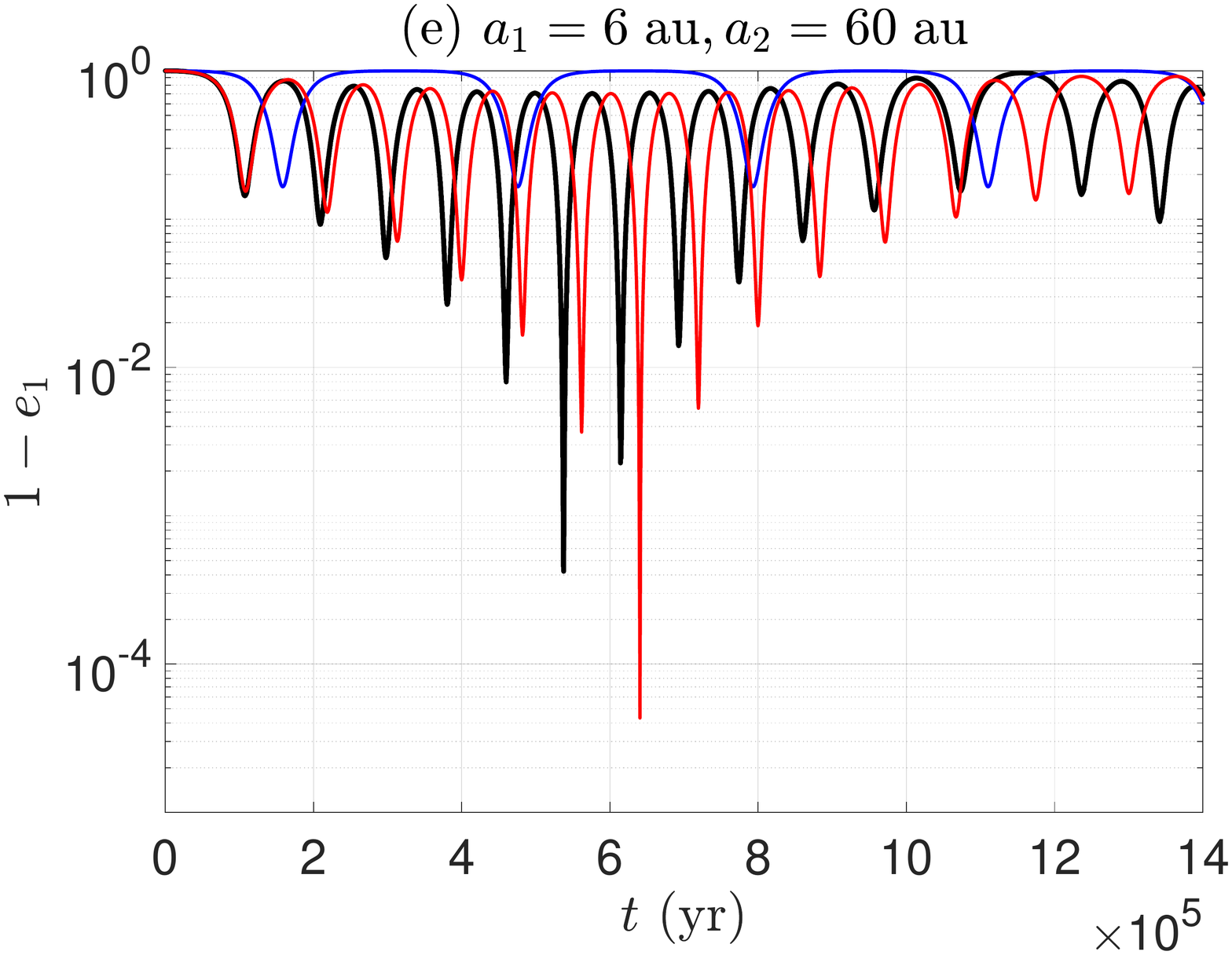}
\includegraphics[width=0.48\textwidth]{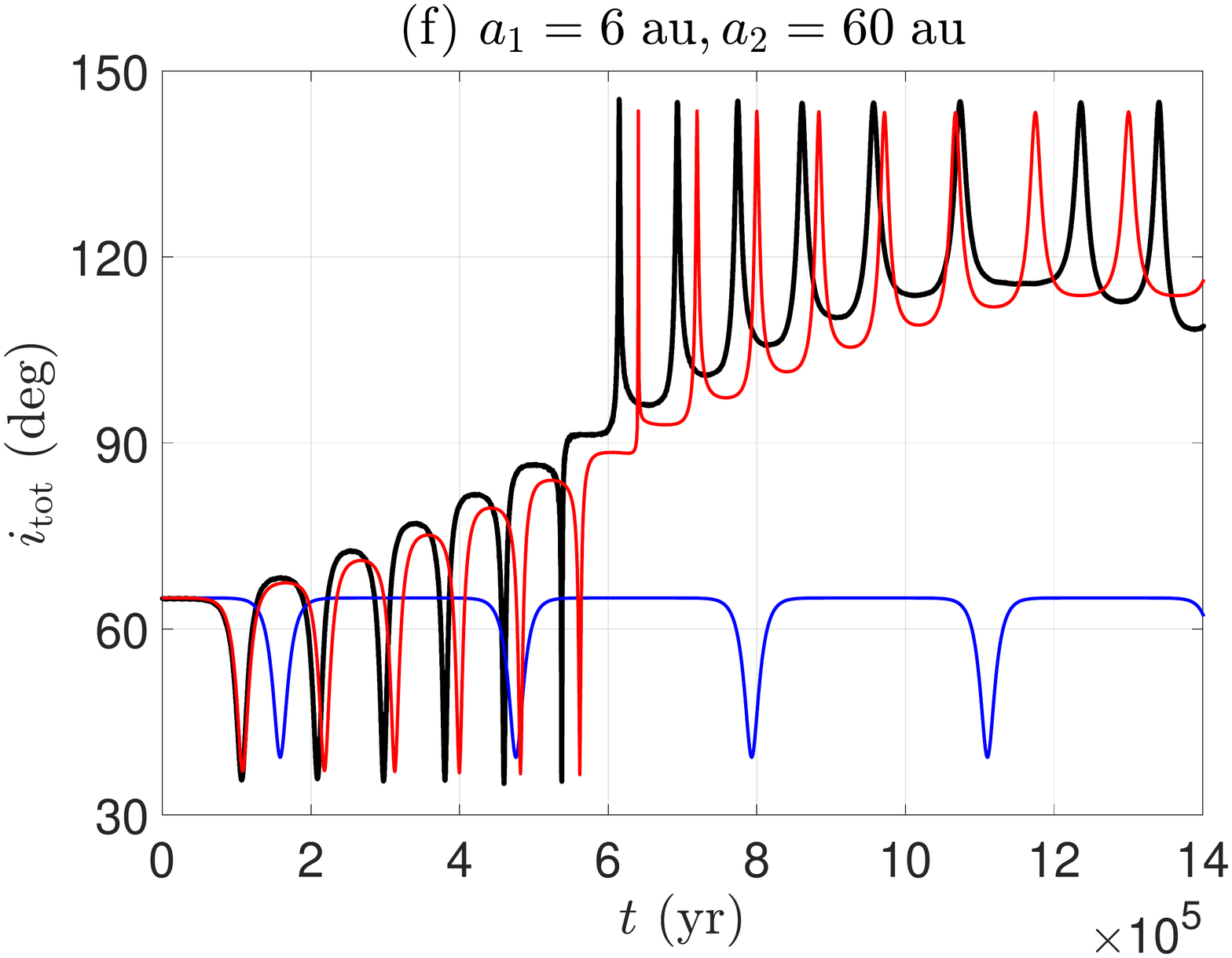}
\caption{Time evolutions of eccentricities of the inner planet and mutual inclinations under different dynamical models (full $N$-body model, double-averaged models at the quadrupole and octupole approximations). The central star has mass $1 M_{\rm Sun}$, the inner planet has mass $M_{\rm Jup}$ and the outer planet has mass $40 M_{\rm Jup}$. The initial eccentricities are assumed at $e_1 = 0.001$ and $e_2 = 0.6$, the initial mutual inclinations are taken as $i_{\rm tot} = 65^{\circ}$ ($i_1 = 64.7^{\circ}$ and $i_2 = 0.3^{\circ}$), the initial arguments of pericenter are $\omega_1 = \omega_2 = 0$, the initial longitudes of ascending node are $\Omega_1 = \pi$ and $\Omega_2 = 0$ and the initial mean anomalies are taken as $M_1 = M_2 = 0$. The semimajor axes of the inner and outer planets used for numerical simulations are provided at the top of each panel. At the flipping moment, the eccentricities of inner planets reach the maximum close to unity. Note that the evolutions shown in panels (a) and (b) are in agreement with the results in \citet{naoz2011hot} (see Figure 1 in their work).}
\label{Fig0}
\end{figure*}

In order to check the validity of the double-averaged Hamiltonian models, we numerically integrate the equations of motion of the full three-body problem and the double-averaged models at the quadrupole- and octupole-level approximations. Please refer to the caption of Fig. \ref{Fig0} for the detailed settings of initial parameters. It should be noted that osculating elements are used in the full three-body model and mean elements (without short-period oscillations) are required in the double-averaged Hamiltonian model. However, for the current problem, the difference between osculating and mean elements is on the order of ${\cal O} (10^{-3})$, thus we ignore their deviations for the initial settings. When the mass of the inner and outer planets are given, the hierarchy of planetary three-body system is determined by the semimajor axis ratio $\alpha = a_1/a_2$. To see the influence of $\alpha$, three typical cases are taken into account ($\alpha = 0.06, 0.075, 0.1$). Please refer to Fig. \ref{Fig0} for the comparisons among different dynamical models. The results produced in full three-body model are given in black lines, the results obtained in the quadrupole-order Hamiltonian model are shown by blue lines and the ones under the octupole-order Hamiltonian model are given by red lines. For the current example, we can observe that (a) in general, the octupole-order Hamiltonian model agrees with the full three-body model better than the quadrupole-order model, and (b) when the semimajor axis ratio increases (the hierarchy of system decreases), the deviation between the full three-body model and the octupole-order Hamiltonian model increases, meaning that high-order approximations are required to achieve a certain accuracy. Thus, in order to ensure the validity of the octupole-order approximation, it requires that the semimajor axis ratio should be small. On the other hand, in order to ensure the validity of the double-averaging approximation (corresponding to the lowest-order perturbation theory), it requires that the mass of planets is much smaller than that of the central body (in general, planetary systems can satisfy this condition well), otherwise it needs to consider the high-order perturbation \citep{krymolowski1999studies, luo2016double, lei2018modified, hamers2019analytica, hamers2019analyticb, lei2019semi, will2021higher}.

Based on the aforementioned discussions, the validity of the double-averaged Hamiltonian model truncated at the octupole order requires that (a) the semimajor axis ratio between the inner and outer planets should be small and (b) the mass of the perturbing and perturbed objects should be much smaller than that of the central star. Thus, we need to emphasize that the analytical developments given in this study are applicable when these two requirements are satisfied.

Without otherwise stated, the parameters of dynamical system for practical simulations in the entire work are the same as the ones adopted by \citet{naoz2011hot} (see the caption of Figure 1 in their work). The physical parameters are given as follows:
\begin{equation*}
\begin{aligned}
&m_0 = 1 M_{\rm Sun},\quad m_1 = 1 M_{\rm Jup},\quad m_2 = 40 M_{\rm Jup}\\
&a_1 = 6\; {\rm au},\quad a_2 = 100\; {\rm au}
\end{aligned}
\end{equation*}
with $M_{\rm Sun}$ as the Solar mass, $M_{\rm Jup}$ as the mass of Jupiter and $\rm au$ as the astronomical unit. Under the setting of system parameters, we can obtain the derived parameters as $\beta  = 160.4234$, $\epsilon  = 1.9886 \times {10^{ - 2}}$ and ${C_0} = 8.1968 \times {10^{ - 11}}$. In addition, if the initial eccentricities of the inner and outer binaries are taken as $e_{1,0} = 0.001$ and $e_{2,0} = 0.6$ and the initial mutual inclination is taken as $i_{\rm tot} = 65^{\circ}$, we can get the total angular momentum equal to $G_{\rm tot} = 128.7645$. These initial parameters are also the same as the ones taken by \citet{naoz2011hot}.

\section{Quadrupole-order dynamics}
\label{Sect3}

In this section, we discuss the secular dynamics under the quadrupole-level approximation. In particular, we analyse the parameter space where the quadrupole-order resonance (i.e., the ZLK resonance) occurs. Then, we discuss regions of orbit flip caused by quadrupole-order dynamics.

\subsection{Quadrupole-order resonance}
\label{Sect3_1}

Up to the quadrupole order in semimajor axis ratio, the Hamiltonian holds
\begin{equation}\label{Eq6}
\begin{aligned}
{\cal H} =  &- \frac{{{\beta ^3}}}{{4G_1^2G_2^5}}\left\{ {\left( {5 - 3G_1^2} \right)\left( {3{G_a^2} - 4G_1^2G_2^2} \right)} \right.\\
&\left. { + 15\left( {1 - G_1^2} \right)\left( {4G_1^2G_2^2 - {G_a^2}} \right)\cos \left( {2{g_1}} \right)} \right\},
\end{aligned}
\end{equation}
where $G_a = {G_{\rm tot}^2 - G_1^2 - G_2^2}$. In this dynamical model, the angle $g_2$ is a cyclic coordinate, leading to the fact that its conjugate momentum $G_2$ is a motion integral. As a result, the Hamiltonian given by equation (\ref{Eq6}) determines a dynamical model with one degree of freedom $(g_1, G_1)$, depending on the motion integral $G_2$. Under the quadrupole-order model, \citet{hamers2021properties} studied the properties of ZLK effect, including the maximum eccentricities, timescales of eccentricity oscillation and orbit flips.

According to the expression of the total angular momentum
\begin{equation*}
G_{\rm tot}^2 = G_1^2 + G_2^2 + 2{G_1}{G_2}\cos {i_{\rm tot}},
\end{equation*}
and the expression of the angular momentum of the inner binary,
\begin{equation*}
G_1 = \sqrt{1-e_1^2},
\end{equation*}
we can get the allowable domain of the angular momentum of the inner binary $G_1$ as
\begin{equation*}
{G_1} \in \left[ {\left| {{G_{\rm tot}} - {G_2}} \right|,1} \right].
\end{equation*}
In particular, when the mutual inclination $i_{\rm tot}$ is equal to $0$ or $\pi$, $G_1$ takes its minimum $\left| {{G_{\rm tot}} - {G_2}} \right|$, and it takes $1.0$ when the eccentricity $e_1$ is equal to zero. Thus, those spaces with $\left| {{G_{{\rm tot}}} - {G_2}} \right| > 1$ correspond to physically forbidden regions.

\begin{figure*}
\centering
\includegraphics[width=0.48\textwidth]{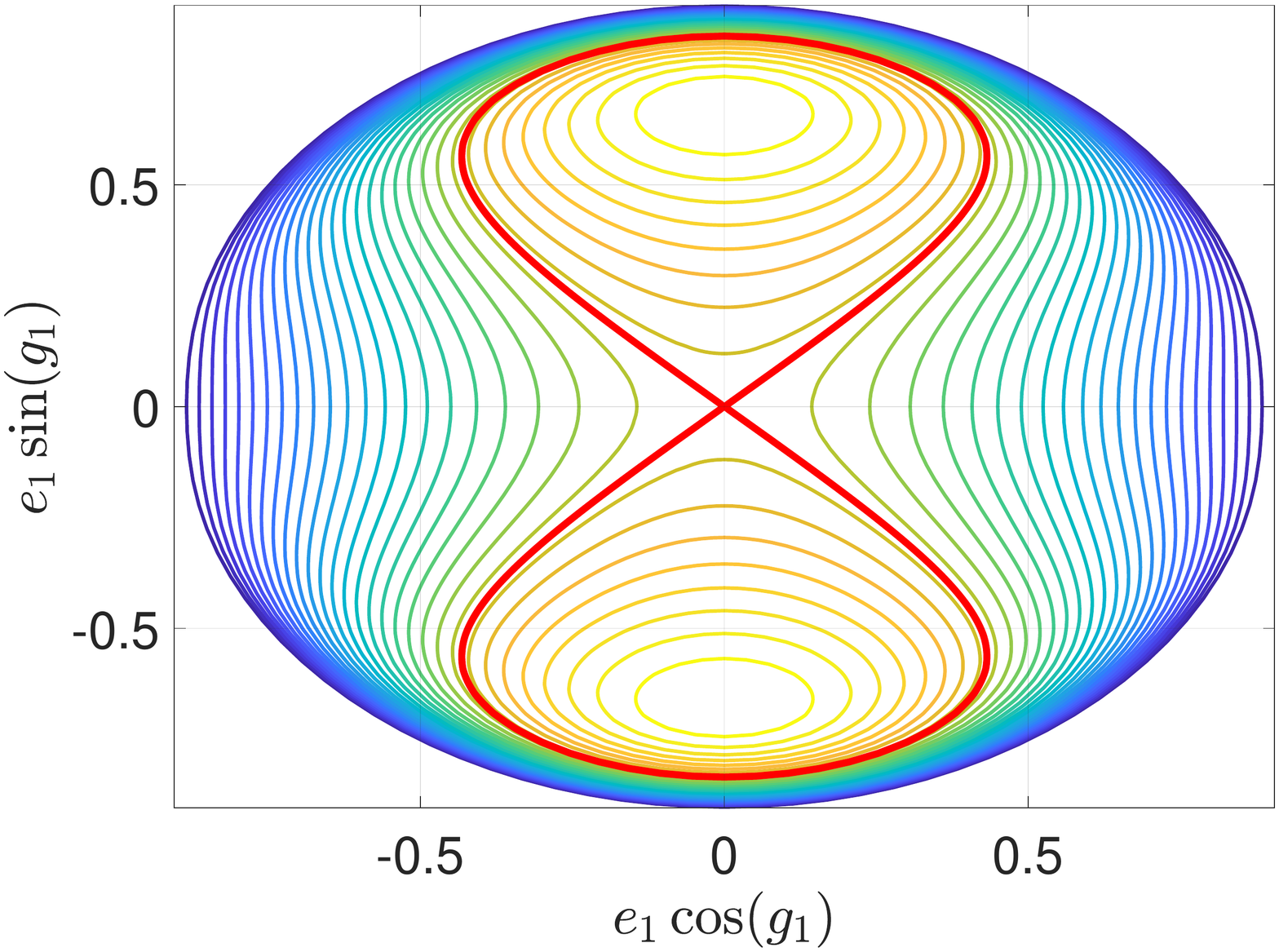}
\includegraphics[width=0.48\textwidth]{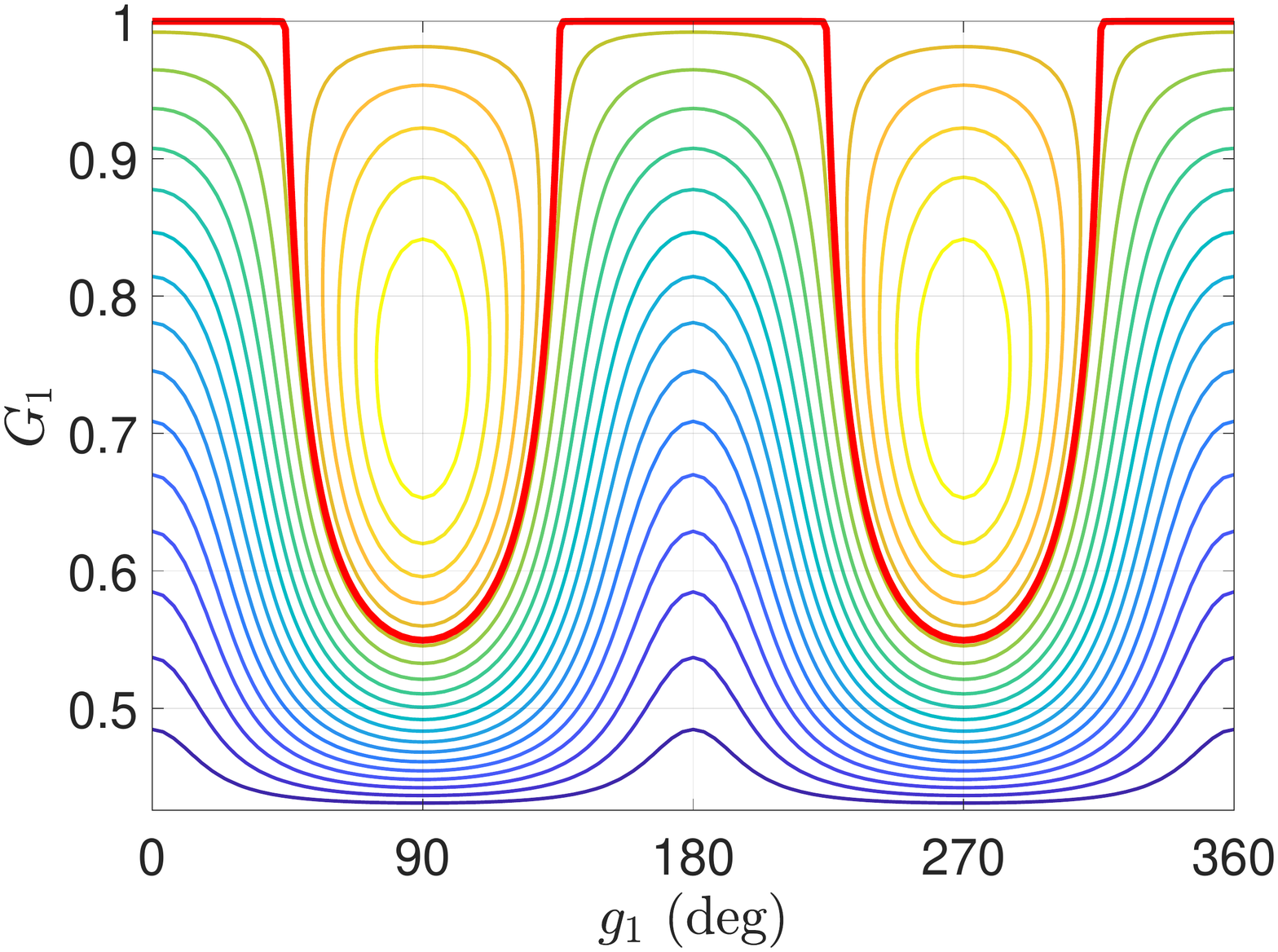}
\caption{Level curves of quadrupole-level Hamiltonian ${\cal H} = {{\cal H}_2}$ (i.e., phase portraits) shown in the $(e_1\cos{g_1},e_1\sin{g_1})$ space (\emph{left panel}) and in the $(g_1, G_1)$ space (\emph{right panel}). The conserved parameters of this example are given by $G_2 = 128.3387$ and $G_{\rm tot} = 128.7645$. The red lines represent the dynamical separatrices, dividing rotating ZLK cycles from librating ZLK cycles. It is observed that the ZLK centres are located at $2g_1 = \pi$. The zero-eccentricity point corresponds to the saddle point.}
\label{Fig1}
\end{figure*}

The dynamical model represented by equation (\ref{Eq6}) is integrable and the dynamical structures can be revealed by phase portraits (i.e., level curves of Hamiltonian in the phase space). Figure \ref{Fig1} shows the phase portraits in the $(e_1\cos{g_1},e_1\sin{g_1})$ space (see the left panel) and in the $(g_1, G_1)$ space (see the right panel). From Fig. \ref{Fig1}, we can observe that (a) the ZLK resonance can take place in this case and its centre is located at $2g_1 = \pi$, (b) the origin (i.e., zero-eccentricity point) correspond to the saddle point, and (c) the ZLK rotating cycles and the ZLK librating cycles are divided by the dynamical separatrix, corresponding to the level curve passing through the saddle point (please see the red line in Fig. \ref{Fig1}). Let us denote the Hamiltonian of the separatrix by ${\cal H}_{\rm sep}$. From the viewpoint of Hamiltonian, those cycles with ${\cal H}_2 > {\cal H}_{\rm sep}$ are of ZLK libration and those cycles with ${\cal H}_2 < {\cal H}_{\rm sep}$ are of ZLK circulation.

\begin{figure*}
\centering
\includegraphics[width=0.48\textwidth]{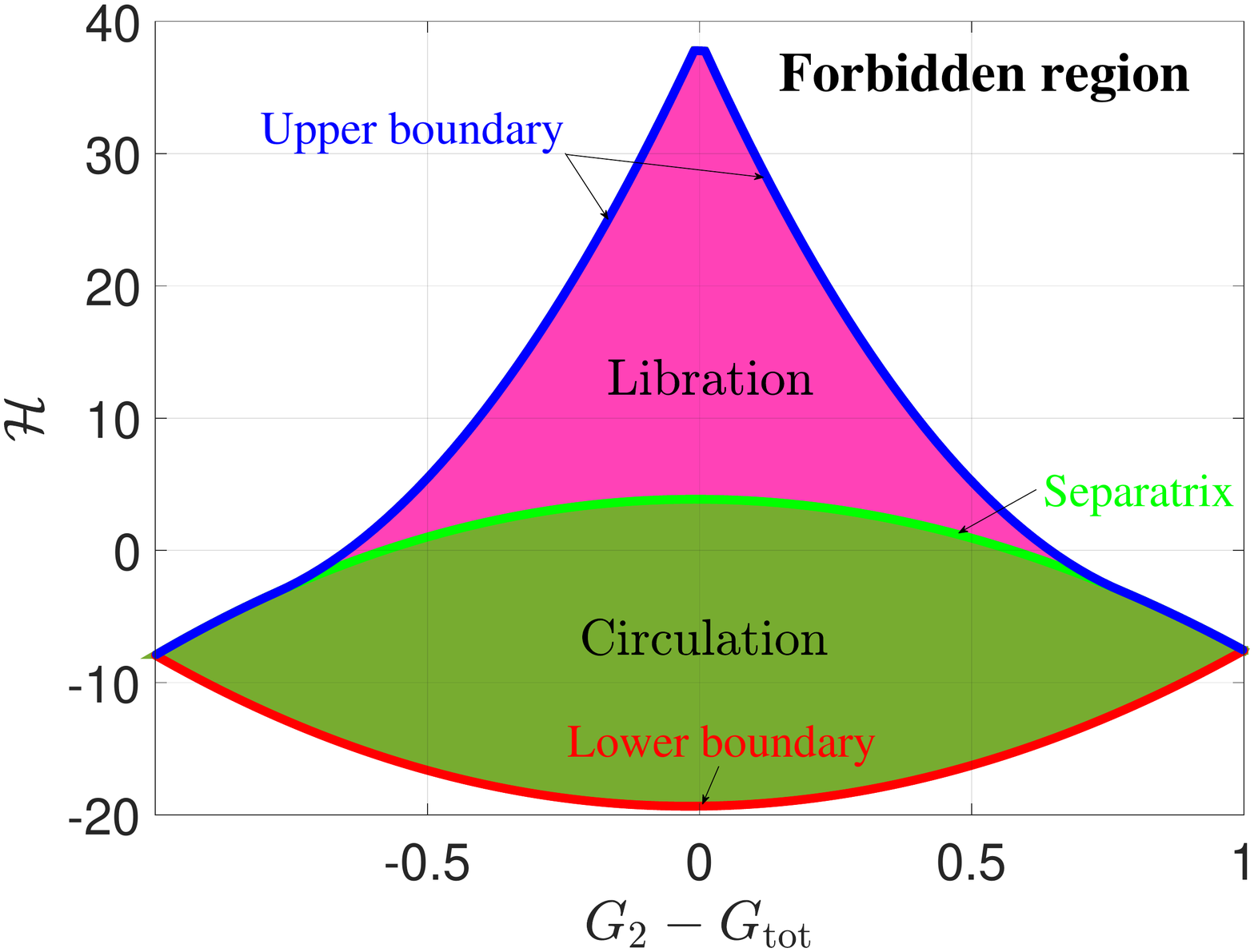}
\includegraphics[width=0.48\textwidth]{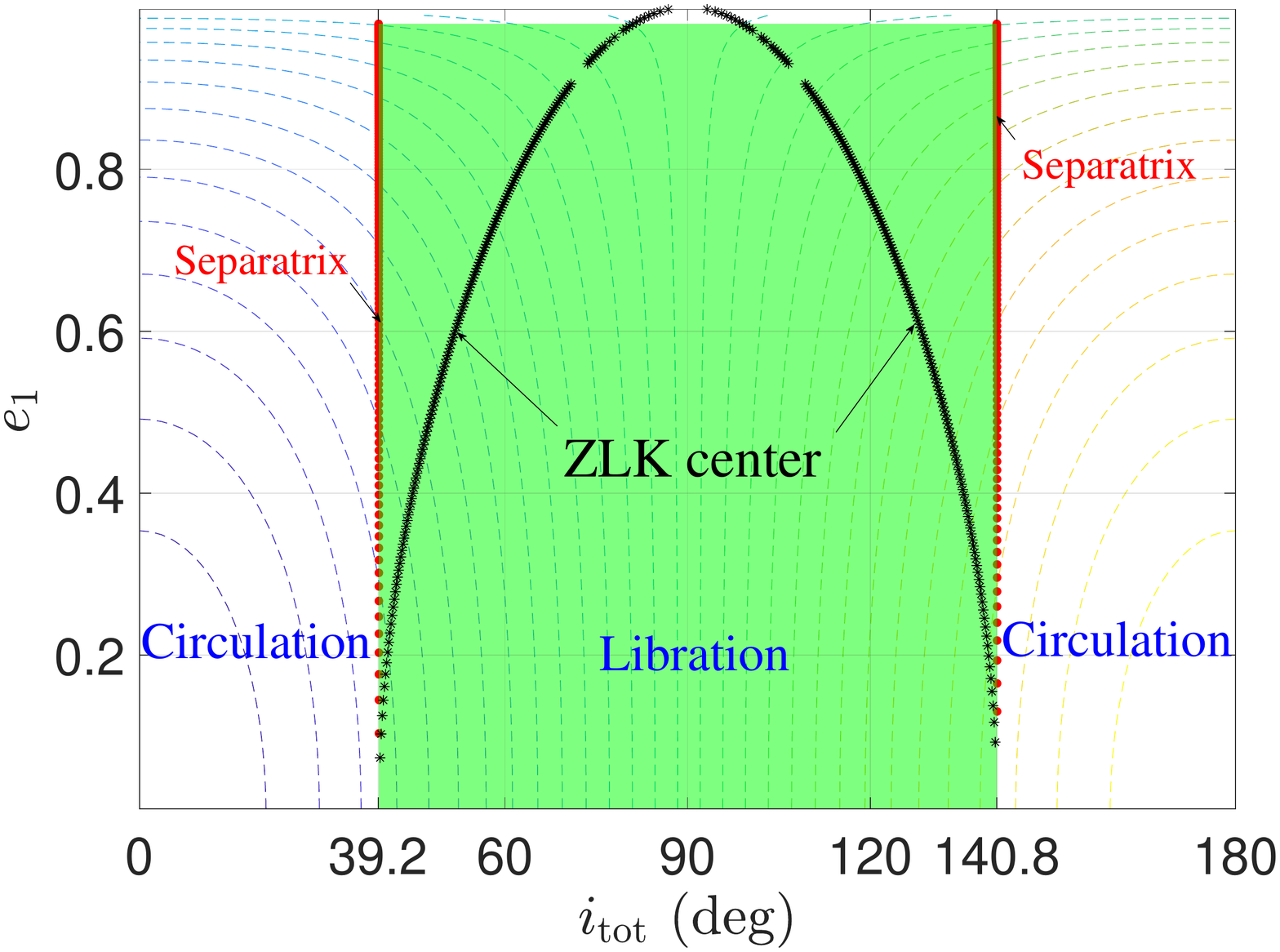}
\caption{Under the quadrupole-level Hamiltonian model, the libration and circulation domains are shown in the $(G_2 - G_{\rm tot}, {\cal H})$ space (\emph{left panel}) and in the $(i_{\rm tot}, e_1)$ space (\emph{right panel}) when the total angular momentum is at $G_{\rm tot} = 128.7645$. In the left panel, the ZLK centre is located at the upper boundary (as the maximum of Hamiltonian takes place at the ZLK centre). The circulation region is bounded by the lower boundary (red curve) and separatrix (green curve), and the libration region is bounded by the separatrix (green curve) and the upper boundaries (blue curves). The regions outside the boundaries (white regions) correspond to physically forbidden domains. In the right panel, the libration centres are marked by black stars, and the resonant width is evaluated at $2g_1 = \pi$ (corresponding to the angle of the ZLK centre). It is observed that the left separatrix is located at $i_{\rm tot} \approx 39.2^{\circ}$ and the right one is placed at $i_{\rm tot} \approx 140.8^{\circ}$. The region bounded by the left and right separatrices is of ZLK libration (shaded region) and the other regions are of ZLK circulation.}
\label{Fig2}
\end{figure*}

Under the quadrupole-order dynamical model determined by equation (\ref{Eq6}), there are three conserved quantities: the Hamiltonian ${\cal H}$, the total angular momentum $G_{\rm tot}$ and the angular momentum of the outer binary $G_2$. When these parameters are given (if it is not inside the forbidden region), there are two types of motion modes: ZLK circulation and ZLK libration. It is mentioned that the classification of dynamical regimes for the ZLK librating and rotating cycles at the quadrupole-level approximation in the test-particle limit is discussed in the $(c_1, c_2)$ space \citep{lidov1962evolution, broucke2003long}, in the $({\cal H}, |H|)$ space \citep{sidorenko2018eccentric} and in the $({\cal H}, i_*)$ and $(i,e)$ spaces \citep{lei2021structures}. In these previous studies, $c_1$ and $c_2$ are two motion integrals (energy and angular momentum), ${\cal H}$ is the quadrupole-order Hamiltonian, $H = \cos{i_*}$ is the vertical component of angular momentum, $e$ and $i$ are the eccentricity and inclination of test particle.

Similarly, let us analyse the distribution of libration and circulation regions of quadrupole-order resonance (ZLK resonance) under non-restricted hierarchical planetary systems in the parameter space spanned by the conserved quantities $(G_2, {\cal H}, G_{\rm tot})$.

Firstly, let us derive the lower boundary of Hamiltonian when $G_2$ and $G_{\rm tot}$ are provided. From the right panel of Fig. \ref{Fig1}, we can see that the minimum of Hamiltonian happens at $2g_1 = 0$. In the case with $2g_1 = 0$, the Hamiltonian can be written as a function of $G_1$ in the following form,
\begin{equation*}
\begin{aligned}
{\cal H} =  &- \frac{{{\beta ^3}}}{{4G_1^2G_2^5}}\left\{ {\left( {5 - 3G_1^2} \right)\left( {3{G_b^2} - 4G_1^2G_2^2} \right)} \right.\\
&\left. { + 15\left( {1 - G_1^2} \right)\left( {4G_1^2G_2^2 - {G_b^2}} \right)} \right\}
\end{aligned}
\end{equation*}
where $G_b$ is given by
\begin{equation*}
G_b = {G_{\rm tot}^2 - G_1^2 - G_2^2}.
\end{equation*}
Taking the first derivative of ${\cal H}$ with respect to $G_1$, we can get
\begin{equation*}
\frac{{\partial {\cal H}}}{{\partial {G_1}}} > 0
\end{equation*}
which shows that ${\cal H}$ is an increasing function of $G_1$ in the case of $2g_1 = 0$. Thus, the minimum of ${\cal H}$ takes place at
\begin{equation*}
G_1 = G_{1,\min} = \left| G_{\rm tot} - G_2 \right|
\end{equation*}
which corresponds to $i_{\rm tot} = 0$ or $i_{\rm tot} = \pi$. Consequently, when $G_2$ and $G_{\rm tot}$ are given, the minimum of ${\cal H}$ is expressed by
\begin{equation}\label{Eq7}
\begin{aligned}
{{\cal H}_{\min}} =  &- \frac{{{\beta ^3}}}{{4G_{1,\min }^2G_2^5}}\left\{ {\left( {5 - 3G_{1,\min }^2} \right)\left( {3G_c^2 - 4G_{1,\min }^2G_2^2} \right)} \right.\\
&+ 15\left( {1 - G_{1,\min }^2} \right)\left. {\left( {4G_{1,\min }^2G_2^2 - G_c^2} \right)} \right\}
\end{aligned}
\end{equation}
where $G_c$ is given by
\begin{equation*}
{G_c} = G_{\rm tot}^2 - G_{1,\min }^2 - G_2^2.
\end{equation*}
The minimum Hamiltonian given by equation (\ref{Eq7}) as a function of $G_2$ and $G_{\rm tot}$ provides the lower boundary in the $(G_2, G_{\rm tot}, {\cal H})$ space.

Secondly, let us analyse the Hamiltonian at the dynamical separatrix. From the left panel of Fig. \ref{Fig1}, we can see that the dynamical separatrix corresponds to the level curve of Hamiltonian passing through the saddle point at $e_1 = 0$ (or $G_1 = 1$). Thus, when $G_2$ and $G_{\rm tot}$ are given, the Hamiltonian of separatrix can be written as
\begin{equation}\label{Eq8}
{{\cal H}_{\rm sep}} =  - \frac{{{\beta ^3}}}{{2G_2^5}}\left[ {3{{\left( {G_{\rm tot}^2 - G_2^2 - 1} \right)}^2} - 4G_2^2} \right].
\end{equation}
The Hamiltonian given by equation (\ref{Eq8}) as a function of $G_2$ and $G_{\rm tot}$ provides the dynamical separatrix in the $(G_2, G_{\rm tot}, {\cal H})$ space.

Finally, let us discuss the upper boundary of the Hamiltonian. From the phase portraits shown in Fig. \ref{Fig1}, we can observe that the maximum Hamiltonian takes places at the ZLK centre. The ZLK centre corresponds to the stable equilibrium points of the dynamical model, satisfying the following stationary conditions \citep{kozai1962secular}:
\begin{equation*}
\frac{{\partial {\cal H}}}{{\partial {g_1}}} = 0, \quad \frac{{\partial {\cal H}}}{{\partial {G_1}}} = 0.
\end{equation*}
The first condition implies $2 g_1 = \pi$. Considering $2 g_1 = \pi$, the second condition becomes
\begin{equation}\label{Eq9}
G_{1,*}^6 - \frac{1}{8}\left( {8G_{\rm tot}^2 + 4G_2^2 + 5} \right)G_{1,*}^4 + \frac{5}{8}{\left( {G_{\rm tot}^2 - G_2^2} \right)^2} = 0.
\end{equation}
By solving equation (\ref{Eq9}), it is possible to obtain the angular momentum of the inner binary $G_1 = G_{1,*}$ at the ZLK centre. The ZLK centre is located at $(2g_1 = \pi, G_1 = G_{1,*})$.

As a result, when $G_2$ and $G_{\rm tot}$ are given, the maximum of Hamiltonian can be expressed by
\begin{equation}\label{Eq10}
\begin{aligned}
{{\cal H}_{\max }} =  &- \frac{{{\beta ^3}}}{{4G_{1,*}^2G_2^5}}\left\{ {\left( {5 - 3G_{1,*}^2} \right)\left( {3G_d^2 - 4G_{1,*}^2G_2^2} \right)} \right.\\
&\left. { - 15\left( {1 - G_{1,*}^2} \right)\left( {4G_{1,*}^2G_2^2 - G_d^2} \right)} \right\}
\end{aligned}
\end{equation}
where $G_d$ is given by
\begin{equation*}
{G_d} = G_{\rm tot}^2 - G_{1,*}^2 - G_2^2.
\end{equation*}
The maximum Hamiltonian given by equation (\ref{Eq10}) as a function of $G_2$ and $G_{\rm tot}$ provides the upper boundary in the $(G_2, G_{\rm tot}, {\cal H})$ space.

In summary, when $G_2$ and $G_{\rm tot}$ are given, we can get that (a) the Hamiltonian takes the minimum at $(2g_1 = 0, G_1 = \left| G_{\rm tot} - G_2 \right|)$ or at $(2\omega_1 = 0, 2 i_{\rm tot} = 0)$ and the lower boundary of Hamiltonian is given by equation (\ref{Eq7}), (b) the dynamical separatrix passes through the zero-eccentricity point and the Hamiltonian of separatrix is given by equation (\ref{Eq8}), and (c) the Hamiltonian takes the maximum at the ZLK centre $(2g_1 = \pi, G_1 = G_{1,*})$ where $G_{1,*}$ is obtained by solving equation (\ref{Eq9}), and the upper boundary of Hamiltonian is given by equation (\ref{Eq10}).

Figure \ref{Fig2} reports the distribution of libration and circulation regions when the total angular momentum is fixed at $G_{\rm tot} = 128.7645$. Practical simulations indicate that the results with different levels of $G_{\rm tot}$ are qualitatively similar.

In the left panel of Fig. \ref{Fig2}, the lower boundary, dynamical separatrix and upper boundary for the quadrupole-order dynamics are presented in the $(G_2 - G_{\rm tot}, {\cal H})$ space. In particular, the lower boundary expressed by equation (\ref{Eq7}) is shown by red curve, the dynamical separatrix expressed by equation (\ref{Eq8}) is shown in green curve and the upper boundary expressed by equation (\ref{Eq10}) is shown in blue curve. The regions bounded by the lower boundary and separatrix correspond to the circulation zones, the regions bounded by the separatrix and the upper boundary correspond to the libration zones, and those regions outside the boundaries are physically forbidden zones where the forbidden condition $\left|G_2 - G_{\rm tot}\right|>1$ is satisfied. According to the previous discussion, we know that the ZLK centres are located at the upper boundary.

In the right panel of Fig. \ref{Fig2}, the distributions of libration and circulation regions are plotted in the $(i_{\rm tot}, e_1)$ space. It is noted that resonant width is evaluated at $2g_1 = \pi$ (i.e., the angle of ZLK centre). In particular, the ZLK centres are shown in black stars, and the dynamical separatrices are shown in red dots. The level curves of the motion integral $G_2$ are shown by dashed lines. It is observed that the left separatrix is located at $i_{\rm tot} \approx 39.2^{\circ}$ and the right separatrix is located at $i_{\rm tot} \approx 140.8^{\circ}$, meaning that ZLK resonance may happen in the inclination interval $\left[39.2^{\circ}, 140.8^{\circ}\right]$, which is consistent with the standard ZLK resonance at the quadrupole-level approximation in the test-particle limit \citep{lidov1962evolution, kozai1962secular}.

\begin{figure*}
\centering
\includegraphics[width=0.6\textwidth]{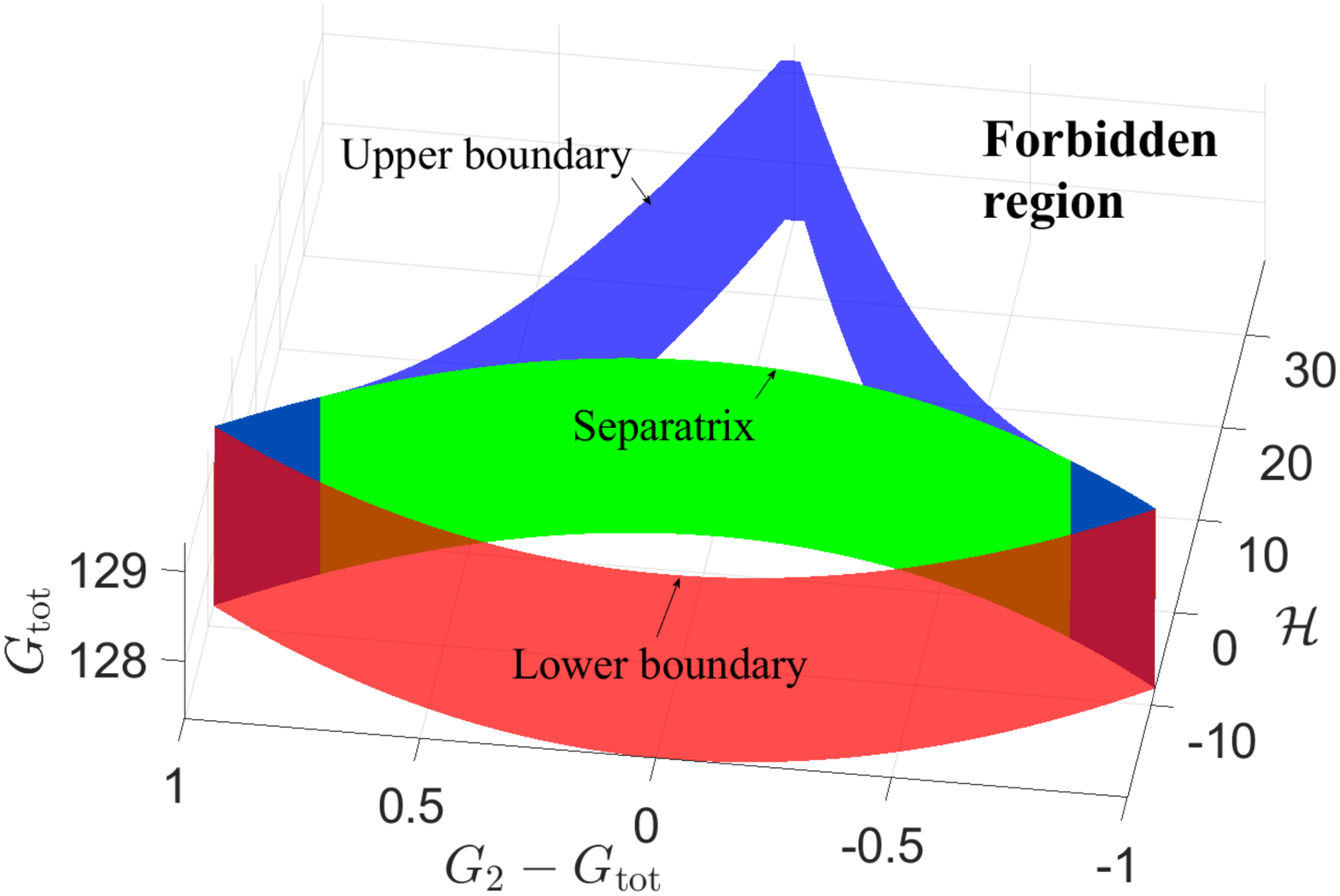}
\caption{Under the quadrupole-level Hamiltonian model, the libration and circulation domains shown in the $({\cal H}, G_2 - G_{\rm tot}, G_{\rm tot})$ space. The circulation region is bounded by the lower boundary (red surface) and separatrix (green surface), and the libration region is bounded by the separatrix (green surface) and the upper boundary (blue surface). The regions outside the lower and upper boundaries correspond to physically forbidden domains.}
\label{Fig3}
\end{figure*}

The general case is considered in Fig. \ref{Fig3}, where the lower boundary, separatrix and upper boundary are reported in the $(G_2 - G_{\rm tot}, {\cal H}, G_{\rm tot})$ space. The red surface stands for the lower boundary, the green surface stands for the dynamical separatrix and the blue surface represents the upper boundary. The circulation region is bounded by the lower boundary and separatrix, and the libration region is bounded by the separatrix and the upper boundary.

\subsection{Orbit flips caused by the quadrupole-order dynamics}
\label{Sect3_2}

\begin{figure*}
\centering
\includegraphics[width=0.48\textwidth]{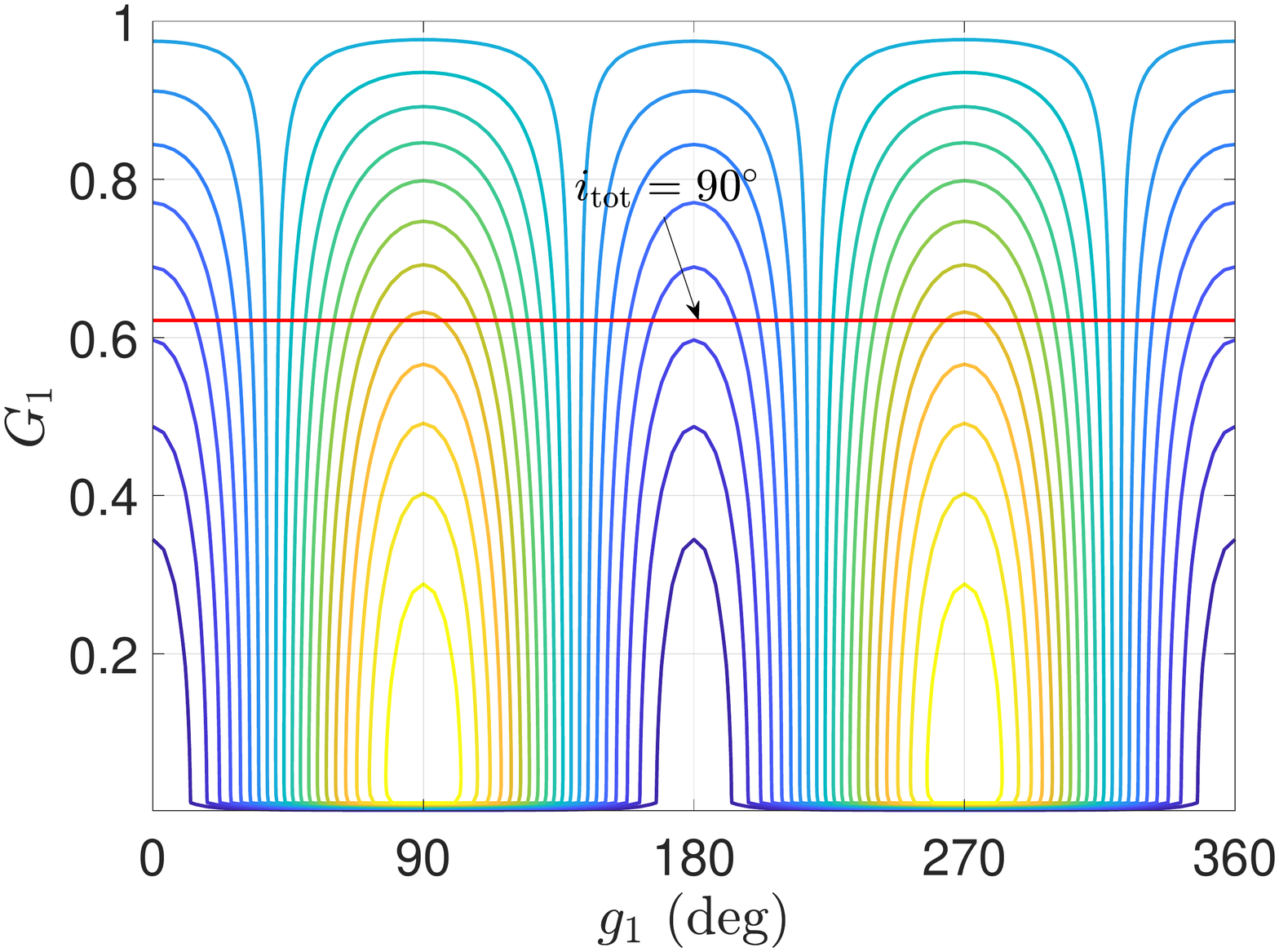}
\includegraphics[width=0.48\textwidth]{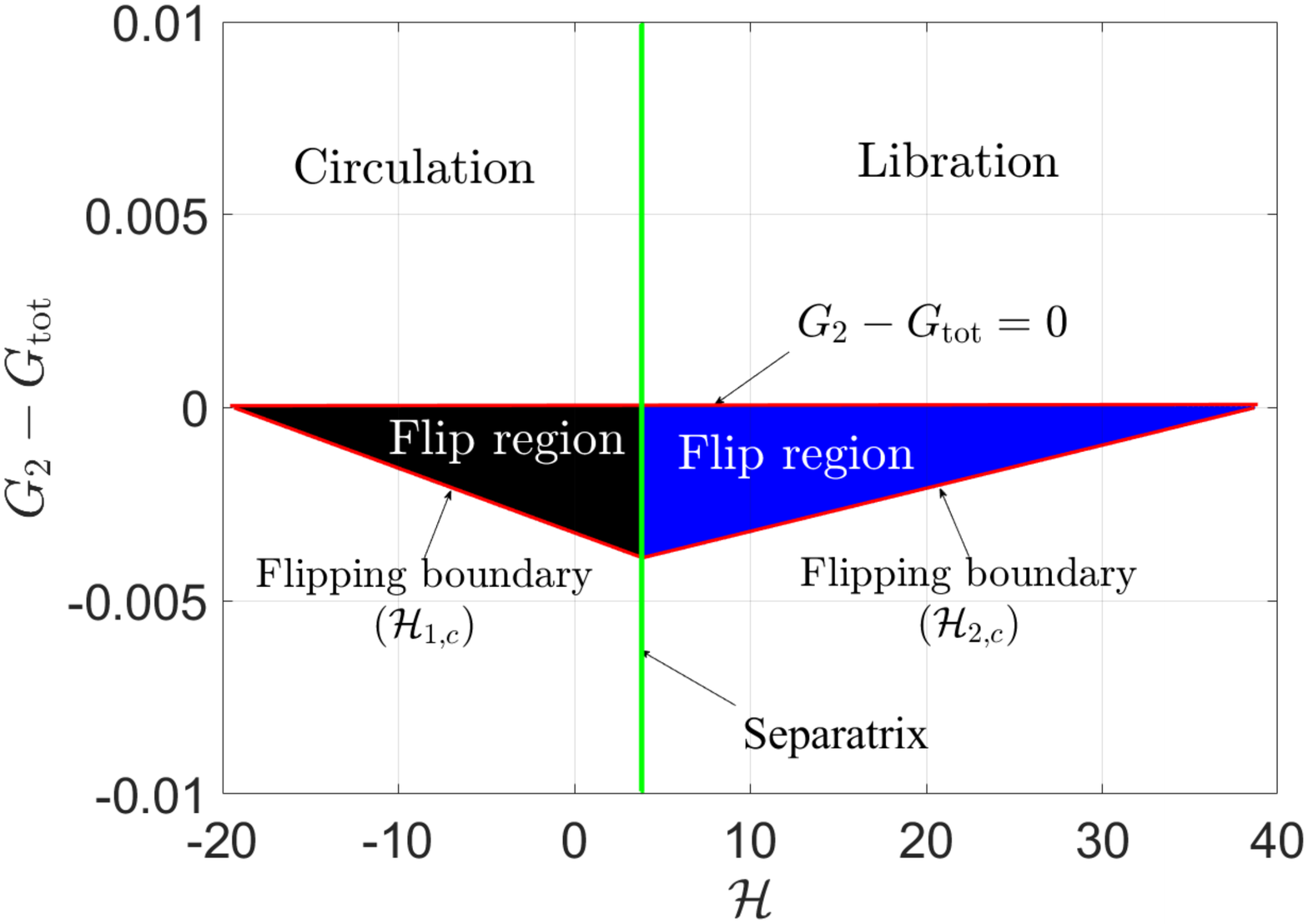}
\caption{Phase portrait of the quadrupole-level Hamiltonian model specified by $G_{\rm tot} = 128.7645$ and $G_2 = 128.7630$ (\emph{left panel}) and flipping regions caused by the quadrupole-order dynamics shown in the (${\cal H},G_2 - G_{\rm tot}$) space for the case of $G_{\rm tot} = 128.7645$ (\emph{right panel}). In the left panel, the location of $i_{\rm tot} = 90^{\circ}$ is marked by red line. In the right panel, flipping orbits corresponding to rotating ZLK cycles are marked in black, flipping orbits corresponding to librating ZLK cycles are marked in blue, and boundaries of flipping region are shown in red lines. The green line shows the dynamical separatrix, dividing the circulation region from the libration one.}
\label{Fig4}
\end{figure*}

In this section, the dynamics of orbit flip is discussed under the quadrupole-level approximation. Flipping orbits are referred to as the ones with mutual inclination $i_{\rm tot}$ switching between prograde and retrograde \citep{naoz2013secular, hamers2021properties}. The critical condition of orbit flip is $i_{\rm tot} = 90^{\circ}$, at which it holds
\begin{equation*}
G_{\rm tot}^2 = G_1^2 + G_2^2.
\end{equation*}
It implies that the critical angular momentum of the inner binary at the flipping moment is
\begin{equation*}
G_{1,c} = \sqrt{G_{\rm tot}^2 - G_2^2},
\end{equation*}
which requires $G_{\rm tot} \ge G_2$. It means that orbit flips may occur in the parameter space with $G_{\rm tot} \ge G_2$, showing that $G_{\rm tot} - G_2 = 0$ provides a boundary of flipping region (see the right panel of Fig. \ref{Fig4}).

To show the flipping behaviors, an example with $G_{\rm tot} = 128.7645$ and $G_2 = 128.7630$ (corresponding to $i_{\rm tot}^0 = 90.1356^{\circ}$ and $e_2 = 0.5965$) is taken. The phase portrait is shown in the left panel of Fig. \ref{Fig4}. For this example, the critical angular momentum of the inner binary is $G_{1,c} = 0.6215$, at which the mutual inclination between the inner and outer binaries' orbits is $i_{\rm tot} = 90^{\circ}$, as shown in red line. It is observed that those ZLK cycles with $G_1 > G_{1,c}$ (or $i_{\rm tot} > 90^{\circ}$) can flip between prograde and retrograde. In the following, we analyse the flipping region in the parameter space spanned by the conserved quantities $(G_2, G_{\rm tot}, {\cal H})$.

Firstly, let us discuss the distribution of flipping orbits corresponding to rotating ZLK cycles. Without loss of generality, the boundary of flipping region is evaluated at $2g_1 = 0$. From the left panel of Fig. \ref{Fig4}, we can see that, under the condition of $2g_1 = 0$, those ZLK cycles with $G_1 \in \left(G_{1,c}, 1\right)$ can achieve flips. Substituting $(g_1 = 0, G_1 = G_{1,c})$ and $(g_1 = 0, G_1 = 1)$ into the quadrupole-level Hamiltonian, we can obtain the lower boundary of flipping orbits corresponding to rotating ZLK cycles as
\begin{equation}\label{Eq11}
\begin{aligned}
{{\cal H}_{1,c}} =&  - \frac{{{\beta ^3}}}{{4G_{1,c}^2G_2^5}}\left\{ {\left( {5 - 3G_{1,c}^2} \right)\left( {3G_e^2 - 4G_{1,c}^2G_2^2} \right)} \right.\\
&\left. { + 15\left( {1 - G_{1,c}^2} \right)\left( {4G_{1,c}^2G_2^2 - G_e^2} \right)} \right\}
\end{aligned}
\end{equation}
where $G_e$ is given by
\begin{equation*}
{G_e} = G_{\rm tot}^2 - G_{1,c}^2 - G_2^2,
\end{equation*}
and the upper boundary of orbit flip (corresponding to the separatrix between rotation and libration) as
\begin{equation}\label{Eq12}
{{\cal H}_{\rm sep}} =  - \frac{{{\beta ^3}}}{{2G_2^5}}\left[ {3{{\left( {G_{\rm tot}^2 - G_2^2 - 1} \right)}^2} - 4G_2^2} \right].
\end{equation}

Secondly, let us discuss the distribution of flipping orbits corresponding to librating ZLK cycles. Without loss of generality, we assume the angle at $2g_1 = \pi$ (angle of the ZLK centre). From the left panel of Fig. \ref{Fig4}, it is observed that, under the condition of $2g_1 = \pi$, those ZLK cycles with $G_1 \in \left(G_{1,c}, 1\right)$ can realize flips. Substituting $(2g_1 = \pi, G_1 = G_{1,c})$ and $(2g_1 = \pi, G_1 = 1)$ into the quadrupole-level Hamiltonian, we can get the lower boundary of orbit flip corresponding to librating ZLK cycles as
\begin{equation}\label{Eq13}
\begin{aligned}
{{\cal H}_{2,c}} = & - \frac{{{\beta ^3}}}{{4G_{1,c}^2G_2^5}}\left\{ {\left( {5 - 3G_{1,c}^2} \right)\left[ {3G_e^2 - 4G_{1,c}^2G_2^2} \right]} \right.\\
&\left. { - 15\left( {1 - G_{1,c}^2} \right)\left[ {4G_{1,c}^2G_2^2 - G_e^2} \right]} \right\}.
\end{aligned}
\end{equation}
and the upper boundary is the same as equation (\ref{Eq12}).

For the case of $G_{\rm tot} = 128.7645$, the right panel of Fig. \ref{Fig4} reports the flipping regions under the quadrupole-order model in the $({\cal H}, G_2 - G_{\rm tot})$ space. In particular, the flipping orbits corresponding to rotating ZLK cycles are shown in black and the ones corresponding to librating ZLK cycles are shown in blue. The line of separatrix (${\cal H}_{\rm sep}$) between circulation and libration is marked in green and the boundaries (${\cal H}_{1,c}$ and ${\cal H}_{2,c}$) are shown in red. As discussed above, orbit flips can happen in the parameter space with $G_2 \le G_{\rm tot}$, thus $G_2 - G_{\rm tot} = 0$ corresponds to another boundary of flipping region (also shown in red line). It is observed that (a) flipping orbits are distributed in the space with $G_2 \le G_{\rm tot}$ and the range in the direction of $G_2 - G_{\rm tot}$ is very narrow (smaller than 0.0039), (b) the flipping region on the left side of the separatrix (black region) is of ZLK circulation (i.e., the flipping orbits correspond to rotating ZLK cycles) and the one on the right side of the separatrix (blue region) is of ZLK libration (i.e., the flipping orbits correspond to librating ZLK cycles), and (c) the area of the flipping region inside ZLK resonance is larger than that outside ZLK resonance.

\begin{figure*}
\centering
\includegraphics[width=0.48\textwidth]{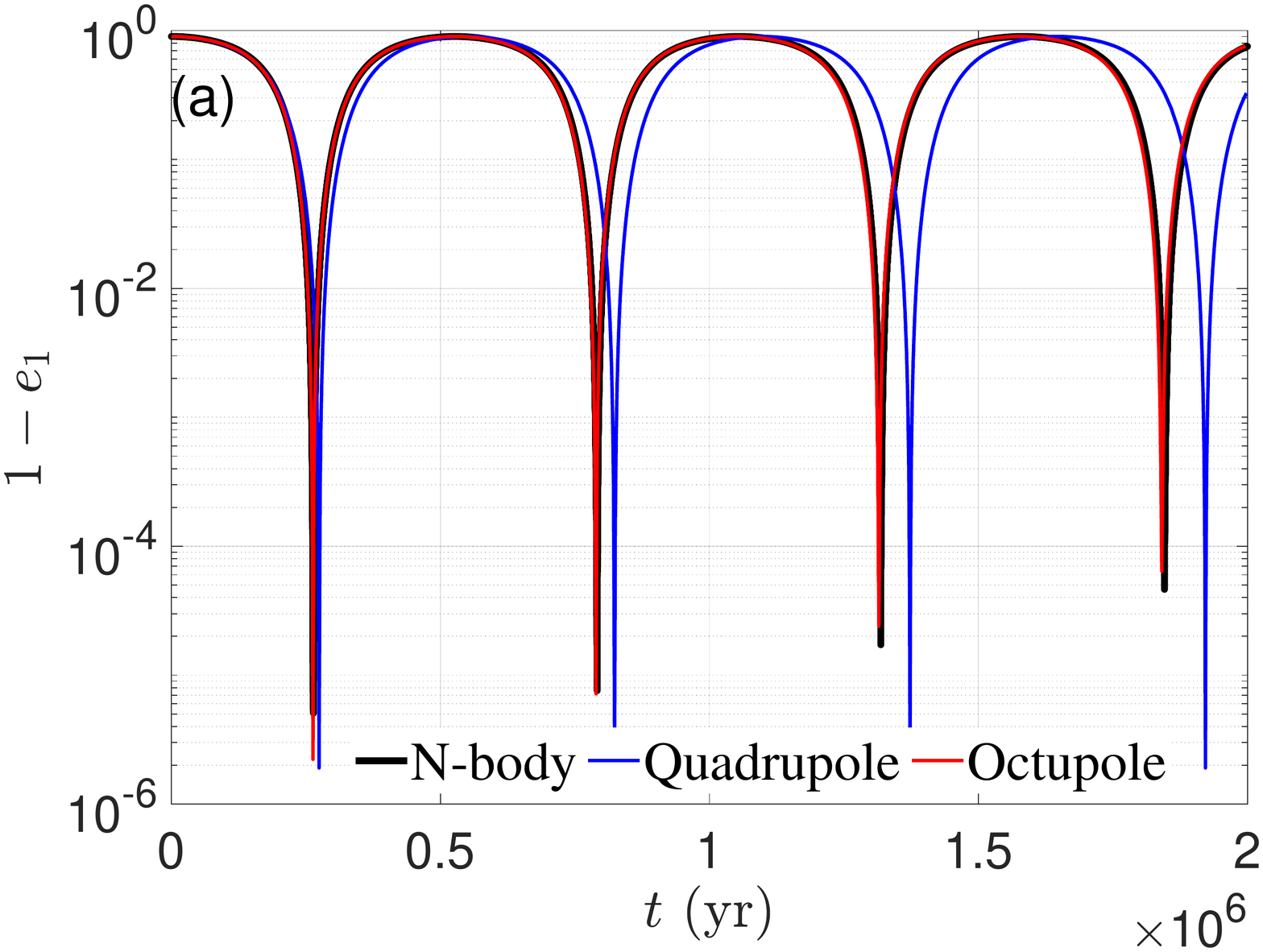}
\includegraphics[width=0.48\textwidth]{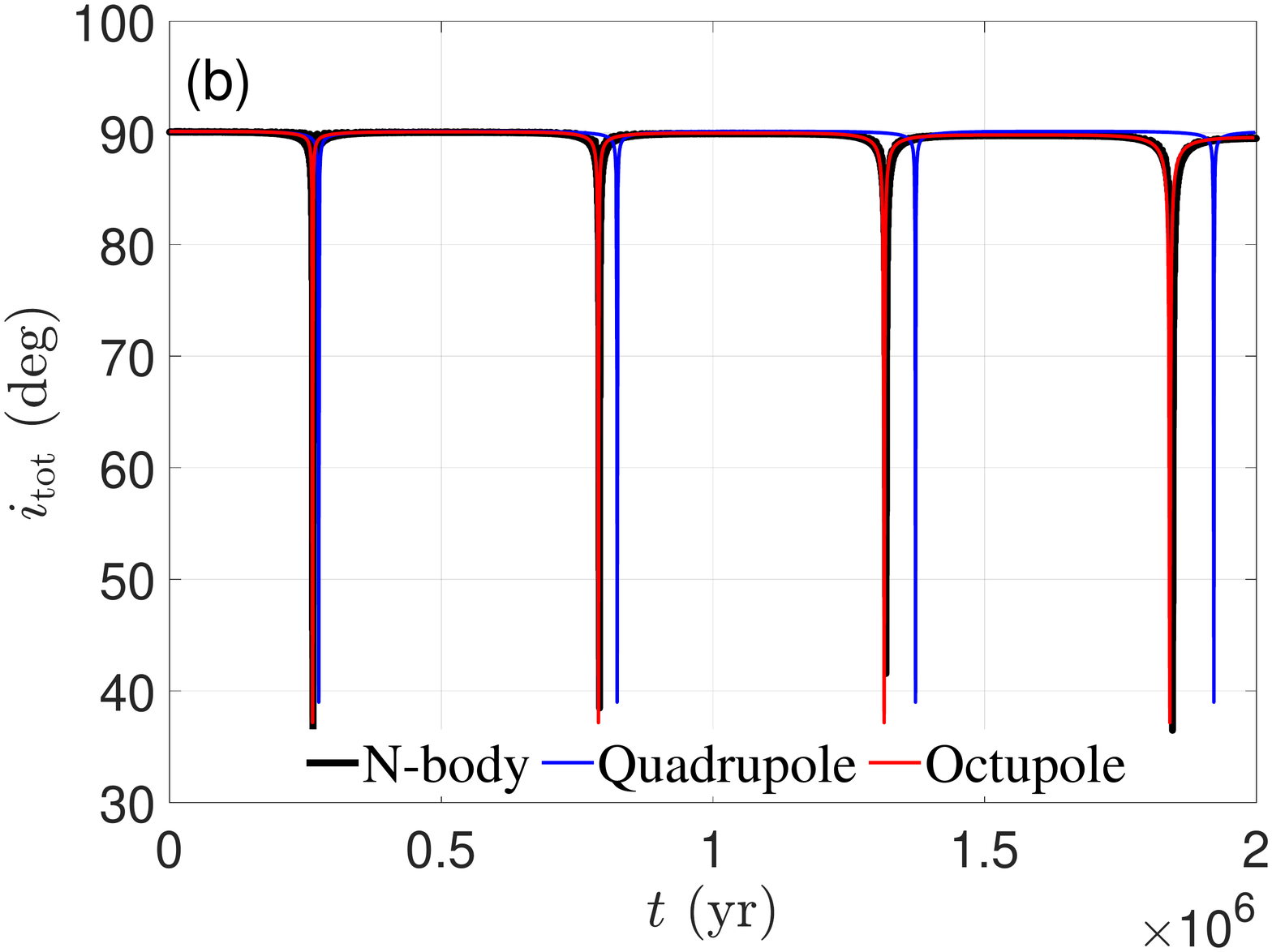}
\caption{Time evolutions of the eccentricities and mutual inclinations for a representative example of flipping orbits dominated by the quadrupole-order dynamics, produced under different dynamical models. The initial eccentricities are taken as $e_1 = 0.1$, the initial angular momentum of the outer planet is taken as $G_2 = 128.7630$, the initial arguments of pericenter are $\omega_1 = \omega_2 = 0$(meaning that this example is located in the ZLK circulation zone shown in the right panel of Fig. \ref{Fig4}).}
\label{Fig4_1}
\end{figure*}

Taking an example from the ZLK circulation region shown in the right panel of Fig. \ref{Fig4}, we performed numerical integrations under different dynamical models. The time evolutions of the eccentricities and mutual inclinations are reported in Fig. \ref{Fig4_1} (please refer to the caption for the initial setting), which shows that the orbit can flip from retrograde to prograde, and vice versa. The results produced from different dynamical models can agree well, implying that the flipping behaviours of this example is dominated by the quadrupole-order dynamics.

\section{Dynamics of octupole-order resonance}
\label{Sect4}

Up to the octupole order in semimajor axis ratio, the Hamiltonian is composed of the quadrupole-order term and the octupole-order term, denoted by
\begin{equation}\label{Eq14}
{\cal H}\left( {{g_1},{g_2},{G_1},{G_2}} \right) = {{\cal H}_2}\left( {{g_1},{G_1},{G_2}} \right) + {{\cal H}_3}\left( {{g_1},{g_2},{G_1},{G_2}} \right)
\end{equation}
where the dynamical model determined by ${\cal H}_2$ is integrable and the associated dynamics are investigated in the previous section. Under the quadrupole-level approximation, only the ZLK resonance occurs in the phase space. With the inclusion of the octupole-order term, the resulting dynamical model is not integrable and the octupole-order resonances appear. In this section, our purpose is to study the dynamics of octupole-order resonance by means of perturbative treatments \citep{henrard1986perturbation, henrard1990semi}.

From the viewpoint of perturbative treatments, we take the quadrupole-order term in the Hamiltonian (integrable part) as the unperturbed dynamical model and take the octupole-order term in the Hamiltonian as the perturbation to the quadrupole-order dynamics \citep{henrard1990semi}. The magnitude of perturbation is measured by the small parameter $\epsilon$. In order to study octupole-order resonances by using perturbative treatments, we introduce a new set of canonical variables $(g_1^*,g_2^*,G_1^*,G_2^*)$. After transformation, the quadrupole-order Hamiltonian ${\cal H}_2$ is independent on the angular coordinates $g_1^*$ and $g_2^*$ (here ${\cal H}_2$ is called the kernel function of the Hamiltonian model). Based on such a transformation, it is possible to formulate the resonant Hamiltonian for octupole-order resonances by means of averaging technique. Averaging approximation corresponds to the lowest-order perturbation theory.

\subsection{Nominal location of octupole-order resonance}
\label{Sect4_1}

Under the quadrupole-order Hamiltonian flow, we introduce the following transformation,
\begin{equation}\label{Eq15}
\begin{aligned}
g_1^* &= {g_1} - {\rho _1}\left( {t,G_1^*,G_2^*} \right),\quad G_1^* = \frac{1}{{2\pi }}\int\limits_{0}^{2\pi } {{G_1}{\rm d}{g_1}}\\
g_2^* &= {g_2} - {\rho _2}\left( {t,G_1^*,G_2^*} \right),\quad G_2^* = {G_2}
\end{aligned}
\end{equation}
which is canonical with the generating function
\begin{equation}\label{Eq16}
S_1\left( {{g_1},{g_2},G_1^*,G_2^*} \right) = {g_2}G_2^* + \int {{G_1}\left( {{{\cal H}_2}\left( {G_1^*,G_2^*} \right),{g_1},G_2^*} \right){\rm d}{g_1}}.
\end{equation}
In equation (\ref{Eq15}), $G_1^*$ is called Arnold action, which corresponds to the area in the phase space bounded by the rotating ZLK cycle (divided by $2\pi$), and ${\rho _1}$ and ${\rho _2}$ are periodic functions with the same period of the rotating ZLK cycle. Let us denote the period of the rotating ZLK cycle as $T$. The new angle $g_1^*$ is a linear function of time, given by $g_1^* = \frac{2\pi}{T} t$.

According to the generating function given by equation (\ref{Eq16}), we can get the transformation between old and new sets of angles as
\begin{equation*}
g_1^* = \frac{{\partial S_1}}{{\partial G_1^*}},\quad g_2^* = \frac{{\partial S_1}}{{\partial G_2^*}}
\end{equation*}
showing that the periodic functions ${\rho _1}$ and ${\rho _2}$ can be expressed as
\begin{equation*}
\begin{aligned}
{\rho _1} &= {g_1} - \frac{\partial }{{\partial G_1^*}}\int {{G_1}\left( {{{\cal H}_2}\left( {G_1^*,G_2^*} \right),{g_1},G_2^*} \right){\rm d}{g_1}}\\
&= \int {\left[ {\frac{{\partial {{\cal H}_2}}}{{\partial {G_1}}} - \left\langle {\frac{{\partial {{\cal H}_2}}}{{\partial {G_1}}}} \right\rangle } \right]{\rm d}t}\\
{\rho _2} &=  - \frac{\partial }{{\partial G_2^*}}\int {{G_1}\left( {{{\cal H}_2}\left( {G_1^*,G_2^*} \right),{g_1},G_2^*} \right){\rm d}{g_1}}\\
&= \int {\left[ {\frac{{\partial {{\cal H}_2}}}{{\partial {G_2}}} - \left\langle {\frac{{\partial {{\cal H}_2}}}{{\partial {G_2}}}} \right\rangle } \right]{\rm d}t}.
\end{aligned}
\end{equation*}
with
\begin{equation*}
\begin{aligned}
\left\langle {\frac{{\partial {{\cal H}_2}}}{{\partial {G_1}}}} \right\rangle  &= \frac{1}{T}\int\limits_0^T {\frac{{\partial {{\cal H}_2}}}{{\partial {G_1}}}{\rm d}t} = {\dot g}_1^*,\\
\left\langle {\frac{{\partial {{\cal H}_2}}}{{\partial {G_2}}}} \right\rangle  &= \frac{1}{T}\int\limits_0^T {\frac{{\partial {{\cal H}_2}}}{{\partial {G_2}}}{\rm d}t} = {\dot g}_2^*.
\end{aligned}
\end{equation*}
At the moment of $t=0$ and $t=T$, it holds
\begin{equation*}
\begin{aligned}
{\rho _1}\left( {0,G_1^*,G_2^*} \right) &= {\rho _1}\left( {{T},G_1^*,G_2^*} \right) = 0,\\
{\rho _2}\left( {0,G_1^*,G_2^*} \right) &= {\rho _2}\left( {{T},G_1^*,G_2^*} \right) = 0
\end{aligned}
\end{equation*}
which shows that the old and new sets of angles are the same at the initial instant and at one period \citep{henrard1990semi}.

\begin{figure*}
\centering
\includegraphics[width=0.48\textwidth]{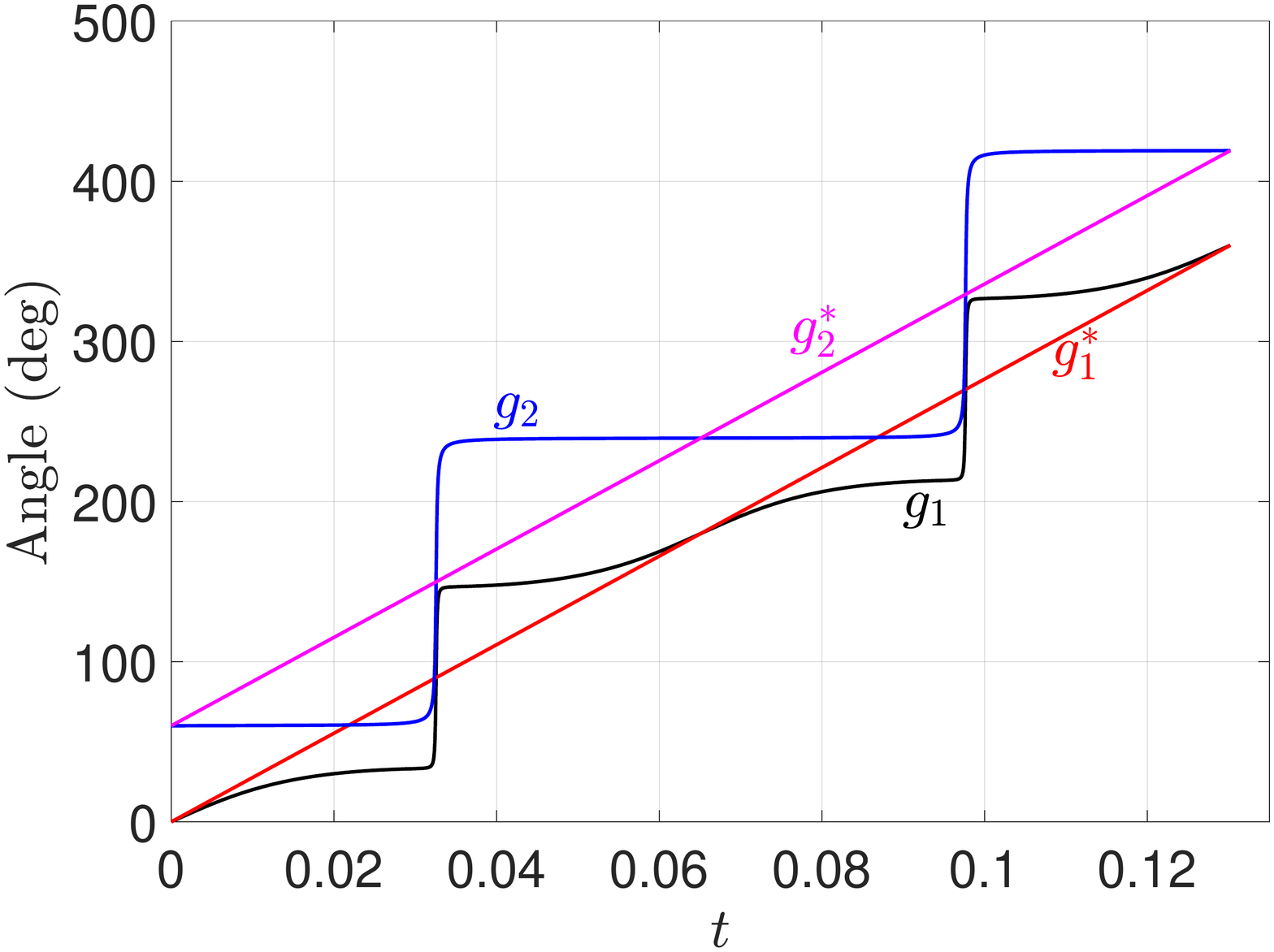}
\includegraphics[width=0.48\textwidth]{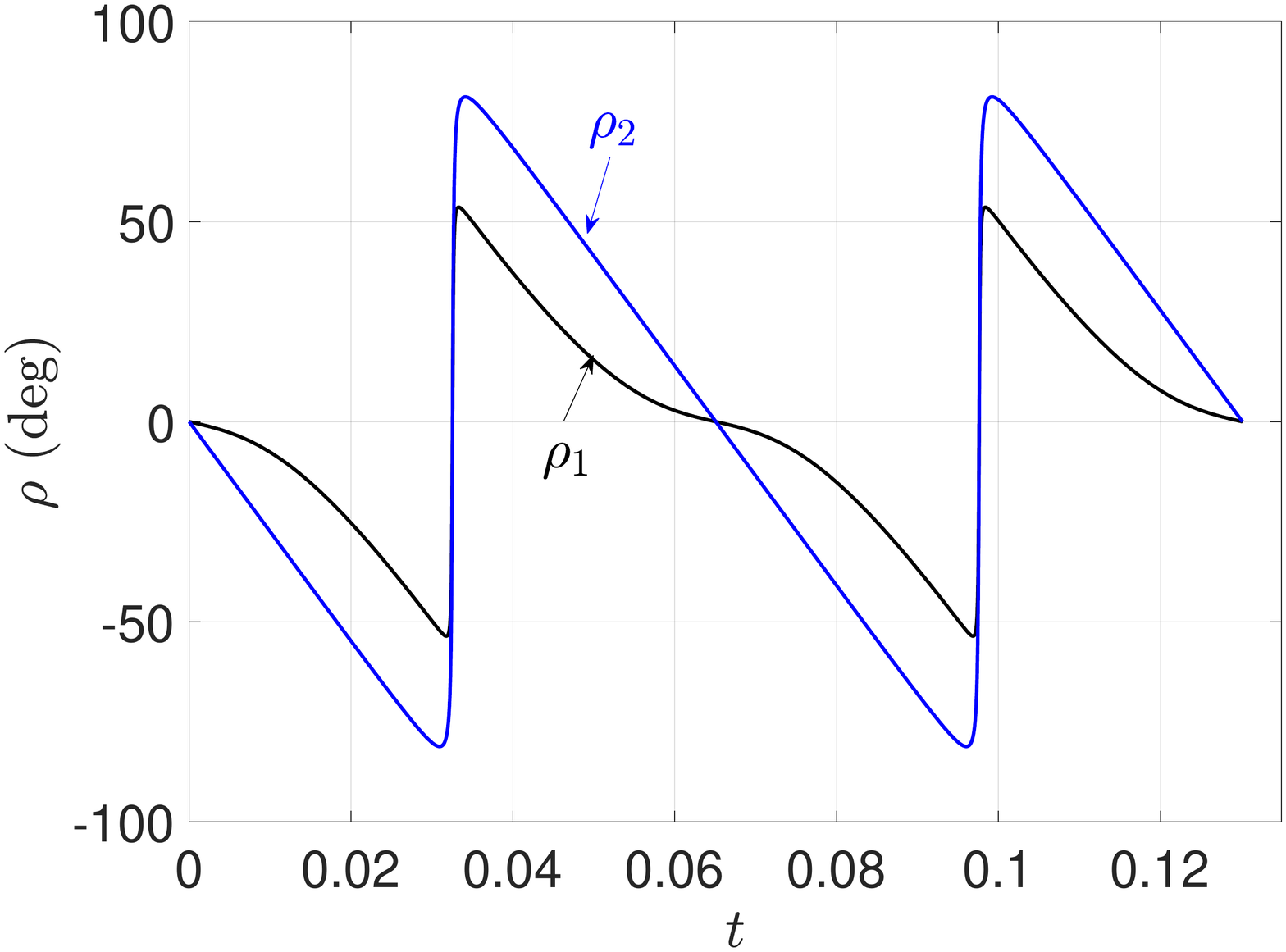}
\caption{The old and new sets of angles $(g_1, g_2)$ and $(g_1^*, g_2^*)$ during one period of the rotating ZLK cycle (\emph{left panel}) and their differences $\rho_1 = g_1 - g_1^*$ and $\rho_2 = g_2 - g_2^*$ as functions of time (\emph{right panel}). It is observed that the periodic functions $\rho_1$ and $\rho_2$ are equal to zero when the time $t$ is equal to $0$, $T/4$, $T/2$, $3T/4$ and $T$, meaning that at these special instants the old and new sets of angles are consistent.}
\label{Fig5}
\end{figure*}

Figure \ref{Fig5} shows the time histories of $g_{1,2}$ and $g_{1,2}^*$ (see the left panel) and time histories of $\rho_1$ and $\rho_2$ (see the right panel). It is observed that (a) $g_{1,2}^*$ are linear functions of time, (b) $g_{1,2}$ have the same period equal to $T$, (c) $\rho_{1,2}$ are periodic functions with the same period of the rotating ZLK cycle, and (d) $\rho_{1,2}$ are equal to zero when the time is at $t=0$, $T/4$, $T/2$, $3T/4$ and $T$.

Through such a canonical transformation,
\begin{equation*}
\left( {{g_1},{g_2},{G_1},{G_2}} \right) \leftrightarrow \left( {g_1^*,g_2^*,G_1^*,G_2^*} \right)
\end{equation*}
the quadrupole-level Hamiltonian becomes \citep{henrard1990semi}
\begin{equation}\label{Eq17}
{{\cal H}_2}\left( {{g_1},{G_1},{G_2}} \right) = {{\cal H}_2}\left( {G_1^*,G_2^*} \right)
\end{equation}
which shows that, under the quadrupole-level approximation, the actions $G_1^*$ and $G_2^*$ are constant and the angular coordinates $g_1^*$ and $g_2^*$ are linear functions of time.

The quadrupole-order Hamiltonian ${\cal H}_2 (G_1^*, G_2^*)$ produces the fundamental frequencies, given by
\begin{equation}\label{Eq18}
\begin{aligned}
\dot g_1^* &= \frac{{\partial {{\cal H}_2}}}{{\partial G_1^*}} = \frac{1}{T}\int\limits_0^T {\frac{{\partial {{\cal H}_2}}}{{\partial {G_1}}}{\rm d}t} = \frac{2\pi}{T},\\
\dot g_2^* &= \frac{{\partial {{\cal H}_2}}}{{\partial G_2^*}} = \frac{1}{T}\int\limits_0^T {\frac{{\partial {{\cal H}_2}}}{{\partial {G_2}}}{\rm d}t} = \frac{g_2 (T)}{T}.
\end{aligned}
\end{equation}
Thus, the nominal location of octupole-order resonance between $g_1^*$ and $g_2^*$ can be identified by the following resonant condition,
\begin{equation}\label{Eq19}
{k_1}\dot g_1^* + {k_2}\dot g_2^* = 0
\end{equation}
where $k_1$ and $k_2$ are integers.

\begin{figure*}
\centering
\includegraphics[width=0.48\textwidth]{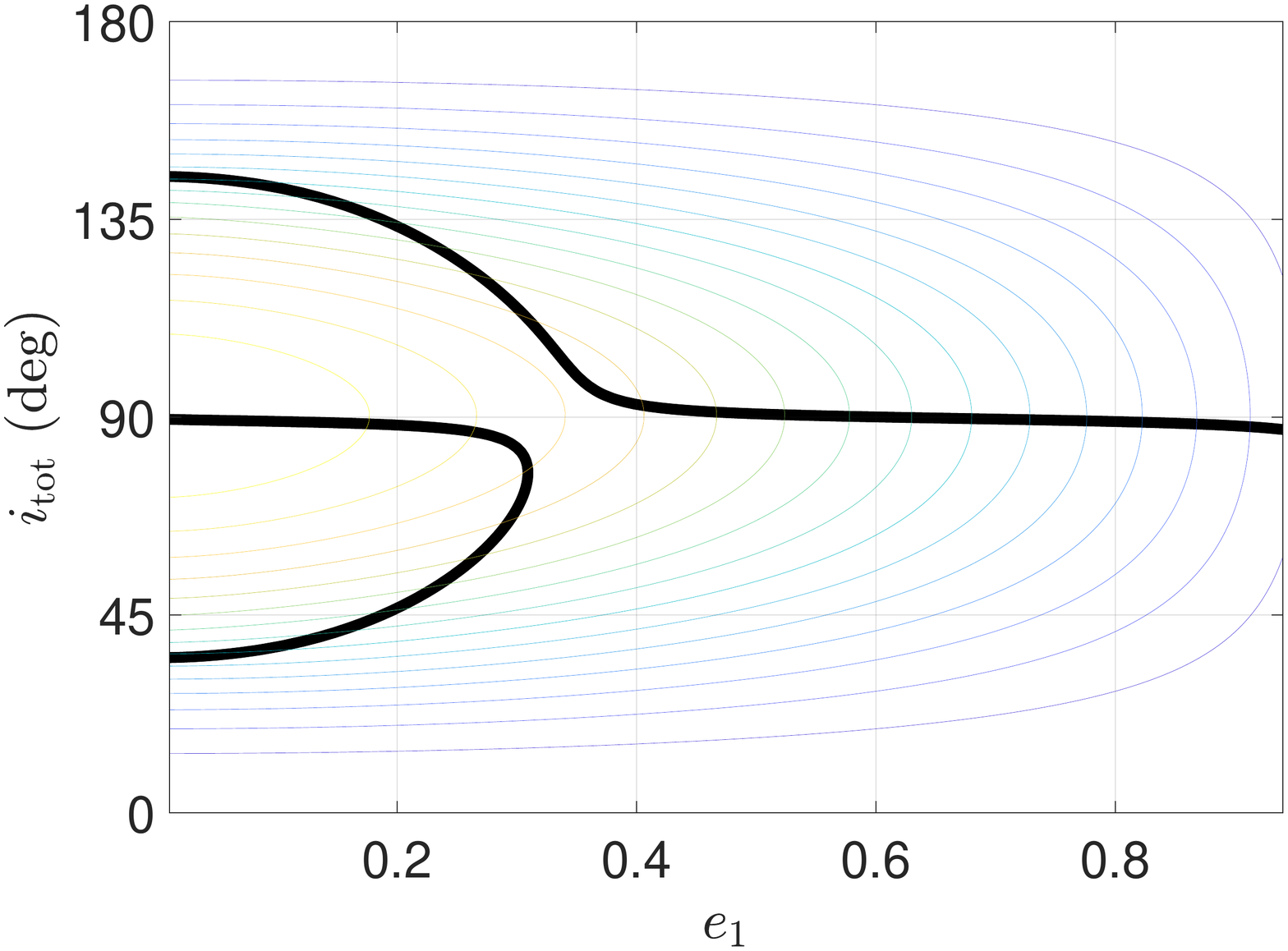}
\includegraphics[width=0.48\textwidth]{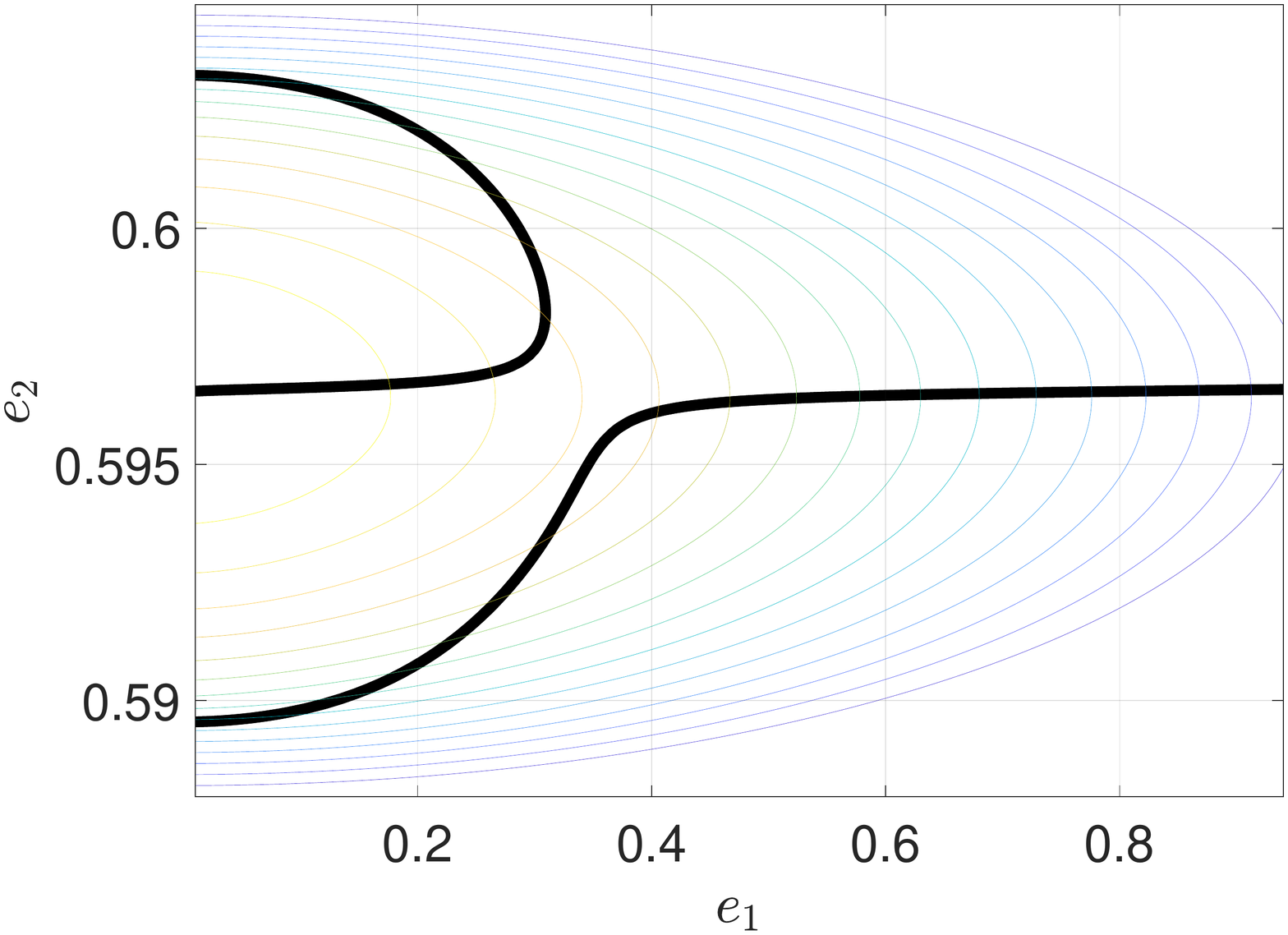}
\caption{Nominal location of octupole-order resonance determined by the fundamental frequencies (i.e., solutions of the resonant equation ${\dot g}_2^*-{\rm sign}(\cos{i_{\rm tot}}) {\dot g}_1^* = 0$) as well as the level curves of the motion integral $\Sigma_2$ shown in the $(e_1, i_{\rm tot})$ space (\emph{left panel}) and in the $(e_1,e_2)$ space (\emph{right panel}). For this example, the total angular momentum is taken as $G_{\rm tot} = 128.7645$.}
\label{Fig6}
\end{figure*}

In Fig. \ref{Fig6}, a representative example with $G_{\rm tot} = 128.7645$ is considered. The nominal location of the secular resonance associated with ${\dot g}_2^*-{\rm sign}(\cos{i_{\rm tot}}) {\dot g}_1^* = 0$ is shown in the $(e_1, i_{\rm tot})$ space (see the left panel) and in the $(e_1,e_2)$ space (see the right panel). It shows that the octupole-order resonance with critical argument of $\sigma = g_2^* - {\rm sign}(\cos{i_{\rm tot}}) g_1^*$ happens in the considered parameter space. There are two branches of libration centres: one branch is located in the low-eccentricity region and the other one occupies the entire range of eccentricity.

Let us discuss the physical essence of the critical argument $\sigma$. According to the general definition of the longitude of pericentre \citep{shevchenko2016lidov}, it holds
\begin{equation*}
\varpi_1 = \Omega_1 + {\rm sign}(\cos{i_1}) \omega_1
\end{equation*}
for the inner binary and
\begin{equation*}
\varpi_2 = \Omega_2 + {\rm sign}(\cos{i_2}) \omega_2
\end{equation*}
for the outer binary. The difference of longitude of pericentre between inner and outer binaries can be written as
\begin{equation*}
\begin{aligned}
\Delta \varpi =& \Omega_2 - \Omega_1 + {\rm sign}(\cos{i_2}) \omega_2 - {\rm sign}(\cos{i_1}) \omega_1\\
=&\pi + {\rm sign}(\cos{i_2}) \omega_2 - {\rm sign}(\cos{i_1}) \omega_1.
\end{aligned}
\end{equation*}
For the dynamical model discussed in the present work, the outer binary holds the dominant angular momentum (i.e., $H_2 \gg H_1$), meaning that $i_2 \approx 0$ and $i_1 \approx i_{\rm tot}$. As a result, the difference of longitude of pericentre $\Delta \varpi$ can be further written as
\begin{equation*}
\begin{aligned}
\Delta \varpi =& \pi + \omega_2 - {\rm sign}(\cos{i_{\rm tot}}) \omega_1\\
=&\pi + g_2 - {\rm sign}(\cos{i_{\rm tot}}) g_1
\end{aligned}
\end{equation*}
Thus, we can get that the critical argument is $\sigma = g_2^* - {\rm sign}(\cos{i_{\rm tot}}) g_1^* = \varpi_2^* - \varpi_1^* - \pi$. It means that the octupole-order resonance with critical argument of $\sigma$ is the so-called apsidal resonance.

In the next subsection, the resonant Hamiltonian associated with $\sigma = g_2^* - {\rm sign}(\cos{i_{\rm tot}}) g_1^*$ is formulated in order to study the resonant dynamics of apsidal resonance.

\subsection{Hamiltonian model of octupole-order resonance}
\label{Sect4_2}

Under the new set of canonical variables $\left( {g_1^*,g_2^*,G_1^*,G_2^*} \right)$, the Hamiltonian up to the octupole order becomes
\begin{equation}\label{Eq20}
{\cal H}\left( {g_1^*,g_2^*,G_1^*,G_2^*} \right) = {{\cal H}_2}\left( {G_1^*,G_2^*} \right) +  {{\cal H}_3}\left( {g_1^*,g_2^*,G_1^*,G_2^*} \right).
\end{equation}
From the viewpoint of perturbative treatments, the quadrupole-order term ${{\cal H}_2}\left( {G_1^*,G_2^*} \right)$ determines the unperturbed dynamical model (the corresponding Hamiltonian is called the Kernel function), and the octupole-order term ${{\cal H}_3}\left( {g_1^*,g_2^*,G_1^*,G_2^*} \right)$ plays the role of perturbation to the quadrupole-order dynamics.

In order to study the dynamics of octupole-order resonance, we introduce the following canonical transformation,
\begin{equation}\label{Eq21}
\begin{aligned}
{\sigma _1} &= g_2^* - {\rm sign}\left( {\cos i_{\rm tot}^{}} \right)g_1^*,\quad {\Sigma _1} = G_2^*\\
{\sigma _2} &= g_1^*,\quad {\Sigma _2} = G_1^* + {\rm sign}\left( {\cos i_{\rm tot}} \right)\left( {G_2^* - {G_{\rm tot}}} \right)
\end{aligned}
\end{equation}
with the generating function,
\begin{equation*}
S_2\left( {g_1^*,g_2^*,{\Sigma _1},{\Sigma _2}} \right) = g_2^*{\Sigma _1} - g_1^*\left[ {{\rm sign}\left( {\cos {i_{\rm tot}}} \right){\Sigma _1} + {\Sigma _2}} \right].
\end{equation*}
Under such a transformation, the Hamiltonian can be further written as
\begin{equation}\label{Eq22}
{\cal H}\left( {\sigma_1, \sigma_2, \Sigma_1, \Sigma_2} \right) = {{\cal H}_2}\left( {\Sigma_1, \Sigma_2} \right) + {{\cal H}_3}\left( {\sigma_1, \sigma_2, \Sigma_1, \Sigma_2} \right),
\end{equation}
where $\sigma_1$ is the resonant angle.

When the planets are inside a resonance, the resonant angle $\sigma_1$ becomes a long-period variable, while the angle $\sigma_2$ is a short-period variable. This is a typical separable Hamiltonian model \citep{henrard1990semi}. In order to study the resonant dynamics, it is reasonable to remove those short-period effects by means of averaging technique (corresponding to the lowest-order perturbation theory) to formulate the resonant Hamiltonian as follows:
\begin{equation}\label{Eq23}
{{\cal H}^*}\left( {{\sigma _1},{\Sigma _1},{\Sigma _2}} \right) = \frac{1}{{2\pi }}\int\limits_0^{2\pi } {{\cal H}\left( {{\sigma _1},{\sigma _2},{\Sigma _1},{\Sigma _2}} \right){\rm d}{\sigma _2}},
\end{equation}
Under the resonant model specified by the Hamiltonian of equation(\ref{Eq23}), the angle $\sigma_2$ becomes a cyclic variable, so that its conjugate momentum $\Sigma_2$ becomes a constant of motion. Thus, the dynamical model given by the resonant Hamiltonian is of one degree of freedom, depending on the motion integral $\Sigma_2$. Please refer to Fig. \ref{Fig6} for the level curves of the motion integral $\Sigma_2$ in the $(e_1, i_{\rm tot})$ space and in the $(e_1,e_2)$ space.

In the long-term evolution, the Hamiltonian ${\cal H}^*$, the total angular momentum $G_{\rm tot}$ and the action $\Sigma_2$ are conserved. In terms of orbit elements, the constant of motion $\Sigma_2$ can be expressed as
\begin{equation*}
\begin{aligned}
\Sigma_2 &= G_1^* +{\rm sign}\left(\cos i_{\rm tot}\right)\left(G_2^* - G_{\rm tot}\right)\\
&=\sqrt{1-e_1^2} + {\rm sign}\left(\cos i_{\rm tot}\right)\left(\beta \sqrt{1-e_2^2} - G_{\rm tot}\right)
\end{aligned}
\end{equation*}
where the total angular momentum is given by
\begin{equation*}
G_{\rm tot}^2 = \left( {1 - e_1^2} \right) + {\beta ^2}\left( {1 - e_2^2} \right) + 2\beta \sqrt {\left( {1 - e_1^2} \right)\left( {1 - e_2^2} \right)} \cos {i_{\rm tot}}.
\end{equation*}
Assuming the eccentricity $e_1$ as zero and taking the eccentricity $e_2$ as its initial value ($e_{2,0}$), we express the constant of motion in the prograde space as
\begin{equation}\label{Eq24}
\begin{aligned}
\Sigma_2 &=1 + \beta \sqrt{1-e_{2,0}^2}\\
&-\sqrt{1 + {\beta ^2}\left( {1 - e_{2,0}^2} \right) + 2\beta \sqrt {1 - e_{2,0}^2} \cos i_{\rm tot}^0}
\end{aligned}
\end{equation}
where $i_{\rm tot}^0$ is the minimum mutual inclination between the orbits of the inner and outer binaries and we express the constant of motion in the retrograde space as
\begin{equation}\label{Eq24_1}
\begin{aligned}
\Sigma_2 = &1 - \beta \sqrt{1-e_{2,0}^2}\\
&+ \sqrt{1 + {\beta ^2}\left( {1 - e_{2,0}^2} \right) + 2\beta \sqrt {1 - e_{2,0}^2} \cos i_{\rm tot}^0}
\end{aligned}
\end{equation}
where $i_{\rm tot}^0$ is the maximum mutual inclination between the orbits of the inner and outer binaries. Equations (\ref{Eq24}) and (\ref{Eq24_1}) provide the correspondence between $\Sigma_2$ and $i_{\rm tot}^0$. Both the minimum mutual inclination and the maximum mutual inclination can be used to characterise the constant of motion $\Sigma_2$. In the following discussion, we often use the minimum mutual inclination $i_{\rm tot}^0$ in the prograde space (or the maximum mutual inclination $i_{\rm tot}^0$ in the retrograde space) to specify the motion integral $\Sigma_2$.

\subsection{Dynamical structures of octupole-order resonance}
\label{Sect4_3}

\begin{figure*}
\centering
\includegraphics[width=0.48\textwidth]{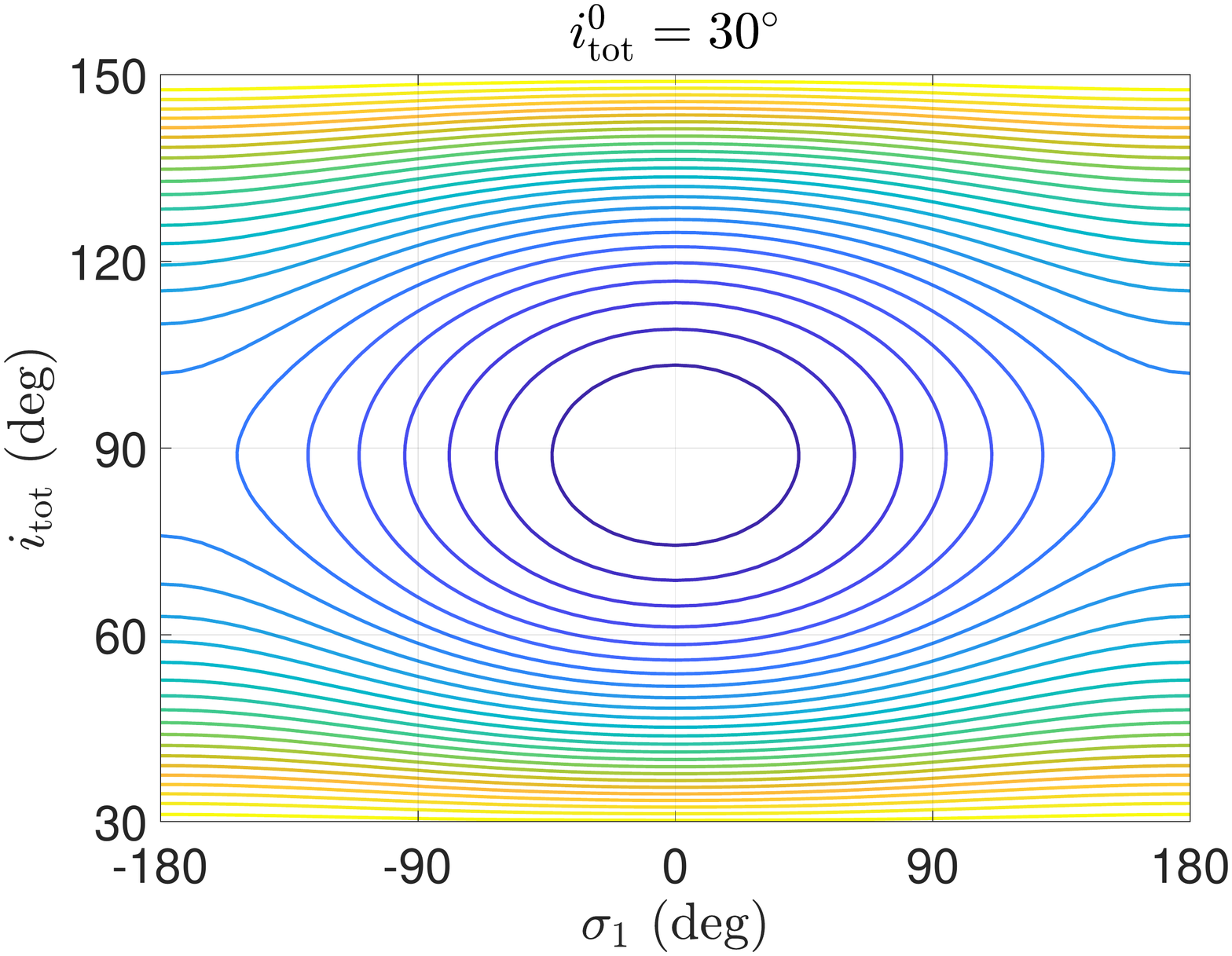}
\includegraphics[width=0.48\textwidth]{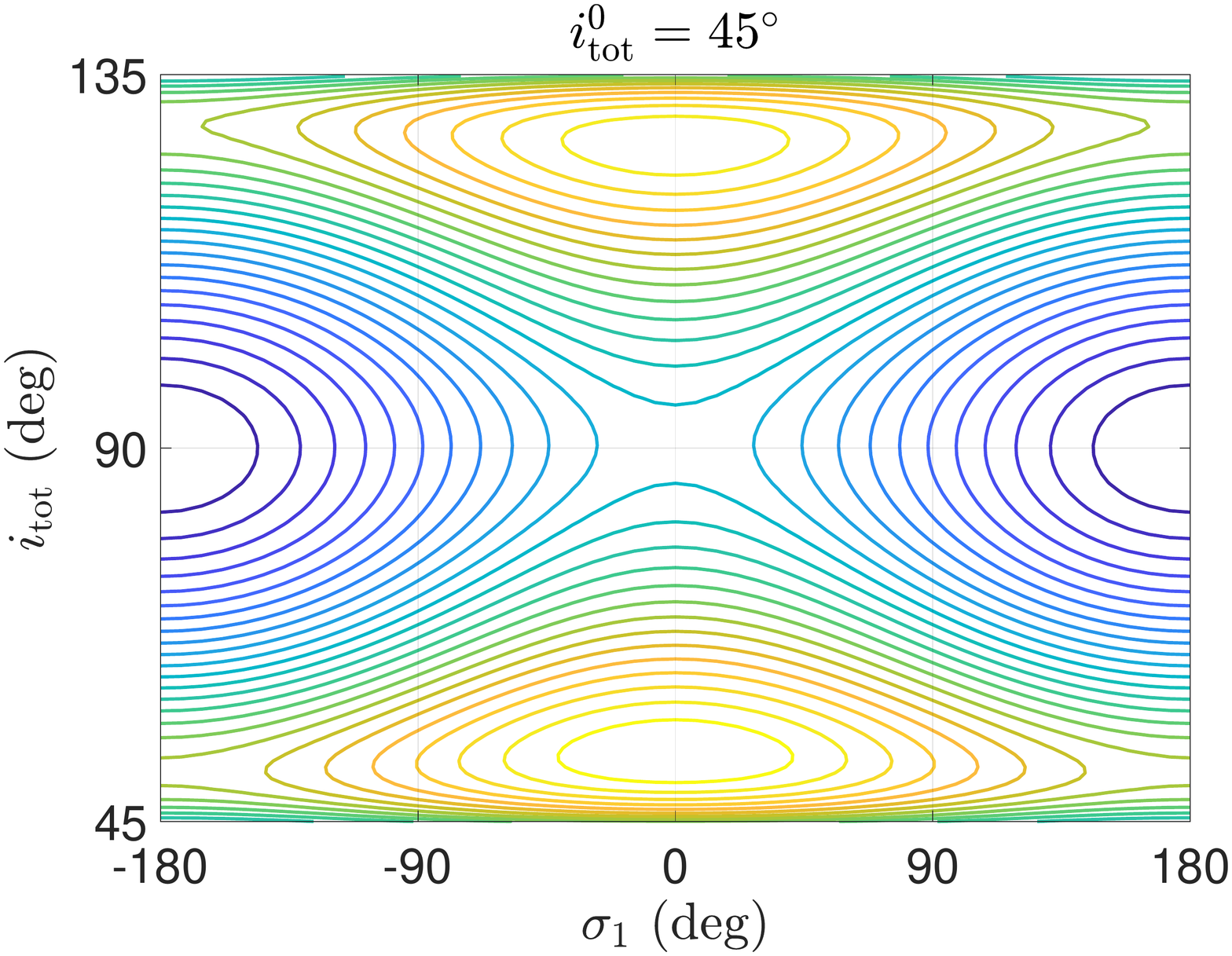}\\
\includegraphics[width=0.48\textwidth]{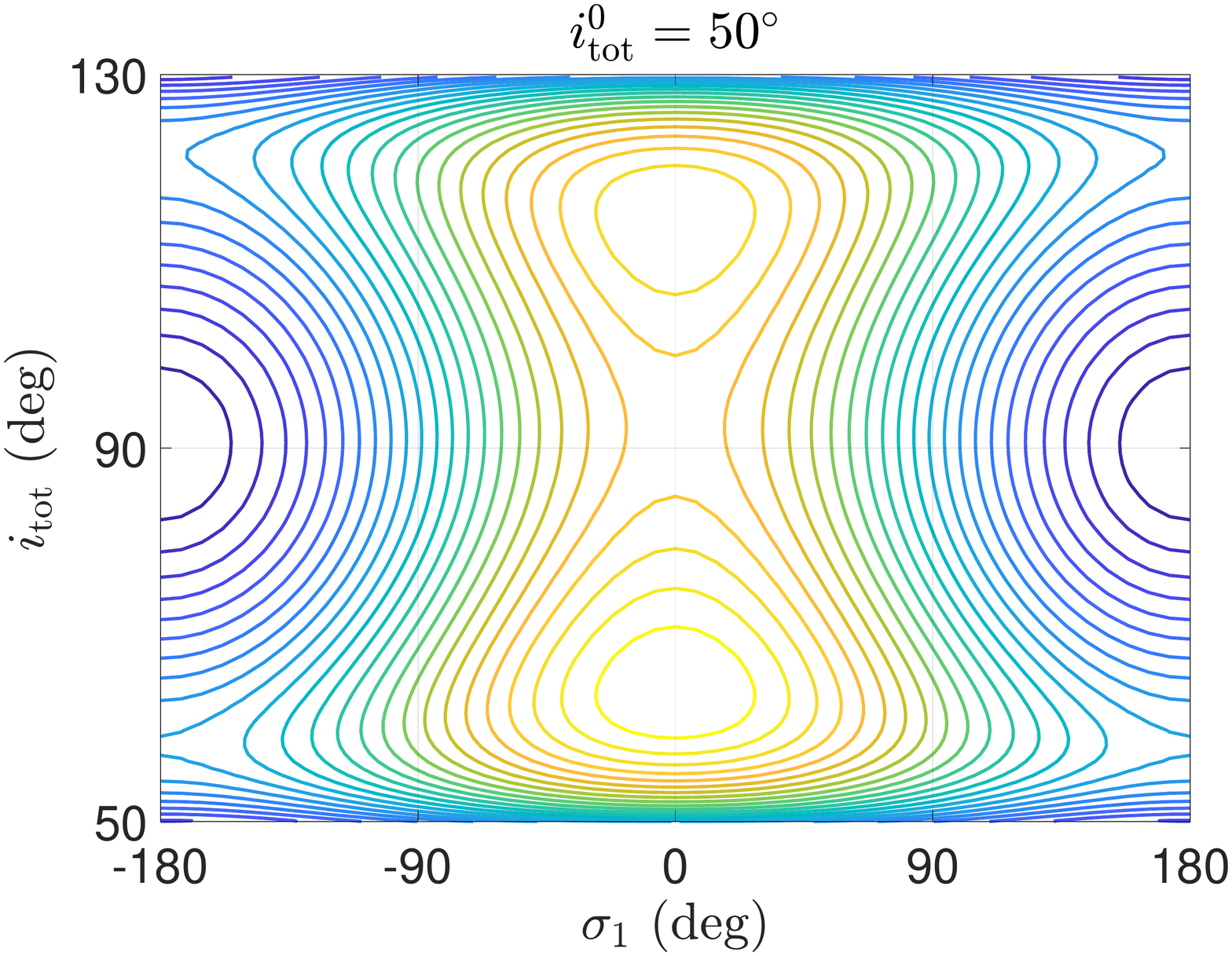}
\includegraphics[width=0.48\textwidth]{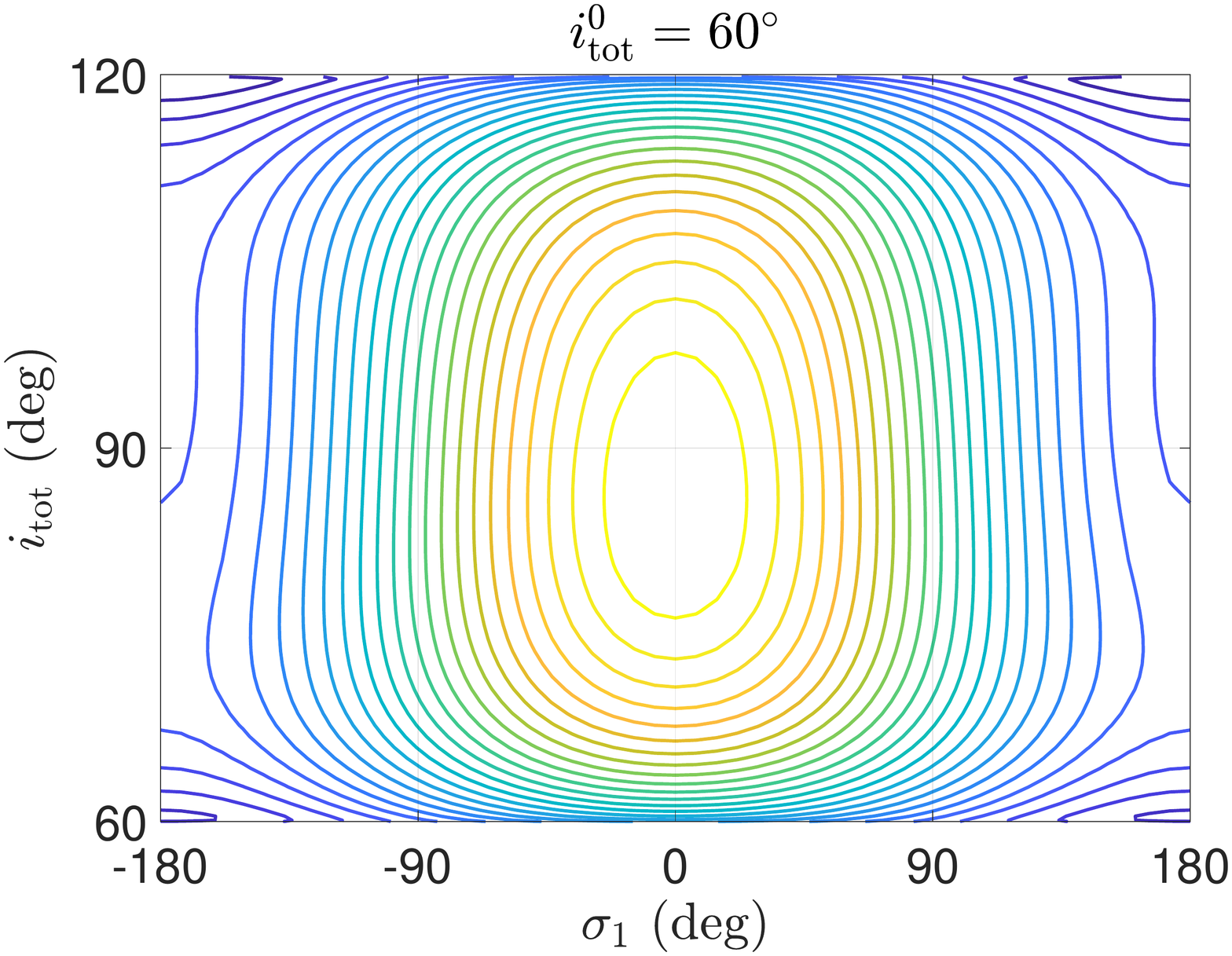}
\caption{Level curves of resonant Hamiltonian in the $(\sigma_1,i_{\rm tot})$ space (i.e., phase portraits). The motion integral $\Sigma_2$ is specified by the minimum mutual inclination $i_{\rm tot}^0$. Practical motions take place along the isolines of resonant Hamiltonian.}
\label{Fig7}
\end{figure*}

\begin{figure*}
\centering
\includegraphics[width=0.48\textwidth]{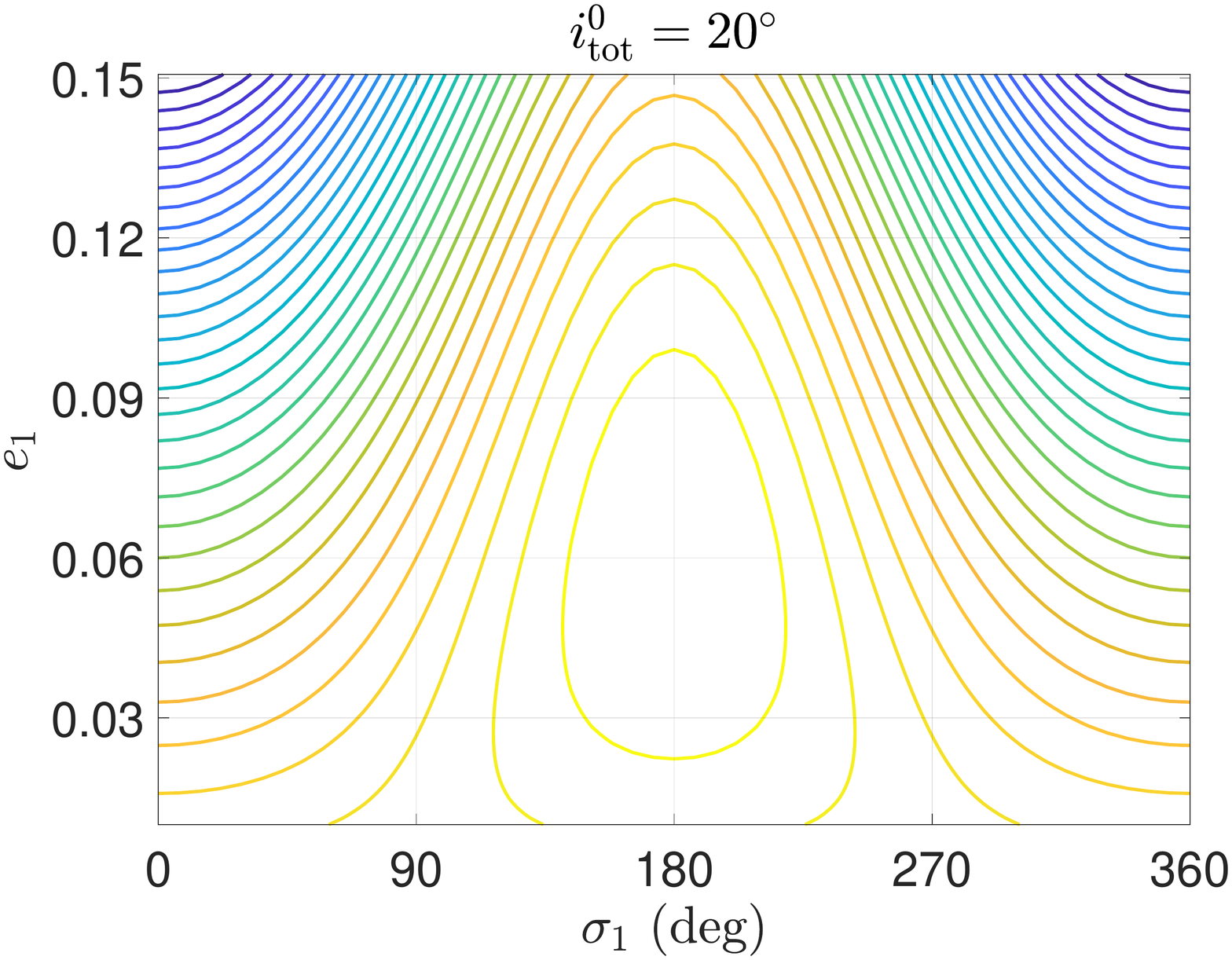}
\includegraphics[width=0.48\textwidth]{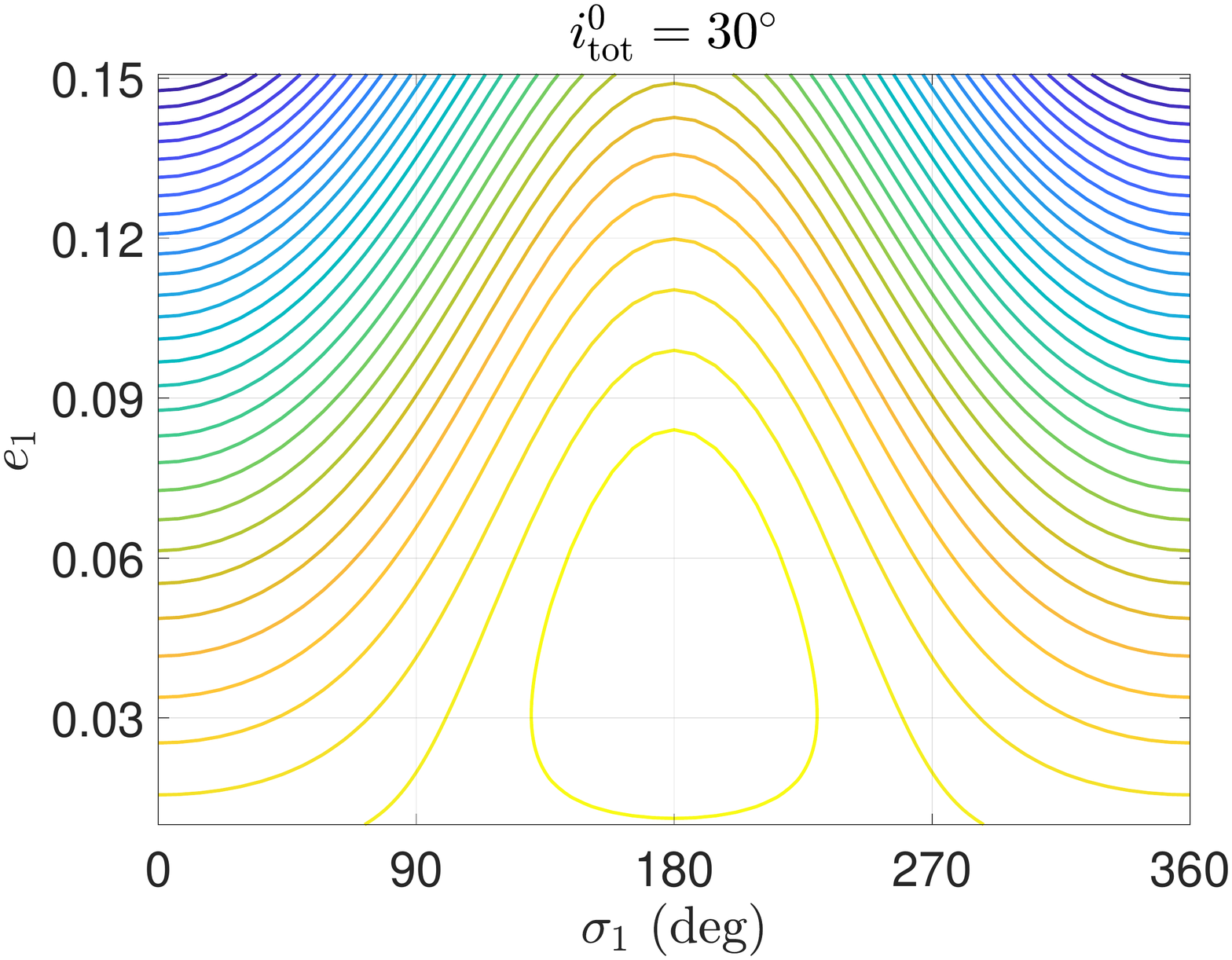}
\caption{Level curves of resonant Hamiltonian shown in the $(\sigma_1,e_1)$ space (i.e., phase portrait) for the low-inclination cases with $i_{\rm tot}^0 = 20^{\circ}$ (\emph{left panel}) and $i_{\rm tot}^0 = 30^{\circ}$ (\emph{right panel}). Practical motions take place along the isolines of resonant Hamiltonian.}
\label{Fig8}
\end{figure*}

Under the resonant model determined by equation(\ref{Eq23}), phase portraits can be produced by plotting level curves of resonant Hamiltonian ${\cal H}^*$ with given motion integral $\Sigma_2$. As discussed above, the constant of motion $\Sigma_2$ can be characterised by $i_{\rm tot}^0$.

In Fig. \ref{Fig7}, four representative (pseudo-) phase portraits with $i_{\rm tot}^0 = 30^{\circ}, 45^{\circ}, 50^{\circ}, 60^{\circ}$ are reported in the $(\sigma_1, i_{\rm tot})$ space. Here $i_{\rm tot}$ is the mutual inclination between inner and outer binaries when the angular variable $g_1$ is equal to zero. In these phase portraits, the level curves stemming from saddle points correspond to the dynamical separatrices, which divides the libration and circulation regions. In addition, dynamical separatrices provide boundaries for libration islands, so they can be used to determine libration zones in the phase space.

When the minimum mutual inclination is at $i_{\rm tot}^0 = 30^{\circ}$ (please refer to the top--left panel of Fig. \ref{Fig7}), the phase portrait has a pendulum-like structure: there is a single island of libration in the phase portrait. The resonant centre is located at $\sigma_1 = 0$ and the saddle point is located at $\sigma_1 = \pi$. There is one separatrix that divides the circulation region from the libration one. The mutual inclinations $i_{\rm tot}$ of the resonant centre and saddle point are close to $90^{\circ}$ but not equal to $90^{\circ}$. Inside this zone, almost all trajectories can flip from prograde to retrograde and back again.

When the minimum mutual inclination is increased up to $i_{\rm tot}^0 = 45^{\circ}$ (please refer to the top--right panel of Fig. \ref{Fig7}), the dynamical structures arising in the phase portrait are complex. There are three islands of libration: one is located at $\sigma_1 = \pi$ and the other two are centred at $\sigma_1 = 0$. There are three saddle points: one is located at $\sigma_1 = 0$ and the other two are located at $\sigma_1 = \pi$. Thus, there are three separatrices, bounding three libration islands. For the current case, those trajectories inside the island of libration centred at $\sigma_1= \pi$ can realize flips between prograde and retrograde. However, those trajectories inside the islands of libration centred at $\sigma_1 = 0$ cannot flip.

When the minimum mutual inclination is further increased up to $i_{\rm tot}^0 = 50^{\circ}$ (please refer to the bottom--left panel of Fig. \ref{Fig7}), dynamical structures are even more complex. There are three resonant centres: one is located at $\sigma_1 = \pi$ and the other two are located at $\sigma_1 = 0$. There are three saddle points: one is located at $\sigma_1 = 0$ and the other two are located at $\sigma_1 = \pi$. There are also three dynamical separatrices, stemming from three different saddle points. The separatrix emanating from the saddle point at $\sigma_1 = 0$ is denoted by the inner separatrix. Evidently, the inner separatrix provides boundaries for the two libration islands centred at $\sigma_1 = 0$. The dynamical separatrices stemming from the remaining two saddle points are called outer separatrices, which provide boundaries for the libration island centred at $\sigma_1 = \pi$. All trajectories inside the island centred at $\sigma_1 = \pi$ can flip from prograde to retrograde and back again. In addition, those phase curves bounded by the inner separatrix and the nearby outer separatrix are also resonant and, in particular, these resonant trajectories can realise flips between prograde and retrograde.

When the minimum mutual inclination is at $i_{\rm tot}^0 = 60^{\circ}$ (please refer to the bottom--right panel of Fig. \ref{Fig7}), there is a main island of libration which is centred at $\sigma_1 = 0$. All the trajectories inside this libration island can realise flips between prograde and retrograde. In addition, two small islands appear at the location of $\sigma_1 = \pi$. However, these latter two islands of libration are easily destroyed in the full dynamical model.

For the low-inclination case, two examples with $i_{\rm tot}^0 = 20^{\circ}$ and $i_{\rm tot}^0 = 30^{\circ}$ are taken into consideration. Figure \ref{Fig8} shows the phase portrait in the $(\sigma_1, e_1)$ space. Here $e_1$ is the eccentricity of the inner binary when the angular variable $g_1$ is equal to zero. Besides the islands of libration shown in the top-left panel of Fig. \ref{Fig7}, we can observe an additional island of libration centred at $\sigma_1 = \pi$ arising in the low-eccentricity region with eccentricities $e_1$ smaller than 0.1.

\begin{figure*}
\centering
\includegraphics[width=0.48\textwidth]{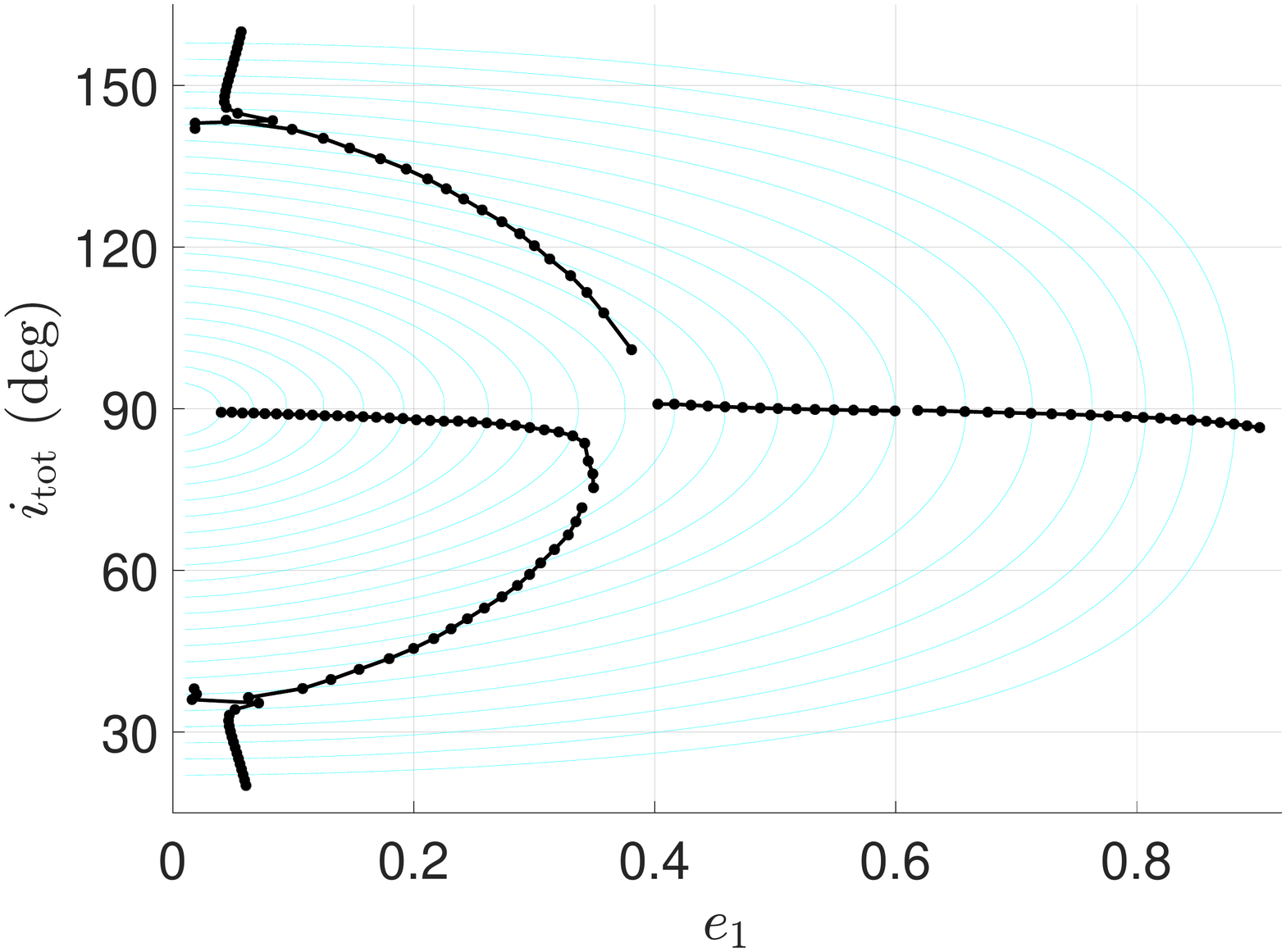}
\includegraphics[width=0.48\textwidth]{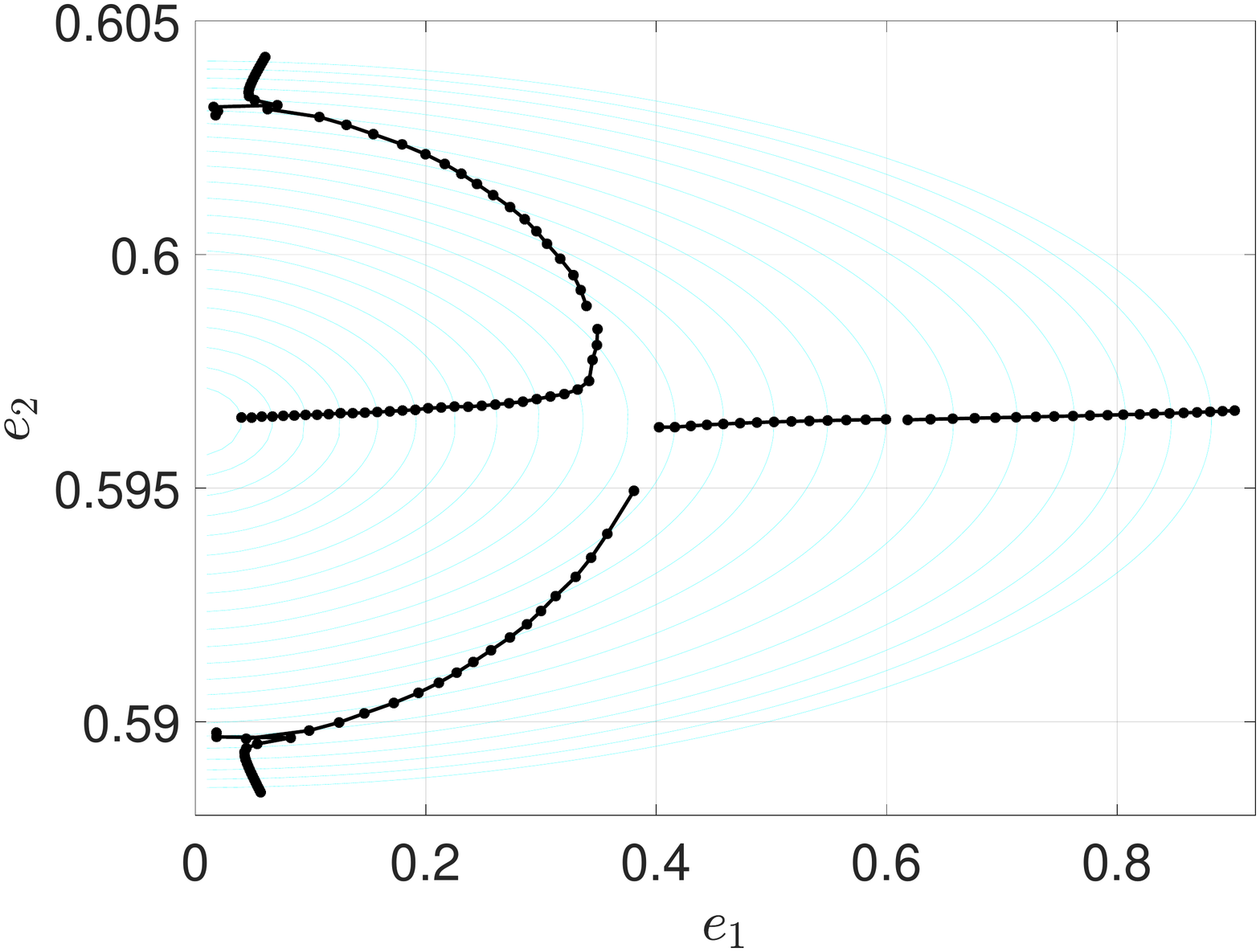}
\caption{Libration centres of octupole-order resonance obtained by analysing phase portraits (black dots). The level curves of $\Sigma_2$ are shown as background. Compared to the nominal location of octupole-order resonance shown in Fig. \ref{Fig6}, two additional branches of libration centre appear in the low-eccentricity region.}
\label{Fig9}
\end{figure*}

Under the resonant model, it is possible to identify the location of resonant centre by solving the following stationary condition,
\begin{equation*}
{\dot \sigma}_1 = \frac{\partial {\cal H}^*}{\partial \Sigma_1} = 0,\quad {\dot \Sigma}_1 = - \frac{\partial {\cal H}^*}{\partial \sigma_1} = 0.
\end{equation*}
In Fig. \ref{Fig9}, the locations of the resonant centre are distributed in the $(e_1, i_{\rm tot})$ space (see the left panel) and in the $(e_1, e_2)$ space (see the right panel). It is noted that $e_1$, $e_2$ and $i_{\rm tot}$ shown in Fig. \ref{Fig9} are the eccentricities and mutual inclination when the angle $g_1$ is equal to zero. In both panels, the level curves of the constant of motion $\Sigma_2$ are shown as background. Besides the two families of libration centres shown in Fig. \ref{Fig6}, there are two additional families of libration centre appearing in the low-eccentricity space. These two new families are present due to the pure effect of the octupole-order Hamiltonian. The distribution of libration centre has no symmetry, which is different from the restricted case \citep{Lei2022dynamical}. Following along the level curves of $\Sigma_2$, the eccentricities ($e_1$ and $e_2$) and the mutual inclinations $i_{\rm tot}$ are in coupled oscillations because of the exchange of orbital angular momentum between the inner and outer binaries. As the outer binary holds the dominant part of the total angular momentum, the variation of $e_2$ is relatively small compared to that of $e_1$.

\begin{figure*}
\centering
\includegraphics[width=0.49\textwidth]{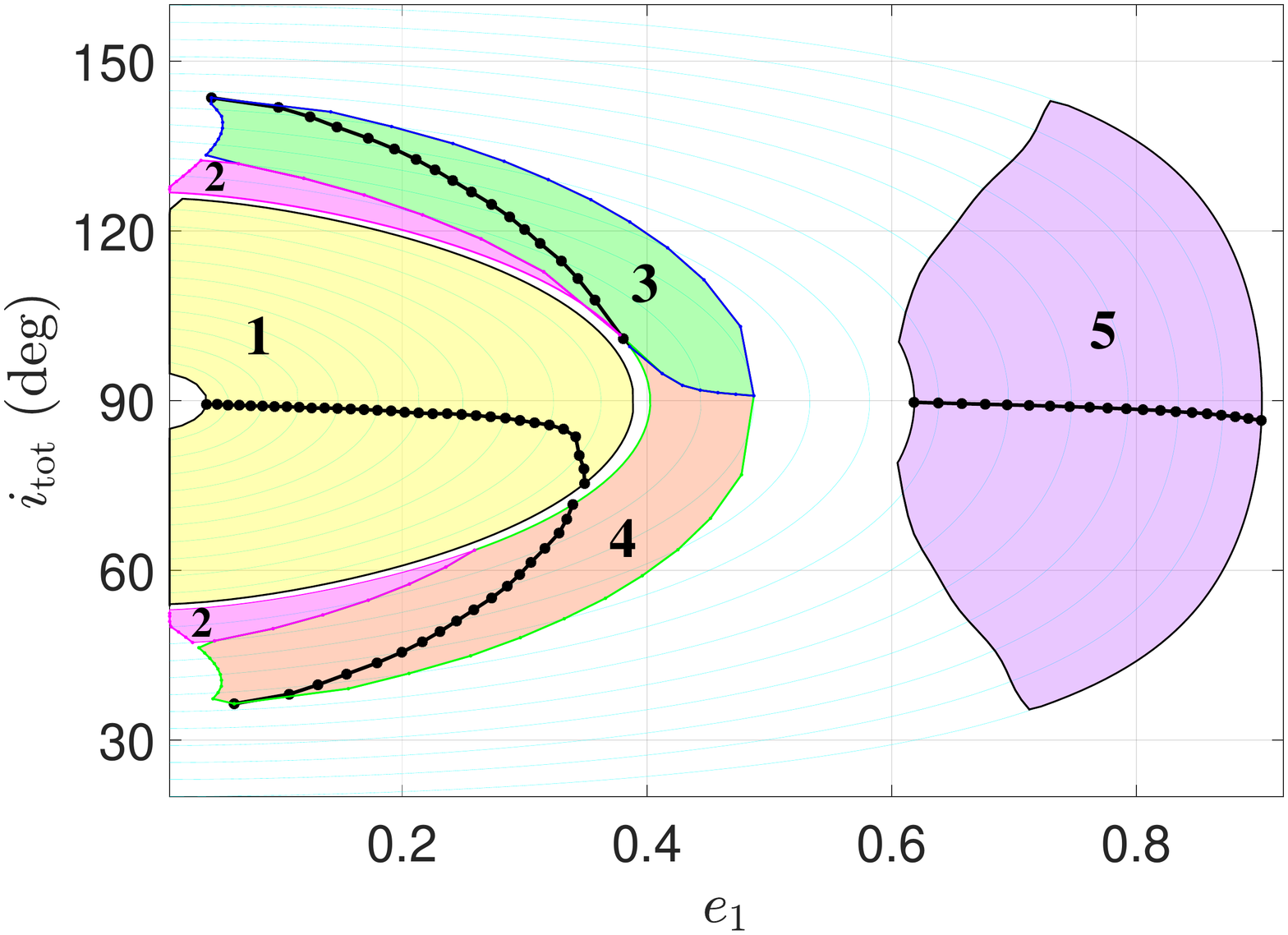}
\includegraphics[width=0.49\textwidth]{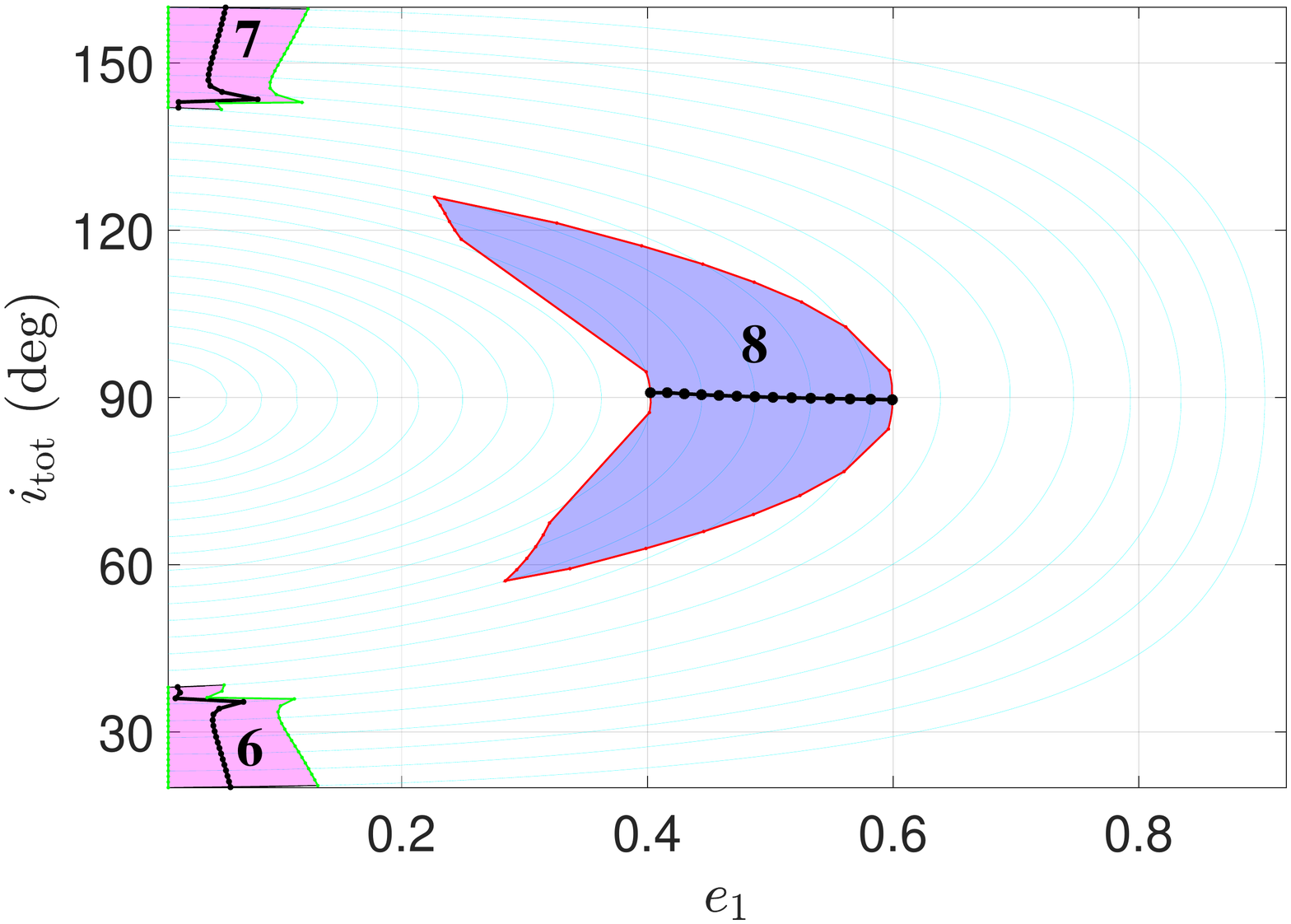}
\caption{Resonant width evaluated at $\sigma_1 = 0$ (\emph{left panel}) and at $\sigma_1 = \pi$ (\emph{right panel}), shown in the $(e_1, i_{\rm tot})$ space. Libration centres are shown in black dots. Libration zones are shown in shaded areas with different colors. The level curves of $\Sigma_2$ are shown as background and resonant width is measured along the isoline of motion integral $\Sigma_2$. For convenience, eight libration zones are denoted by numbers from 1 to 8. At a given motion integral $\Sigma_2$, the resonant motion happens on the isoline of $\Sigma_2$.}
\label{Fig10}
\end{figure*}

In the long-term evolution, the coupled oscillations among $e_1$, $e_2$ and $i_{\rm tot}$ are attributed to the dynamical effect of octupole-order resonance. How large is the libration zones of octupole-order resonance (apsidal resonance) or how strong is the octupole-order resonance (apsidal resonance)? To answer this, we need to evaluate the resonant width. According to the phase portraits shown in Figs \ref{Fig7} and \ref{Fig8}, the dynamical separatrices provide the boundaries for islands of libration. Thus, at a given motion integral, the upper and bottom boundaries can be determined by evaluating the separatrices at the angle of resonant centre. The distance between the boundaries is denoted as resonant width, which corresponds to the maximum size of the island of libration.

The results are reported in Figs \ref{Fig10} for the case of resonant centres at $\sigma_{1,c} = 0$ (see the left panel) and for the case of resonant centres at $\sigma_{1,c} = \pi$ (see the right panel). The libration centres are shown by black dots. For convenience, the level curves of the motion integral $\Sigma_2$ are also shown as background. Resonant width should be measured along the isolines of $\Sigma_2$. There are eight libration zones, which are marked by different colors. For convenience, these libration zones are denoted by numbers from 1 to 8. Evidently, libration zones with numbers from 1 to 5 hold resonant centres at $\sigma_1 = 0$ and the remaining three zones denoted by numbers from 6 to 8 hold resonant centres at $\sigma_1 = \pi$.

Libration zone 1 appears when the minimum mutual inclination $i_{\rm tot}^0$ is greater than $\sim$$53.7^{\circ}$ or the maximum mutual inclination $i_{\rm tot}^0$ is smaller than $\sim$$125.7^{\circ}$. Please refer to the bottom--right panel of Fig. \ref{Fig7} for the representative phase portrait. Inside this zone, the libration centres are located in the region with mutual inclinations smaller than $90^{\circ}$. The zero-eccentricity points $e_1 = 0$ provide the bottom and upper boundaries. It means that the dynamical separatrices are stemming from the points with $e_1 = 0$. Additionally, when $i_{\rm tot}^0$ is approaching $90^{\circ}$, the resonant width in terms of $\Delta e_1$ ($\Delta e_2$ or $\Delta i_{\rm tot}$) decreases. Inside this zone, it is estimated that the maximum variation of $e_1$ is $\sim$0.39 and the maximum variation of $i_{\rm tot}$ is $\sim$$72^{\circ}$. At a given level of $\Sigma_2$ (or the minimum/maximum mutual inclination $i_{\rm tot}^0$), the motion happens on the associated isoline. Thus, it is observed that almost all the trajectories inside zone 1 can flip from prograde to retrograde and back again. However, those trajectories inside a small region near libration centre cannot realise flips.

Libration zone 2 appears in a small interval of $i_{\rm tot}^0$. In the $(e_1, i_{\rm tot})$ space, there are two subregions: one is in the prograde space and the other one is in the retrograde space. Please refer to the bottom--left panel of Fig. \ref{Fig7} for the representative phase portrait. This libration zone is bounded by an inner separatrix and a nearby outer separatrix. The phase curves inside this zone surround the two libration islands centred at $\sigma_1 = 0$ (see the phase portrait for reference). The resonant trajectories inside this zone hold similar shape and behaviours to those horse-shoe trajectories in the co-orbital regions of giant planets in the Solar system. It is known that horse-shoe trajectories surround the libration islands centred at $L_4$ and $L_5$ in phase space \citep{murray1999solar}. In the long-term evolution, the variation of mutual inclination is higher than that of the maximum variation of $i_{\rm tot}$ inside libration zone 1 ($>$$72^{\circ}$). All the trajectories inside zone 2 can realise flips.

Libration zones 3 and 4 occur in the interval of minimum mutual inclination $i_{\rm tot}^0 \in [36.4^{\circ},52.4^{\circ}]$ and in the interval of maximum mutual inclination $i_{\rm tot}^0 \in [127.3^{\circ}, 143.5^{\circ}]$. Please refer to the top--right and bottom--left panels of Fig. \ref{Fig7} for the representative phase portraits. These two zones are bounded and separated by the inner separatrix emanating from the saddle point at $\sigma_1 = 0$. The libration centres of zone 3 are located in the retrograde region and the libration centres of zone 4 are located in the prograde region. In general, libration zone 3 is distributed in the retrograde space and most part of libration 4 is distributed in the prograde space (only a small part of zone 4 is in retrograde region). All the trajectories inside zone 3 cannot realise flips. Whereas, for zone 4, those trajectories inside the zone located in the retrograde space can flip from prograde to retrograde and back again.

Libration zone 5 appear when the minimum mutual inclination $i_{\rm tot}^0$ is smaller than $\sim$$39.5^{\circ}$ or the maximum mutual inclination is greater than $\sim$$140.2^{\circ}$. Inside this zone, the dynamical structures are relatively simple. Please see the top-left panel of Fig. \ref{Fig7} for the representative phase portrait. The libration centres are located in the prograde space. Almost all resonant trajectories inside zone 5 can flip from prograde to retrograde and back again.

Let us move to the libration zones denoted by numbers from 6 to 8, whose resonant centres are at $\sigma_1 = \pi$. Libration zones 6 and 7 are located in the low-eccentricity ($e_1$) regions, and libration zone 8 is located in the intermediate-eccentricity region. In particular, zone 6 is located in the prograde space, zone 7 is located in the retrograde space. For libration zones 6 and 7, it is observed that there are two critical inclinations at $\sim$$36^{\circ}$ and $\sim$$143^{\circ}$, where the resonant width in terms of $\Delta e_1$ is very small. Inside libration 6, the resonant width decreases first and then increases with $i_{\rm tot}$ changing from zero to $\sim$$36^{\circ}$, and when $i_{\rm tot}$ is greater than $\sim$$36^{\circ}$, the resonant width increases with $i_{\rm tot}$ again. Similar behaviour can be found for zone 7. The libration centres inside zone 8 are in the vicinity of $90^{\circ}$. About a half part of zone 8 is in the prograde space and the other part is in the retrograde space. Almost all resonant trajectories inside zone 8 can flip from prograde to retrograde and back again.

\section{Applications to orbit flips}
\label{Sect5}

A trajectory is referred to as a flipping orbit if the mutual inclination $i_{\rm tot}$ can go across the line of $90^{\circ}$. At a given motion integral $\Sigma_2$, the motion happens on the isoline of resonant Hamiltonian. The phase portraits shown in Figs \ref{Fig7} and \ref{Fig8} present all possible types of orbits in phase space and, in particular, those phase curves passing through the line of $i_{\rm tot} = 90^{\circ}$ are flipping orbits. Thus, it is possible to identify the flipping regions by analysing phase portraits\footnote{A systematic study about numerical explorations as well as resonance interpretation for the phenomenon of orbital flips under non-restricted hierarchical planetary systems can be found in Huang et al.~(in prep).}.

\begin{figure*}
\centering
\includegraphics[width=0.49\textwidth]{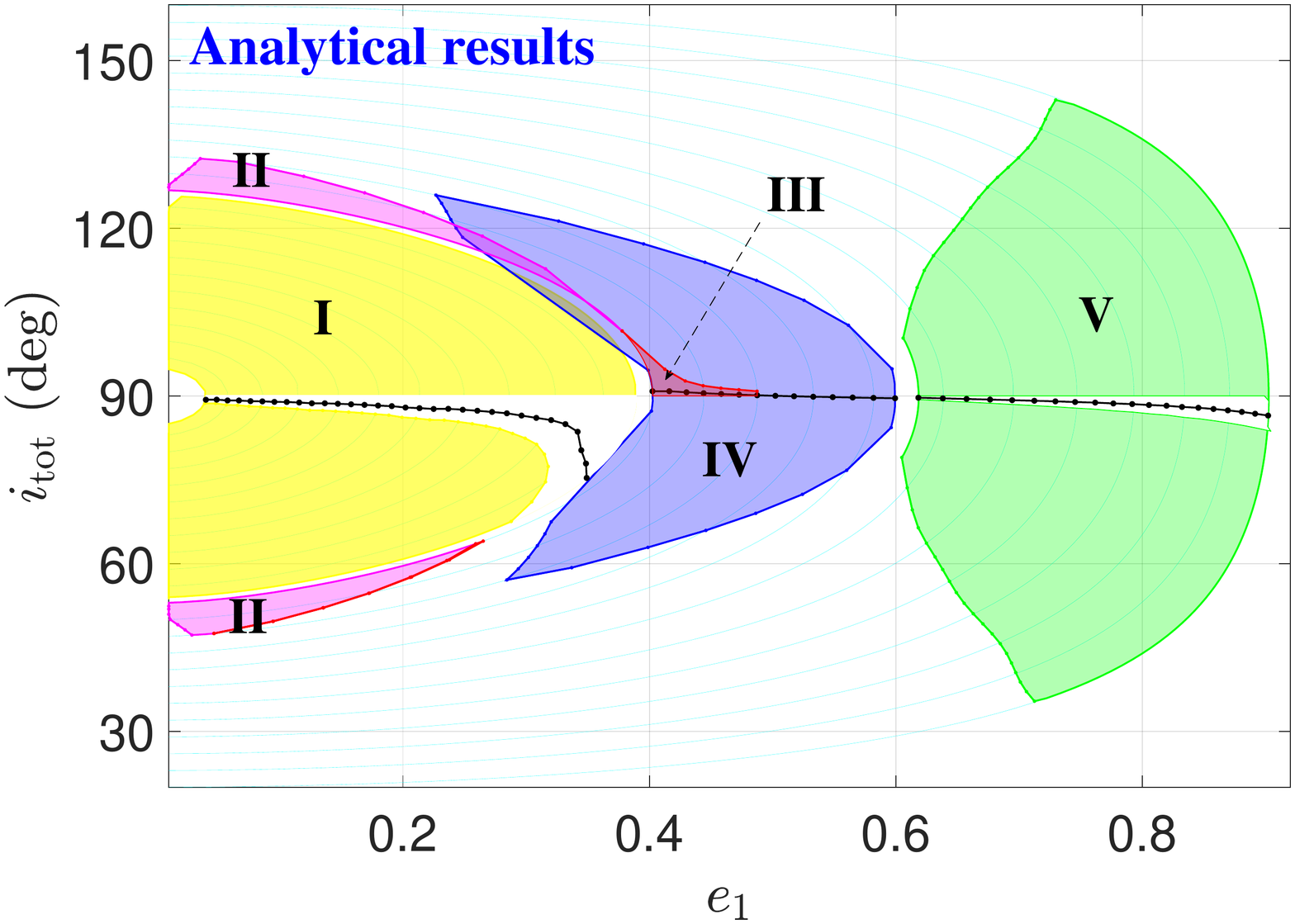}
\includegraphics[width=0.49\textwidth]{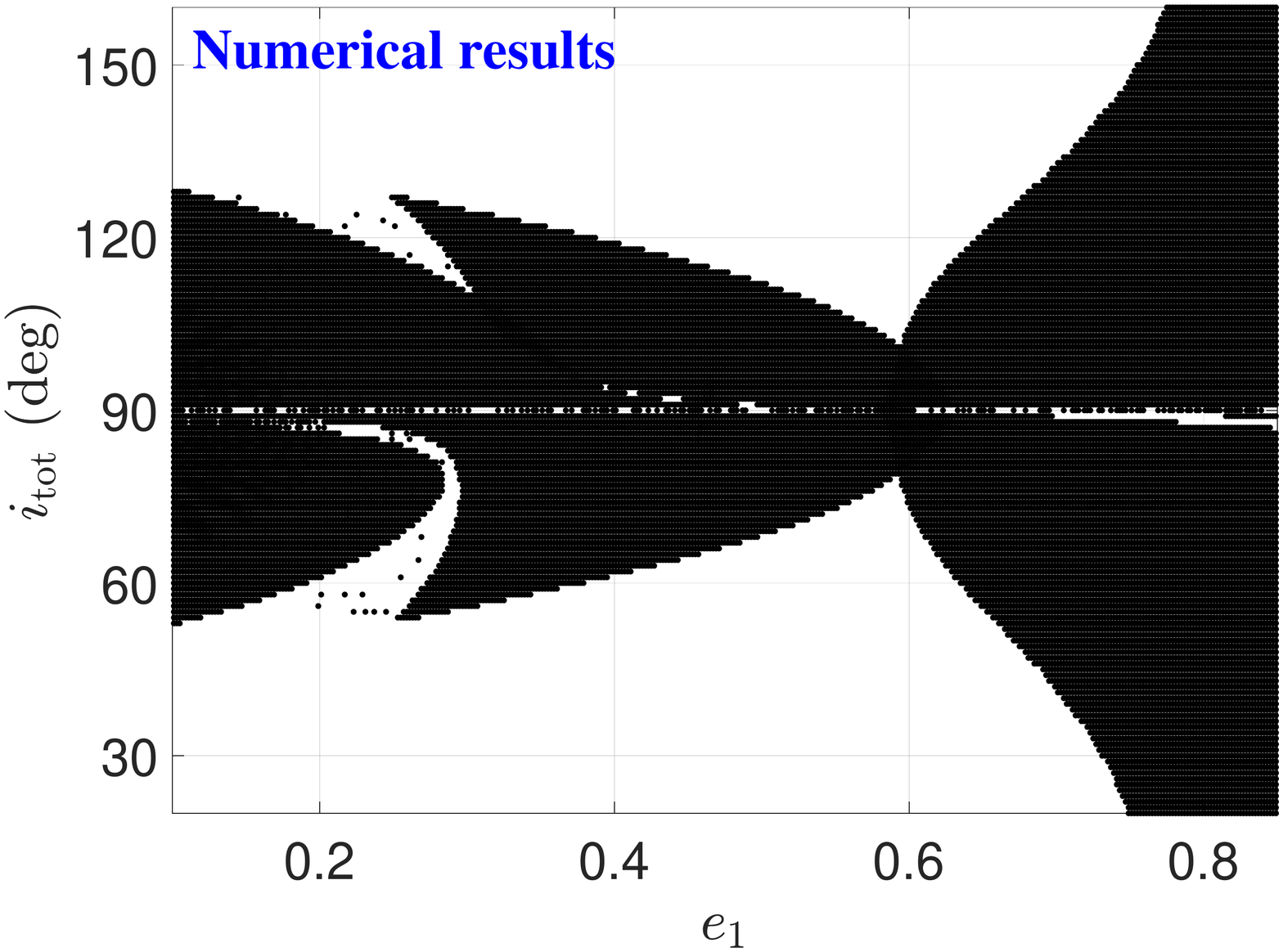}
\caption{Analytical results for libration zones of octupole-order resonance causing orbit flips shown in the $(e_1, i_{\rm tot})$ space (\emph{left panel}) and numerical distribution of flipping orbits in the $(e_1, i_{\rm tot})$ space (\emph{right panel}). In the left panel, libration centres are presented by black dots, and flipping regions are shown in shaded with different colors. There are five flipping regions, denoted by I, II, III, IV and V. Level curves of the motion integral $\Sigma_2$ are shown as background. In the right panel, there are three distinct flipping regions, which are distributed in the low-eccentricity, intermediate-eccentricity and high-eccentricity spaces. In the numerical distribution, some chaotic flipping orbits are found inside the low-eccentricity region. The inconsistence between analytical and numerical results mainly lies in the bottom-left corner of the intermediate-eccentricity flipping region, where the flipping orbits are of circulation.}
\label{Fig11}
\end{figure*}

The analytical results of libration zones causing orbit flips are reported in the left panel of Fig. \ref{Fig11}, where the flipping regions are shown in shaded areas with different colors. There are five flipping regions in the $(e_1, i_{\rm tot})$ space and, for convenience, they are denoted by Roman numbers from I to V. The libration centres are shown in black dots. The level curves of the motion integral $\Sigma_2$ are shown as background.

Flipping region I corresponds to libration zone 1. However, the region in the vicinity of the libration centres shown in libration zone 1 is not included because the resonant trajectories inside it cannot flip. Thus, we can see that the flipping orbits inside region I are resonant trajectories with libration centres at $\sigma_1  = 0$. Flipping region II is exactly equal to libration zone 2. As discussed above, the flipping orbits inside this region have similar dynamical behaviors to those horse-shoe orbits of co-orbital objects. In the phase space, such a kind of resonant orbits surround two libration islands. Flipping region III corresponds to the retrograde part of libration zone 4. Flipping region IV is equal to libration zone 8, which occupies the intermediate-eccentricity space. The critical argument of flipping orbits inside this region are librating around $\sigma_1 = \pi$. Flipping region V corresponds to libration zone 5 but not including the region near the libration centres.

To validate the analytical results, the equations of motion represented by equation(\ref{Eq3_1}) are numerically integrated over 60 million years which is long enough to evaluate the distribution of flipping orbit, and those initial conditions are recorded if the numerical trajectories could flip from prograde to retrograde and back. The initial angles are assumed at $g_1(0) = 0$ and $g_2(0)=0$ or $g_2(0)=\pi$. The numerical results are reported in the right panel of Fig. \ref{Fig11}. Obviously, there are three distinct flipping regions, which are distributed in the low-eccentricity, intermediate-eccentricity and high-eccentricity spaces. Comparing the left and right panels of Fig. \ref{Fig11}, we can see that the analytical and numerical results for flipping regions are qualitatively consistent. According to detailed analysis, we find that the minor inconsistence between the analytical numerical results lies in the following aspects. Firstly, analytical results corresponds to the distribution of regular flipping orbits, but numerical results contain both the regular and chaotic flipping orbits. However, the area of chaotic flipping orbit is very small. Secondly, analytical results corresponds to the distribution of resonant trajectories (because they correspond to libration zones), but in numerical results a small part of flipping orbits are not inside resonance. The non-resonant flipping orbits are mainly distributed in the bottom-left corner of the intermediate-eccentricity flipping region.

\begin{figure*}
\centering
\includegraphics[width=0.48\textwidth]{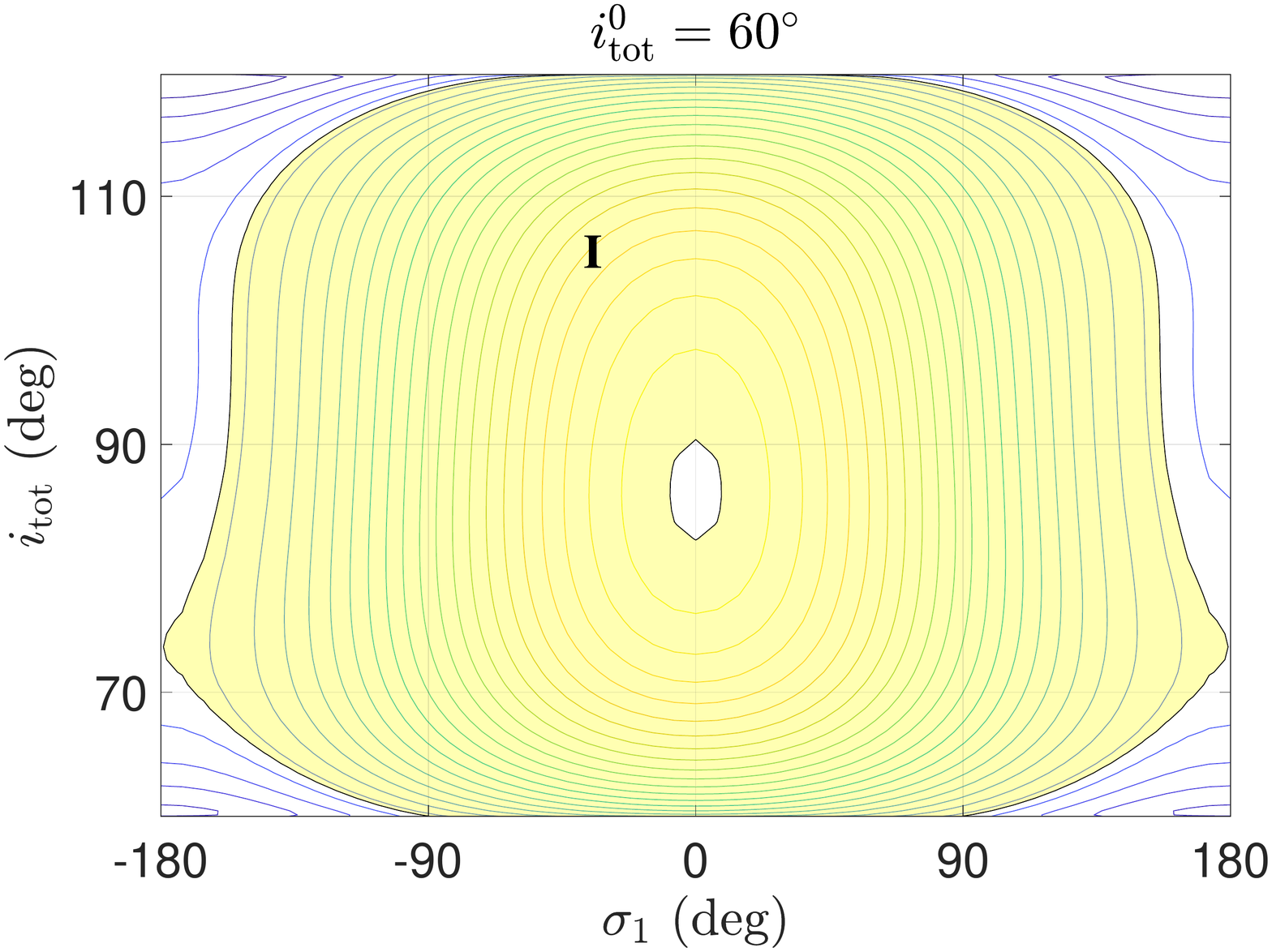}
\includegraphics[width=0.48\textwidth]{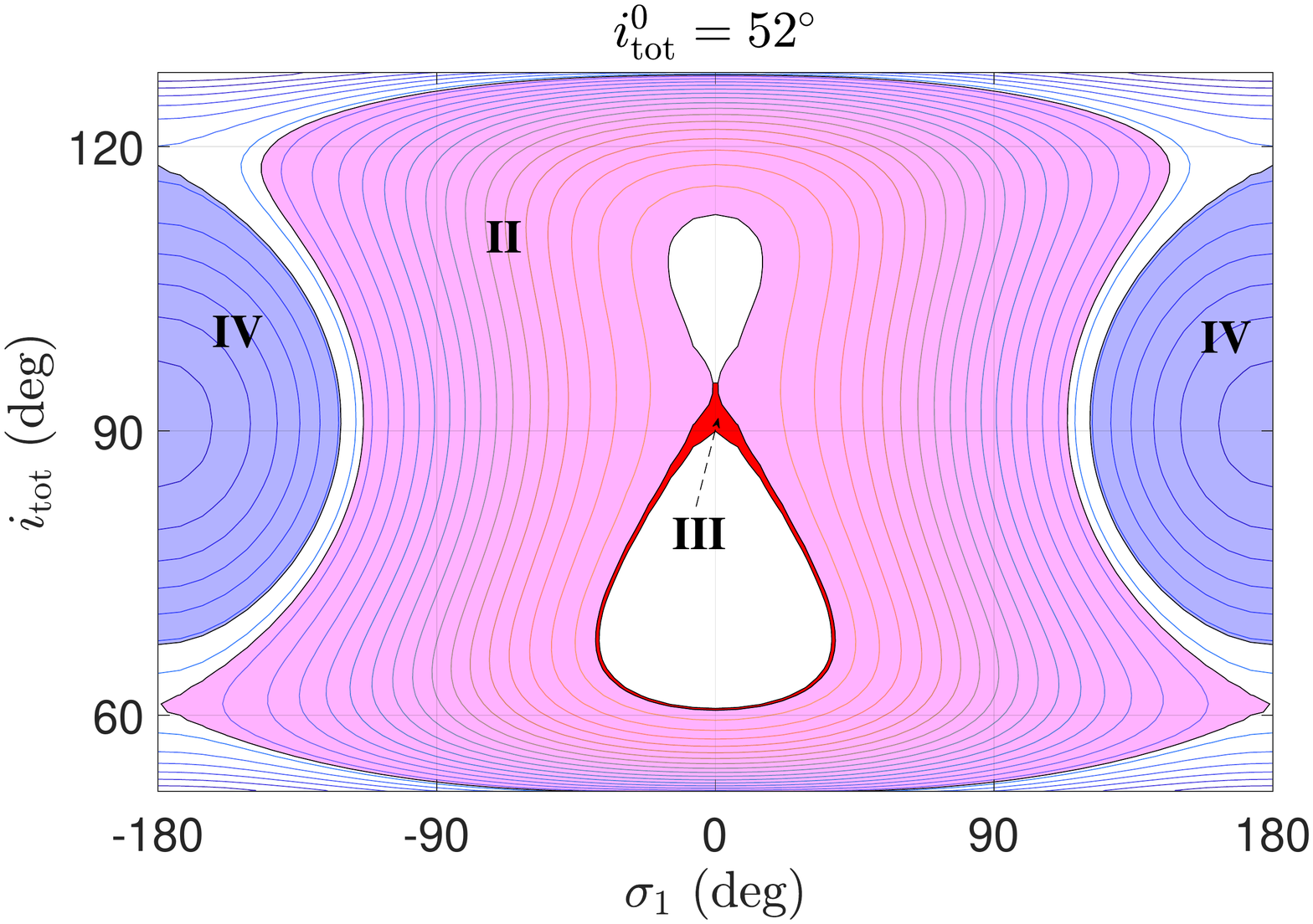}\\
\includegraphics[width=0.48\textwidth]{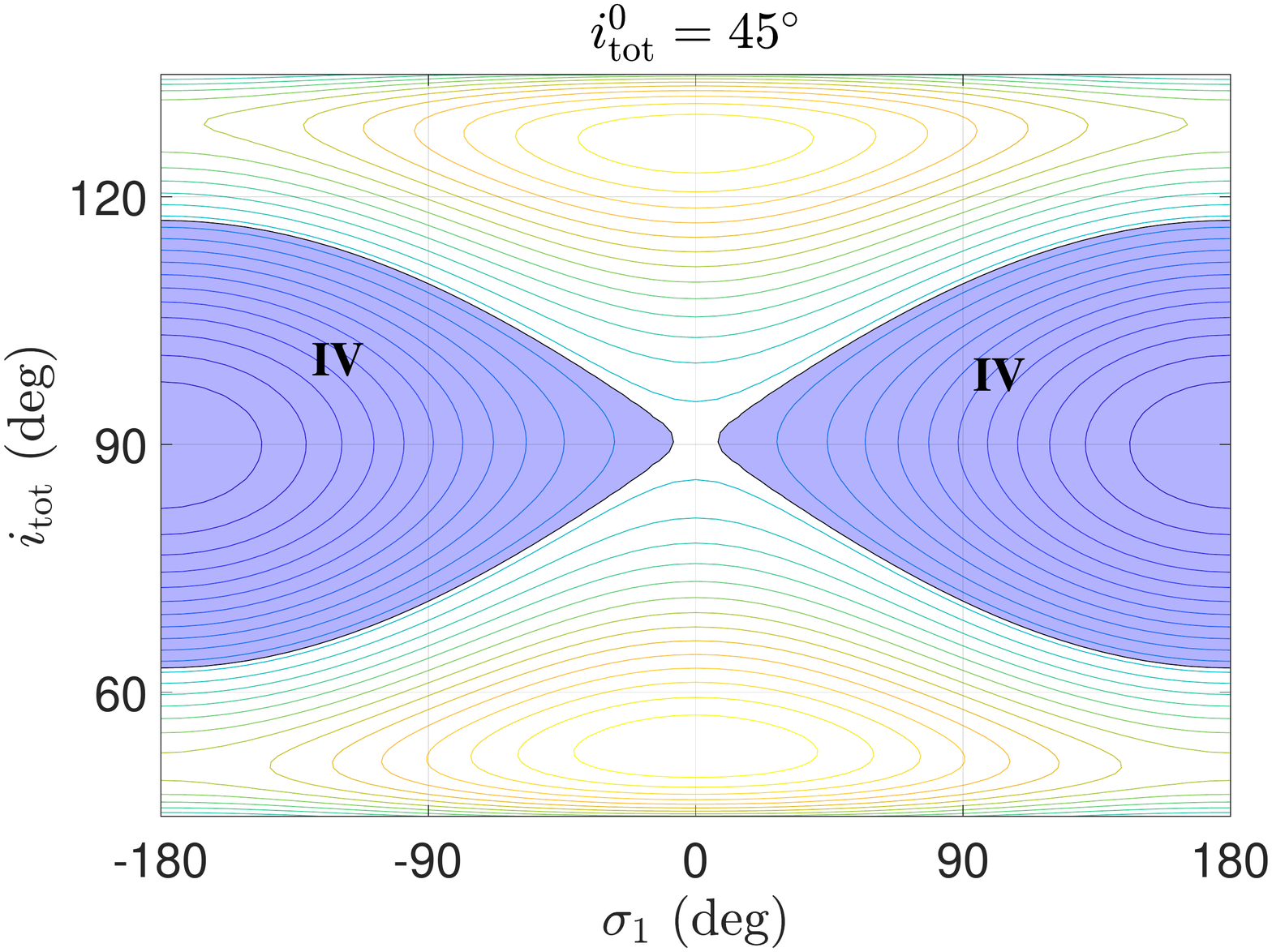}
\includegraphics[width=0.48\textwidth]{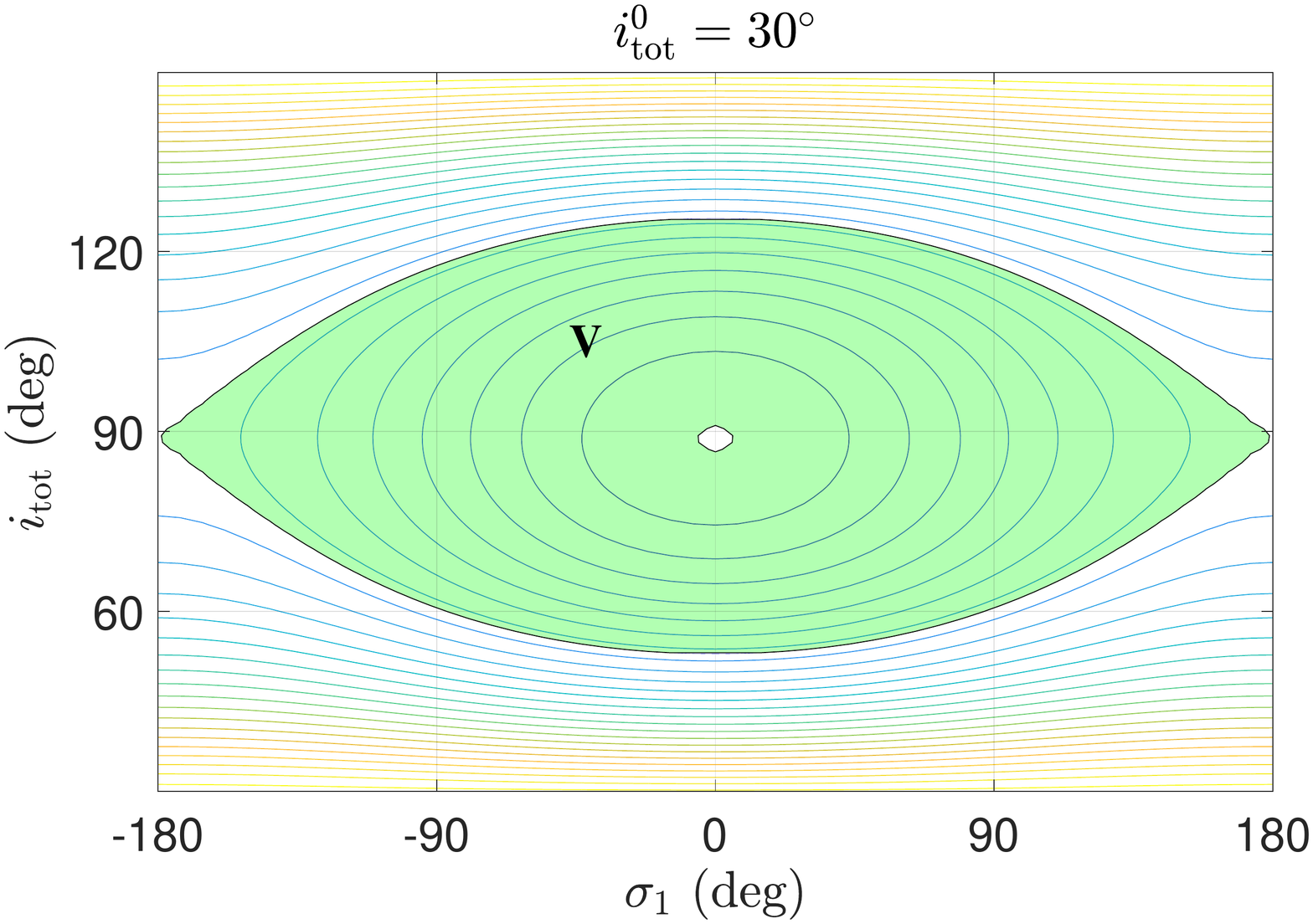}
\caption{Different types of flipping orbit (phase curves shown in the shaded areas) shown in the $(\sigma_1, i_{\rm tot})$ space. The motion integral $\Sigma_2$ is specified by the minimum mutual inclination $i_{\rm tot}^0$. It is noted that the practical trajectories follow along the isolines of resonant Hamiltonian.}
\label{Fig12}
\end{figure*}

In order to make clear the flipping behaviours, let's look at the flipping regions in the phase space by taking several cases of $i_{\rm tot}^0$ as examples. We produce the associated phase portraits, as shown in Fig. \ref{Fig12}. The flipping regions in the phase space are shown in shaded areas with different colors. Flipping orbits inside region I occur in the space with high $i_{\rm tot}^0$ and flipping orbits in region V occur in the space with low $i_{\rm tot}^0$. Please see the top-left panel of Fig. \ref{Fig12} for flipping region I. The central region without flips is removed. The top--right and bottom--left panels show flipping regions II, III and IV. It is observed that flipping region III occupies in a small phase space. The bottom--right panel shows flipping region V. Also, the central region without flips is not included.

\begin{figure*}
\centering
\includegraphics[width=0.48\textwidth]{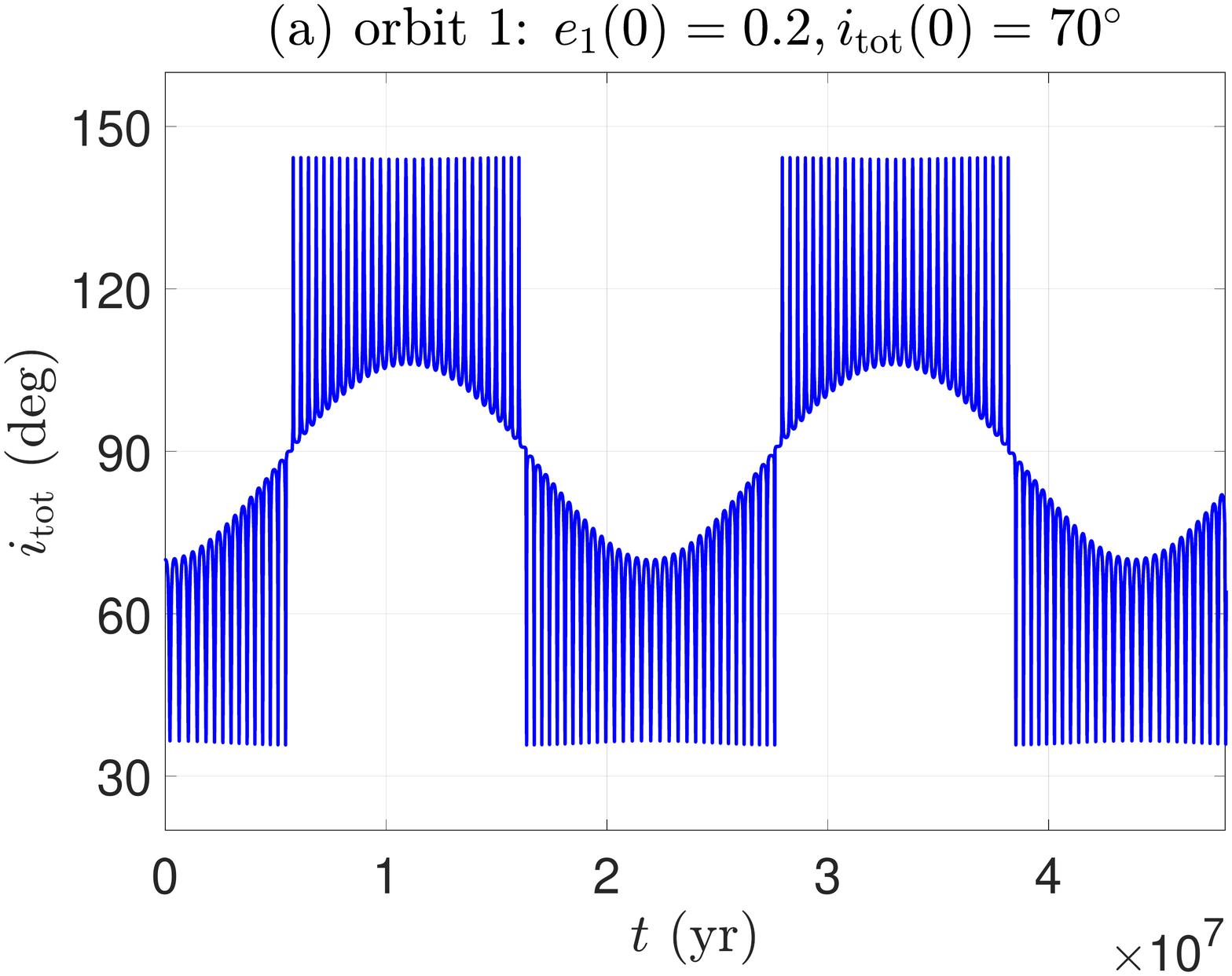}
\includegraphics[width=0.48\textwidth]{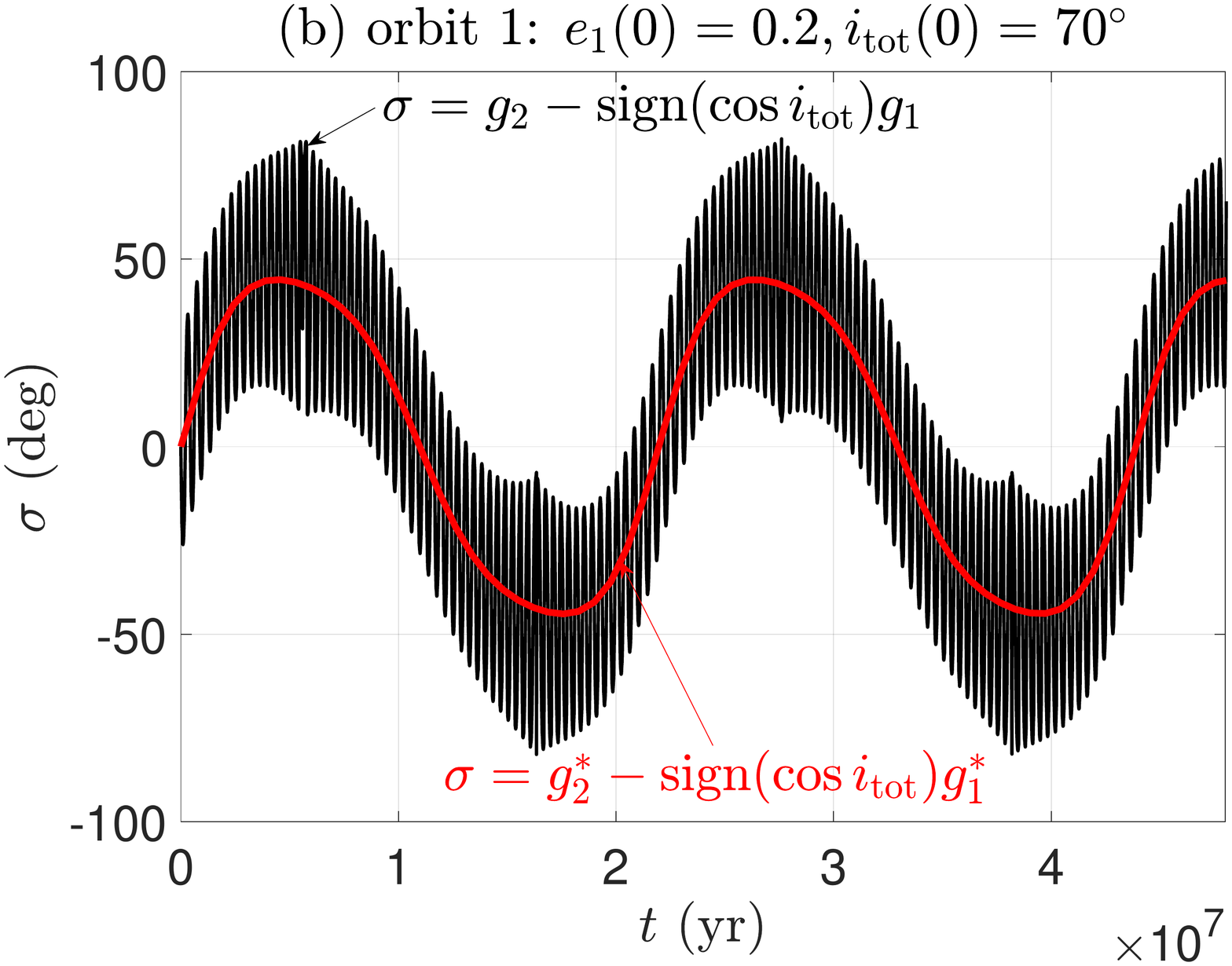}\\
\includegraphics[width=0.48\textwidth]{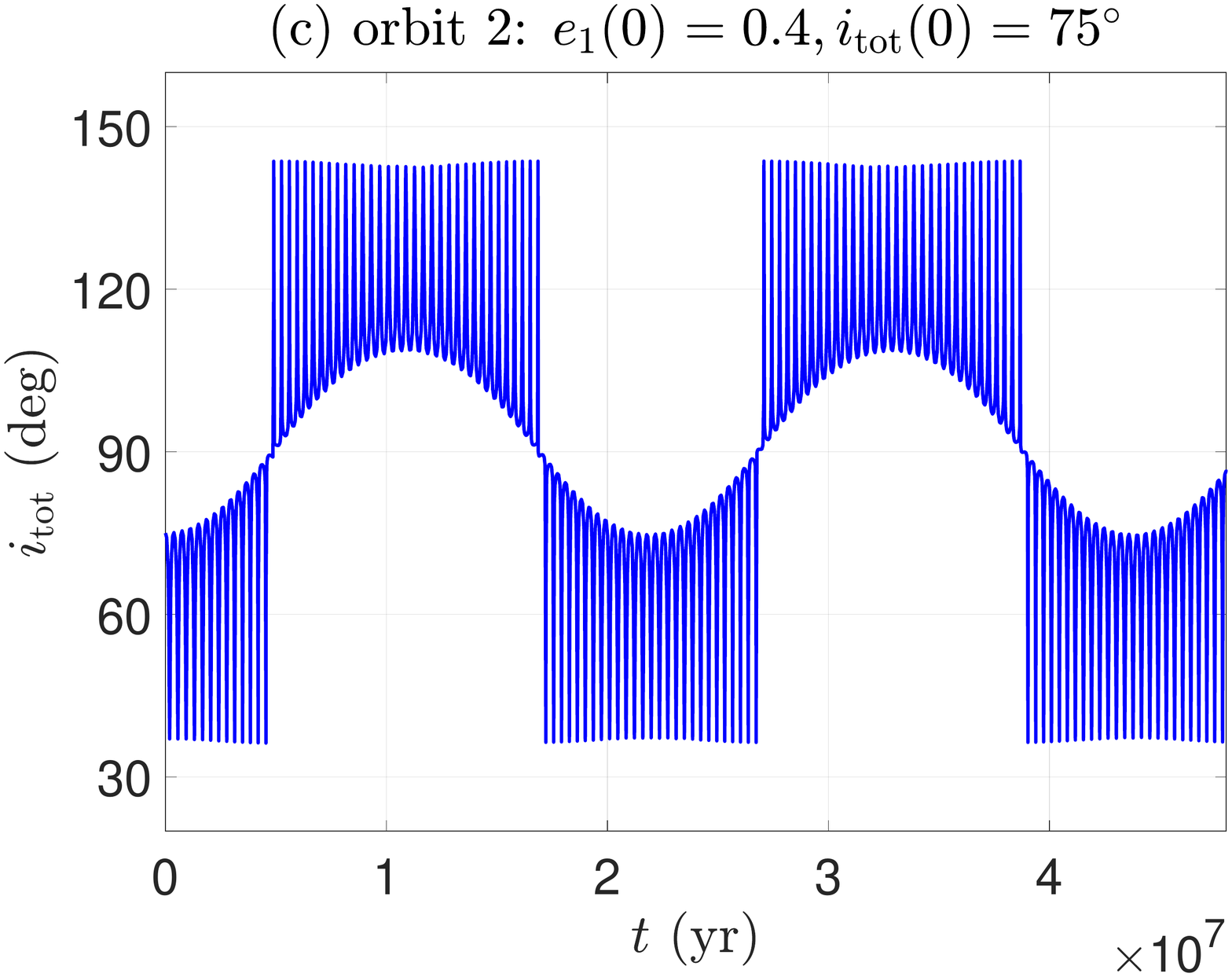}
\includegraphics[width=0.48\textwidth]{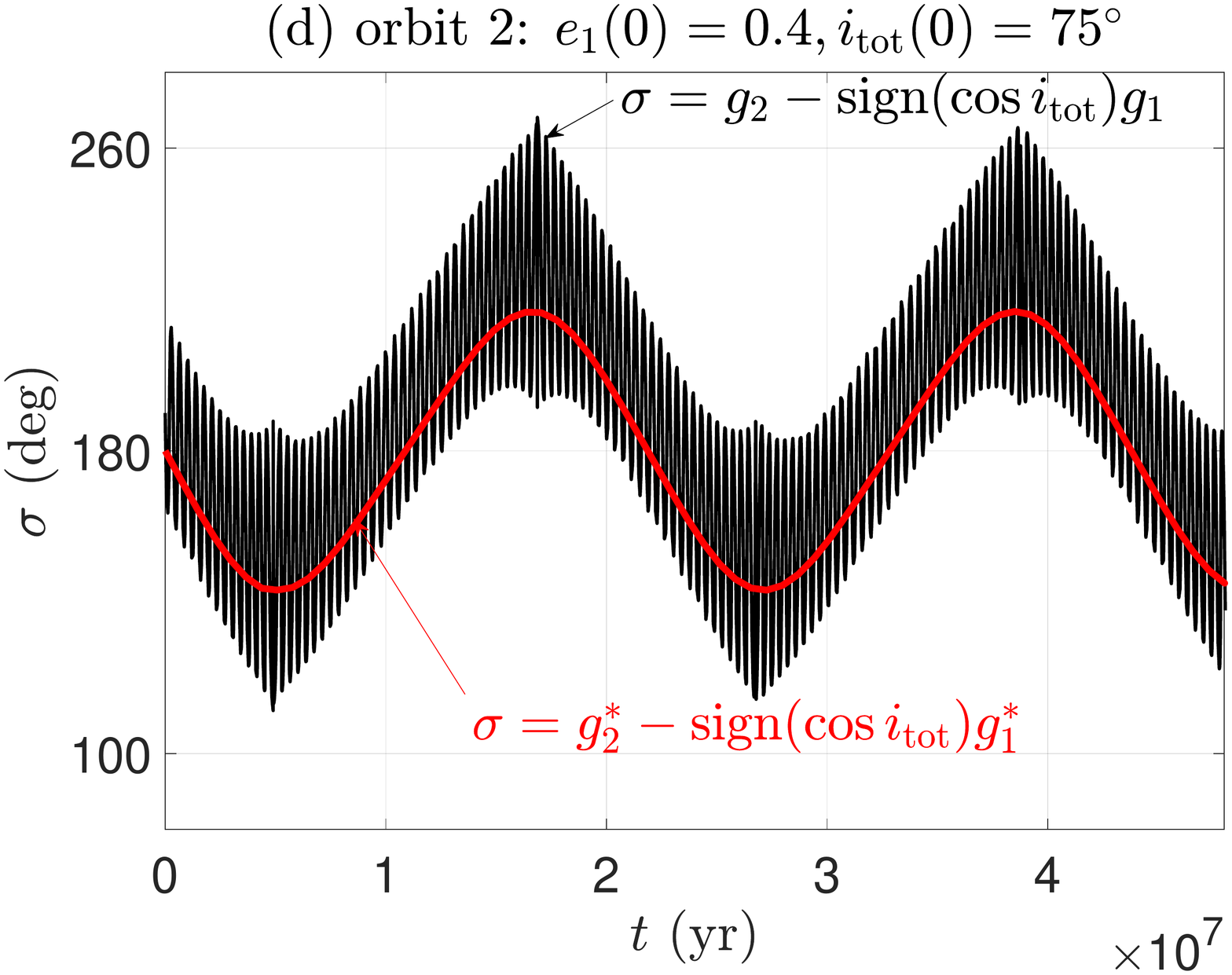}\\
\includegraphics[width=0.48\textwidth]{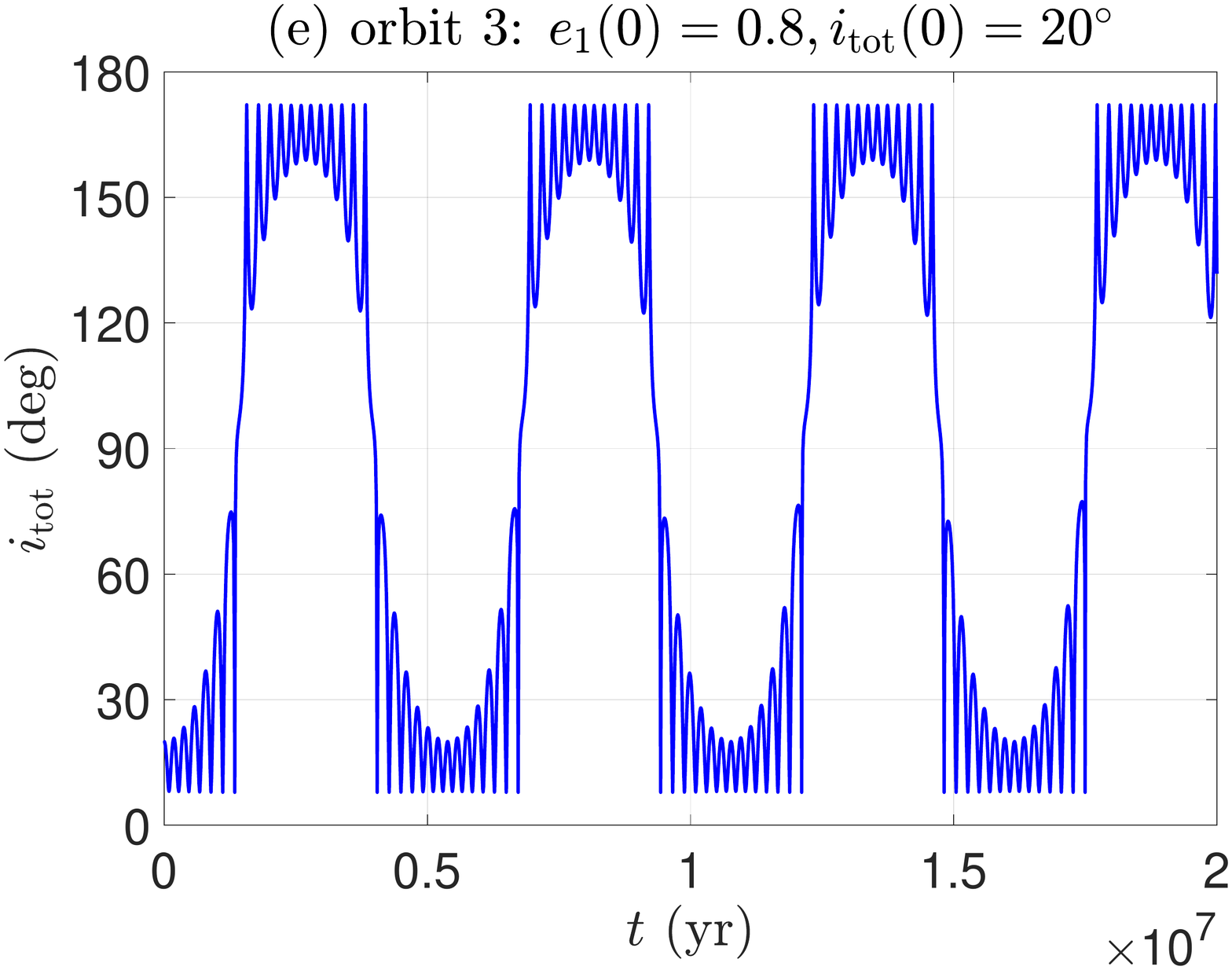}
\includegraphics[width=0.48\textwidth]{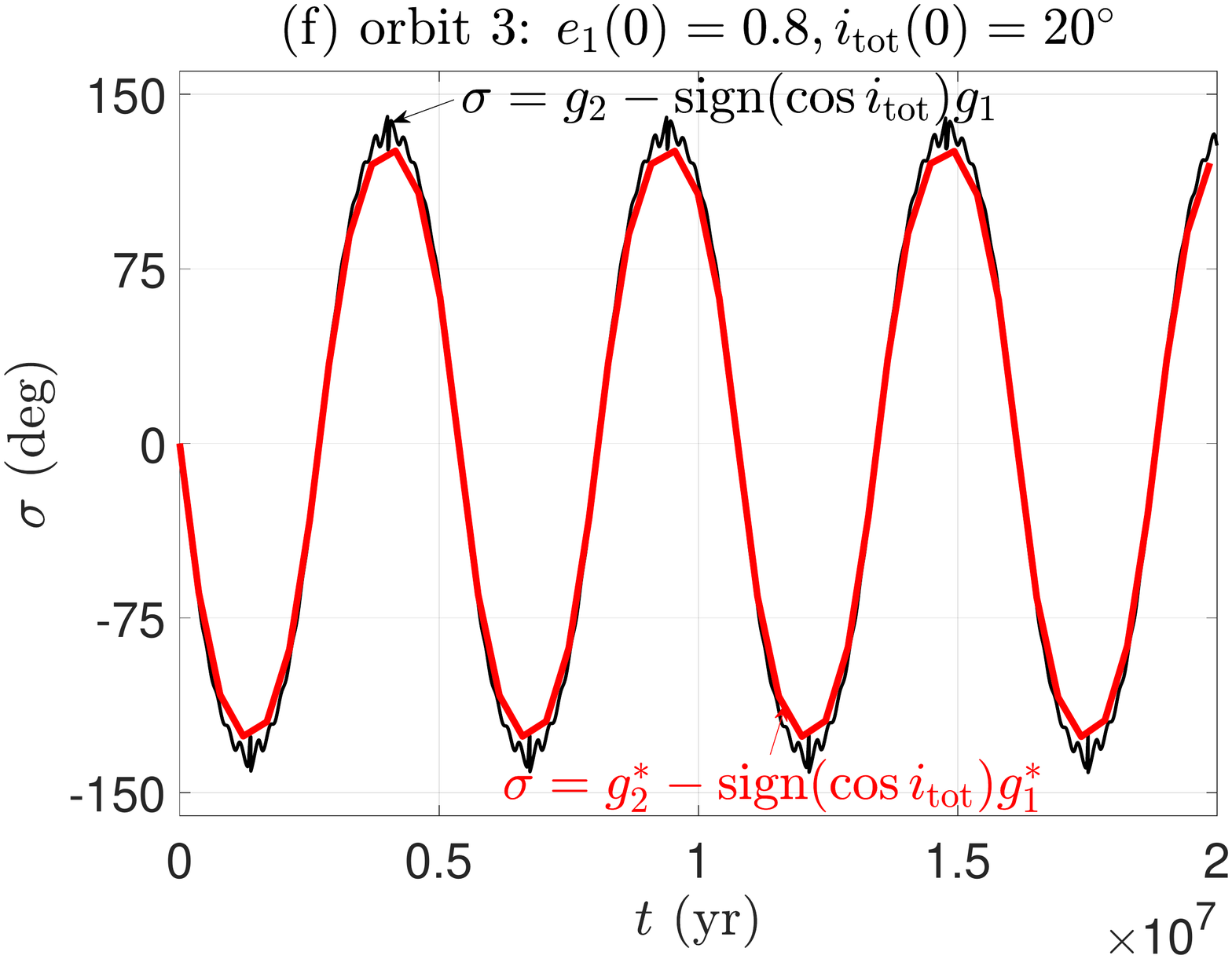}
\caption{Time evolutions of the mutual inclinations and critical arguments for three representative flipping orbits propagated under the Hamiltonian model at the octupole-level approximation. Orbit 1 is located inside flipping region I, orbit 2 is located in flipping region IV and orbit 3 is located in flipping region V. The initial eccentricities and mutual inclinations of flipping orbits are provided at the top of each panel. For orbits 1 and 3, the initial arguments of pericenter are taken as $\omega_1 = \omega_2 = 0$ (corresponding to the resonant center at $\sigma_1 = 0$) and, for orbit 2, the initial arguments of pericenter are assumed at $\omega_1 = 0$ and $\omega_2 = \pi$ (corresponding to the resonant center at $\sigma_1 = \pi$). Regarding the critical argument, both the time histories of $\sigma = g_2 - {\rm sign}(\cos{i_{\rm tot}})g_1$ and $\sigma = g_2^* - {\rm sign}(\cos{i_{\rm tot}})g_1^*$ are presented, where $g_{1,2}$ are the old set of angular coordinates and $g_{1,2}^*$ are the new set of angular coordinates defined by equation (\ref{Eq15}).}
\label{Fig13}
\end{figure*}

To see flipping behaviours, we take three representative examples from three main flipping regions (regions I, IV and V shown in the left panel of Fig. \ref{Fig11}). In particular, orbit 1 is taken from region I with resonant center at $\sigma_1 = 0$, orbit 2 is from region IV with resonant center at $\sigma_1 = \pi$ and orbit 3 is from region V with resonant center at $\sigma_1 = 0$. By numerically integrating the equations of motion under the octupole-order Hamiltonian model, the time evolutions of mutual inclinations as well as critical arguments are reported in Fig. \ref{Fig13} (please refer to the caption for the initial setting). We can see that the flipping timescale of orbit a is about 21.9 million years, flipping timescale of orbit 2 is about 22.2 million years and that of orbit 3 is about 5.38 million years. The panels shown in the right column show the evolutions of critical argument. Here we consider two cases: one is defined with the old set of angular coordinates by $\sigma = g_2 - {\rm sign}(\cos{i_{\rm tot}})g_1$ and the other one is defined with the new set of angular coordinates by $\sigma^* = g_2^* - {\rm sign}(\cos{i_{\rm tot}})g_1^*$. Please refer to equation (\ref{Eq15}) for the transformation between $(g_1,g_2)$ and $(g_1^*,g_2^*)$. It is observed that the curves of $\sigma = g_2 - {\rm sign}(\cos{i_{\rm tot}})g_1$ exhibit short-period oscillations (with timescale of ZLK cycles) along a long-term evolution. While, the curves of $\sigma^* = g_2^* - {\rm sign}(\cos{i_{\rm tot}})g_1^*$ have only long-term evolutions. We can understand that $\sigma^* = g_2^* - {\rm sign}(\cos{i_{\rm tot}})g_1^*$ corresponds to the moving average of $\sigma = g_2 - {\rm sign}(\cos{i_{\rm tot}})g_1$ over one period of ZLK cycles.

\section{Conclusions}
\label{Sect6}

In this work, dynamics of the quadrupole-order resonance as well as the octupole-order resonance are analytically investigated by means of perturbative treatments under the octupole-level approximation in nonrestricted hierarchical planetary systems. In particular, the quadrupole-order resonance corresponds to the ZLK resonance and the octupole-order resonance corresponds to the apsidal resonance. To formulate the dynamical model, the Hamiltonian truncated up to the octupole order in semimajor axis ratio is doubly averaged over the orbital periods of the inner and outer binaries. The resulting averaged Hamiltonian is composed of the quadrupole-order term and the octupole-order term. It determines a two-degree-of-freedom dynamical model, depending on the total angular momentum.

The ZLK resonance is studied under the quadrupole-order dynamical model, where the Hamiltonian ${\cal H} (={\cal H}_2)$, the total angular momentum $G_{\rm tot}$ and the angular momentum of the outer binary $G_2$ are constants of motion. The quadrupole-order dynamical model is integrable. Phase portrait under the quadrupole-order model shows that the centre of the ZLK resonance is located at $2g_1 = \pi$ and the zero-eccentricity points with $e_1 = 0$ (or $G_1 = 1$) are saddle points. Under the quadrupole-order model, we derived analytical expressions for lower boundary, upper boundary and dynamical separatrix for the libration and circulation regions in the parameter space spanned by the conserved quantities $({\cal H}, G_{\rm tot}, G_2)$. The circulation region is bounded by the lower boundary and the separatrix and the libration region is bounded by the separatrix and the upper boundary. The region outside the boundaries corresponds to the physically forbidden zone. Alternatively, the libration and circulation regions are also discussed in the $(i_{\rm tot}, e_1)$ space. It is found that the ZLK resonance may occur when the mutual inclination is greater than $\sim$$39.2^{\circ}$ and smaller than $\sim$$140.8^{\circ}$. This is in line with the conclusion obtained in the test-particle limit. It is known that orbit flips are possible under the quadrupole-order model, which is different from the quadrupole-order dynamics in the test-particle limit. In this work, the flipping orbit is referred to as the one with mutual inclination across the line of $90^{\circ}$. Analytical expressions are derived for boundaries of flipping regions corresponding to rotating ZLK cycles and librating ZLK cycles. It is concluded that (a) orbit flips take place in the parameter space with $G_2 < G_{\rm tot}$ (i.e., in the retrograde space), (b) orbit flips occur in a very small interval of $G_2 - G_{\rm tot}$, and (c) the flipping area corresponding to librating ZLK cycles is larger than that of rotating ZLK cycles.

With inclusion of the octupole-order term in the Hamiltonian, the dynamics becomes complicated. It is noted that the octupole-order Hamiltonian term plays an role of perturbation, which is much smaller than the quadrupole-order Hamiltonian. The octupole-order term has negligible influences upon the dynamics of ZLK resonance, which is dominated by the quadrupole-order Hamiltonian. Thus, for those spaces where the ZLK resonance occurs, it is unnecessary to take the octupole-order Hamiltonian into account. However, for those spaces where the ZLK resonance is absent, the octupole-order resonance appears and it dominates the long-term dynamics. Thus, we focus on the regions filled with rotating ZLK cycles when we are discussing the octupole-order resonances. Without loss of generality, we assume the initial condition of $g_1 = 0$ for rotating ZLK cycles.

To study the dynamical structures of octupole-order resonance, the action-angle transformation is made under the quadrupole-order Hamiltonian flow. After such a canonical transformation, the quadrupole-order Hamiltonian becomes independent on angular coordinates, indicating the conjugate angular momenta are constants of motion and the angles are linear functions of time. The transformed quadrupole-order Hamiltonian yields the fundamental frequencies of system, which can be used to determine the nominal location of secular resonance. It is found that the secular resonance with critical argument of $\sigma = g_2^* - {\rm sign}(\cos{i_{\rm tot}})g_1^*$ takes place in the considered space and there are two branches of libration centre. According to the general definition of longitude of pericentre, we pointed out that the resonance with $\sigma = g_2^* - {\rm sign}(\cos{i_{\rm tot}})g_1^*$ is, in essence, the well-known apsidal resonance.

The resonant Hamiltonian is formulated by means of the lowest-order perturbation theory (i.e., averaging theory). Such an averaging operation gives rise to a new constant of motion. Consequently, the resulting resonant Hamiltonian model is integrable. The global dynamical structures can be analysed by taking advantage of phase portraits. Besides the nominal branches of libration centre, there are two additional branches of libration centre appearing in the low-eccentricity region. By analysing phase portraits, it is found that there are eight libration zones in the considered space. For convenience, they are denoted by numbers from 1 to 8. Thereinto, libration zones denoted by numbers from 1 to 5 hold resonant centres at $\sigma_1 = 0$ and the ones denoted by numbers from 6 to 8 hold resonant centres at $\sigma_1 = \pi$.

It is found the resonant trajectories inside some libration zones can go across the line of $i_{\rm tot} = 90^{\circ}$, meaning that they correspond to flipping orbits. By analysing phase portraits, we identify five flipping regions in the considered space. These flipping regions are part of libration zones, meaning that flipping orbits are inside octupole-order resonance but resonant orbits are not necessarily flipping ones. For convenience, the flipping regions are denoted by Roman numbers from I to V. Thereinto, flipping regions I and II are located in the low-eccentricity space, flipping regions III and IV are distributed in the intermediate-eccentricity space and flipping region V is located in the high-eccentricity space. From the viewpoint of dynamics, flipping regions I, II, III and V correspond to libration zones where the libration centres are at $\sigma_1 = 0$, and the flipping region IV corresponds to the libration zone with libration centres at $\sigma_1 = \pi$. Flipping orbits inside region II hold similar behaviors to horse-shoe trajectories of co-orbital asteroids.

Analytical results of libration zones causing orbit flips are validated by comparing to the numerical distribution of flipping orbits. It is observed that the analytical and numerical results of flipping region are qualitatively consistent.

\section*{Acknowledgments}
This work is supported by the National Natural Science Foundation of China (Nos 12073011).

\section*{Data availability}
The analysis and codes are available upon request.

\bibliographystyle{mn2e}
\bibliography{mybib}


\bsp
\label{lastpage}
\end{document}